\documentclass[11pt,a4paper]{article}
\usepackage[a4paper,margin=2.65cm]{geometry}
\usepackage[T1]{fontenc}
\usepackage[utf8]{inputenc}
\usepackage{lmodern}
\usepackage{amsmath,amssymb,mathtools,amsthm,bm}
\usepackage{microtype}
\usepackage{enumitem}
\usepackage{booktabs}
\usepackage{cite} 
\usepackage{tabularx}
\usepackage{hyperref}
\usepackage[nameinlink,noabbrev]{cleveref}
\allowdisplaybreaks[2]
\newcommand{\R}{\mathbb{R}}
\newcommand{\dd}{\mathrm{d}}
\newcommand{\Div}{\operatorname{div}}
\newcommand{\tr}{\operatorname{tr}}
\newcommand{\grad}{\nabla}
\newcommand{\half}{\tfrac12}

\newcommand{\Acal}{\mathcal A}
\newcommand{\Stoich}{S}
\newcommand{\Ucal}{\mathcal U}
\newcommand{\Fcal}{\mathcal F}
\newcommand{\bv}{\boldsymbol{v}}
\newcommand{\bu}{\boldsymbol{u}}
\newcommand{\bj}{\boldsymbol{j}}
\newcommand{\bq}{\boldsymbol{q}}
\newcommand{\bb}{\boldsymbol{b}}
\newcommand{\bfv}{\boldsymbol{f}}
\newcommand{\bmom}{\boldsymbol{m}}
\newcommand{\bPhi}{\boldsymbol{\Phi}}
\newcommand{\bS}{\boldsymbol{S}}
\newcommand{\bD}{\boldsymbol{D}}
\newcommand{\bI}{\boldsymbol{I}}
\newcommand{\bB}{\boldsymbol{B}}

\newcommand{\blambda}{\boldsymbol{\lambda}}
\newcommand{\bone}{\boldsymbol{1}}
\newcommand{\bP}{\boldsymbol{P}}
\newcommand{\bM}{\boldsymbol{M}}

\newtheorem{proposition}{Proposition}[section]
\newtheorem{theorem}[proposition]{Theorem}
\newtheorem{corollary}[proposition]{Corollary}
\theoremstyle{definition}
\newtheorem{definition}[proposition]{Definition}
\newtheorem{example}[proposition]{Example}
\theoremstyle{remark}
\newtheorem{remark}[proposition]{Remark}
\crefname{proposition}{Proposition}{Propositions}
\Crefname{proposition}{Proposition}{Propositions}
\crefname{theorem}{Theorem}{Theorems}
\Crefname{theorem}{Theorem}{Theorems}
\crefname{corollary}{corollary}{corollaries}
\Crefname{corollary}{Corollary}{Corollaries}
\crefname{definition}{Definition}{Definitions}
\Crefname{definition}{Definition}{Definitions}
\crefname{example}{Example}{Examples}
\Crefname{example}{Example}{Examples}
\crefname{remark}{Remark}{Remarks}
\Crefname{remark}{Remark}{Remarks}
\crefname{section}{Section}{Sections}
\crefname{subsection}{Section}{Sections}
\crefname{appendix}{Appendix}{Appendices}
\crefname{equation}{Equation}{Equations}
\hypersetup{hidelinks,
pdftitle={Continuum thermodynamics of chemically reacting multi-temperature fluid mixtures},
pdfauthor={Dieter Bothe; Tommaso Ruggeri}}
\title{Continuum thermodynamics of chemically reacting\\multi-temperature fluid mixtures}
\author{Dieter Bothe$^{1}$ \quad Tommaso Ruggeri$^{2,3}$\\[0.45em]
\normalsize $^{1}$Department of Mathematics, Technical University of Darmstadt,\\
\normalsize Peter-Gr\"unberg-Str. 10, 64287 Darmstadt, Germany\\
\normalsize $^{2}$Department of Mathematics, University of Bologna, 40123 Bologna, Italy\\
\normalsize Via Saragozza 8, Bologna, Italy\\
\normalsize $^{3}$Accademia Nazionale dei Lincei, Via della Lungara 10, 00165 Rome, Italy}\date{}
\begin{document}

\maketitle

\begin{abstract}
We develop a continuum-thermodynamic model for chemically reacting, multi-velocity and multi-temperature fluid mixtures.
Partial balances of mass, momentum and total energy are closed using a single entropy density depending jointly on the constituent mass and nonkinetic internal-energy densities.
Its partial derivatives determine the constituent temperatures and chemical potentials.
For the chosen entropy flux and separately treated volume deformation, the strengthened entropy principle yields a weighted Euler relation constraining the partial thermodynamic pressures.
We construct thermodynamically consistent constitutive relations for viscosity, heat conduction, Maxwell--Stefan momentum exchange, thermal relaxation and nonlinear reaction kinetics.
The baseline model includes thermo-diffusion and its reciprocal coupling; a reciprocal completion is constructed within a specified isotropic co-factor class. In the companion paper \cite{BotheRuggeriPartII}, we construct a mixed symmetric formulation on the fixed-flux-compatible subclass for coupled or nonlinear constitutive responses satisfying the stated local-representation and entropy-production hypotheses.
An alternative formulation couples chemical conversion and energy redistribution through specified elementary processes.
We identify the material functions required for closure and admissible shifts of constituent energy zeros.
An ionization network with resolved electronic levels yields Saha-type ionization relations and Boltzmann excitation ratios under the stated detailed-balance assumptions, recovering the Saha--Eggert relation at common temperature. An explicit two-level calculation shows how competing electron- and heavy-particle processes maintain stationary populations while transferring energy between groups at different prescribed temperatures.
\end{abstract}

\section{Introduction}
\label{sec:introduction}

In the constituent-balance formulation of fluid mixtures associated with Truesdell~\cite{Truesdell1969}, interpenetrating constituents satisfy partial mass, momentum, and energy balances with interaction productions. We retain these balances while allowing a nonadditive entropy and constitutive dependence on the full mixture state. A systematic common-temperature theory was developed by I.~M\"uller~\cite{Mueller1968,Mueller1985}.

We use the Class-I/II/III hierarchy of Hutter and J\"ohnk~\cite[Section~7.1]{HutterJoehnk2004} and Bothe and Dreyer~\cite{BotheDreyer2015}. All classes retain partial mass balances. Class~I uses one mixture momentum and one mixture energy balance; Class~II resolves partial momenta but retains one mixture energy balance. Both use a common temperature $T$. Class~III also resolves partial energies and permits individual temperatures $T_i$. Class~I allows nonzero diffusion fluxes, for example through Fick--Onsager or Maxwell--Stefan relations; Class~II retains relative-motion dynamics, and Class~III additionally resolves interconstituent thermal nonequilibrium. This hierarchy specifies fields and balances; constitutive relations and reductions require further assumptions.

Our primitive thermal variables are the balanced nonkinetic internal-energy densities $\rho_i e_i$, with entropy conjugates $1/T_i$. Diffusion kinetic energy remains a resolved mechanical contribution. The entropy may depend jointly on all constituent energy and mass densities without being constituent additive. Under the stated stability and positivity assumptions, the temperatures locally parametrize the thermal state at fixed densities.

Multi-temperature continuum models were developed in~\cite{DunwoodyMueller1968,BowenGarcia1970}, followed by energy-equivalent average-temperature and reacting multi-temperature gas models~\cite{GouinRuggeri2008,BisiMartaloSpiga2011Reacting,BisiMartaloSpiga2013Reacting}. In rational extended thermodynamics (RET), Ruggeri and Simi\'c~\cite{RuggeriSimic2007} studied hyperbolic multi-temperature Eulerian mixtures and their single-temperature principal subsystem. Conforto and Martal\`o~\cite{ConfortoMartalo2024} analyse detonation waves in a four-species reacting multi-velocity, multi-temperature Euler model derived from a Grad moment system. Galilean covariance and the entropy principle for mixture-type balance systems were analysed in~\cite{RuggeriSimic2007,ArimaRuggeriSugiyamaTaniguchi2017}. Multi-temperature and dissipative polyatomic mixtures are discussed in~\cite[Chapters~28 and~31]{RuggeriSugiyama2021}.

Rajagopal and Tao~\cite{RajagopalTao1995} develop the broader theory of interacting constituents. For chemically nonreacting binary mixtures at common temperature, Sou\v{c}ek, Pr\r{u}\v{s}a, M\'alek, and Rajagopal~\cite{SoucekEtAl2014} derive restrictions on thermodynamic potentials and fluxes from compatibility of constituent and mixture descriptions. Their setting differs from Class~III.

Kinetic theory complements the continuum construction by determining constitutive relations and exchange coefficients. Desvillettes, Monaco, and Salvarani~\cite{DesvillettesMonacoSalvarani2005} obtain reacting polytropic Euler equations in the hydrodynamic limit from reacting polyatomic mixtures with a continuous microscopic internal-energy variable, which is distinct from separate macroscopic constituent temperatures. For nonreacting polyatomic Euler mixtures, Pavi\'c-\v{C}oli\'c~\cite{PavicColic2019} computes momentum and energy exchange and relates it to near-equilibrium phenomenological coefficients. For nonreacting Euler mixtures, \v{C}oli\'c~\cite{Colic2024} derives nonlinear interaction productions and closures for different velocity and temperature specializations. Such kinetic results, experiments, or other material information are needed to calibrate the continuum model.

We construct a closed continuum-thermodynamic Class-III model for chemically reacting multicomponent fluids, following Bothe and Dreyer~\cite{BotheDreyer2015} and its sharp-interface extensions~\cite{BotheSharpInterface2022,BotheSharpClassII2025}, with the balance-anchoring Principle~(V) introduced in~\cite{BothePolymer2026}. Balances, thermodynamic state, constitutive dependencies, and dissipative mechanisms are specified before entropy exploitation. The minimal \emph{baseline model} retains thermo-diffusion and its reciprocal effect. Further cross-effects compatible with the state, conservation constraints, and material symmetries are admitted through specified coupled blocks with nonnegative entropy production. Entropy-invariant mixing followed by nonnegative closure of the resulting binary products implements these couplings in Section~\ref{subsec:thermodiffusion-mixing} and the reciprocal extension linear in the selected co-factors in Appendix~\ref{app:full-coupling}.

The common entropy determines $1/T_i$ and $-\mu_i/T_i$ as derivatives. The mechanical pressures $P_i=-\tr\bS_i/3$ belong to the isotropic stresses, whereas the thermodynamic pressures $p_i$ are additional state functions defined by the  baseline model's zero-rate scalar stress response. For the selected entropy flux and separately dissipative volume deformation, common expansions and compressions yield an Euler relation constraining $\sum_i p_i/T_i$, but not its individual terms. The reduced entropy production consists of binary products representing mechanisms or coupled blocks selected before closure. Unqualified constituent sums run from $1$ to $N$.

Multi-temperature mass-action and Saha-type relations have been derived under different thermal assumptions~\cite{MorroRomeo1986,MorroRomeo1988,MorroRomeoSaha1988,VanDeSandenEtAl1989,VanDeSandenSchram1991,AnnaloroEtAl2017}. The dependence of two-temperature ionization relations on imposed thermodynamic constraints was emphasized by Giordano and Capitelli~\cite{GiordanoCapitelli2002}. Our alternative source formulation in Section~\ref{subsec:general-event-channels} assigns each resolved elementary reaction or energy-exchange process simultaneous forward and backward rates, a composition change, and zero-sum objective-energy increments per mole of progress. At unequal temperatures, the energy distribution enters the process entropy change and hence the ratio of its directional rates.

The baseline model contains constituent shear and volume viscosity, heat conduction with reciprocal thermo-diffusion, barycentrically constrained Maxwell--Stefan momentum exchange, reciprocal-temperature-driven thermal relaxation, and nonlinear forward--backward reaction kinetics. If reaction-induced momentum transfer is represented separately, it is distinguished from the remaining interaction terms before jointly testing reaction, momentum transfer, and energy transfer against the entropy inequality.

The model is formulated for strictly positive constituent mass densities and temperatures, $\rho_i>0$ and $T_i>0$, within the thermodynamic domain specified below. Boundary constitutive relations and radiation or electromagnetic field balances are outside the present scope.

Section~\ref{sec:balances} gives the balance identities and objective exchanges; Sections~\ref{sec:entropy-principle}--\ref{sec:closure-philosophy} specify and exploit the entropy principle and its constrained force spaces; Sections~\ref{sec:closures}--\ref{sec:closed-model} give the baseline model, alternative sources, and thermodynamic-consistency theorem. Section~\ref{sec:saha-overview} introduces the resolved-level ionization example derived in Appendix~\ref{sec:saha}. Appendix~\ref{app:full-coupling} develops the reciprocal completion within the specified isotropic class linear in the selected co-factors, and Appendix~\ref{app:liu-consistency} recovers the entropy conjugates and residual inequality by Liu's method.

The contribution is a constitutive synthesis for reacting Class-III mixtures with a joint entropy potential, a specified pressure and entropy-flux representation, and explicit conservation-compatible source reconstruction. It extends the common-temperature framework while separating thermodynamic restrictions from kinetic data.   PDE well-posedness and convergence in dynamical reduction limits are not established here. The companion paper~\cite{BotheRuggeriPartII} characterizes the additional constituentwise Gibbs--Duhem conditions for generating potentials of the prescribed constituent Euler fluxes. Under its stated local hypotheses, these conditions select a locally additive-entropy subclass with compatible partial pressures. For this subclass and the  specified constitutive laws of the baseline model, it constructs a complete degenerate symmetric main-field formulation, including zero-storage constitutive equations, and algebraic principal subsystems. Its separate Class-III-to-Class-II construction by constrained entropy maximization applies locally also to nonadditive parent entropies, under the stated domain and attainment assumptions. Neither principal restriction nor entropy maximization alone establishes dynamical invariance or convergence in a relaxation limit.

\section{Balance equations and objective exchange variables}
\label{sec:balances}

\subsection{Constituent balances}

We consider a fluid mixture occupying a region $\Omega\subset\R^3$ and composed of $N$ molecularly miscible constituents $A_1,\ldots,A_N$. For constituent $i$, let
\[
  \rho_i,\qquad \bv_i,\qquad e_i
\]
denote partial mass density, velocity, and nonkinetic internal energy per unit mass of $A_i$, with $\rho_i>0$ throughout the interior active-species domain. For velocity vectors, $\bv_i^2:=\bv_i\cdot\bv_i$. The caloric description includes the resolved formation, binding, ionization, and excitation energies. We call this nonkinetic energy thermal energy. The velocity fields carry kinetic energy, while constituent energies and conservative sources represent chemical energy conversion. A \emph{constituent} is a thermodynamic species: electronic levels may thus be distinct species with equal charge and model mass but different internal-state energies (\cref{sec:saha}). Denote the partial Cauchy stress, nonconvective energy flux, and body force per unit mass by $\bS_i$, $\bq_i$, and $\bb_i$. The partial stresses are symmetric under the nonpolar mechanical assumption in Specification~(M3). Their isotropic--deviatoric decomposition is
\begin{equation}
  \bS_i=-P_i\bI+\bS_i^{\circ},
  \qquad
  P_i:=-\frac13\tr\bS_i,
  \qquad
  \tr\bS_i^{\circ}=0.
\label{eq:stress-decomposition}
\end{equation}
Thus $P_i$ is the mechanical partial pressure. The partial thermodynamic pressures $p_i$ are introduced with the constitutive specification in \cref{sec:model-class-specification}.

Constitutive fluxes and exchange terms may depend on the full mixture state and on gradients, introducing couplings between constituent fields.

The partial balances of mass, momentum, and total energy are
\begin{align}
\partial_t\rho_i+\Div(\rho_i\bv_i)&=r_i,
\label{eq:mass-balance}
\\
\partial_t(\rho_i\bv_i)+\Div(\rho_i\bv_i\otimes\bv_i-\bS_i)
&=\rho_i\bb_i+\bfv_i,
\label{eq:momentum-balance}
\\
\partial_t\big(\rho_i e_i+\half\rho_i\bv_i^2\big)
+\Div\Big[
\big(\rho_i e_i+\half\rho_i|\bv_i|^2\big)\bv_i
-\bS_i\bv_i+\bq_i
\Big]
&=\rho_i\bb_i\cdot\bv_i+k_i.
\label{eq:energy-balance}
\end{align}
Here $r_i$, $\bfv_i$, and $k_i$ are the mass, momentum, and total-energy productions due to interactions among the constituents. Conservation of total mass, momentum, and energy requires
\begin{equation}
  \sum_i r_i=0,
  \qquad
  \sum_i\bfv_i=\boldsymbol 0,
  \qquad
  \sum_i k_i=0.
\label{eq:source-conservation}
\end{equation}

The mass productions result from $N_R$ bidirectional reaction paths,
\begin{equation}
  \sum_i\alpha_i^a A_i\rightleftharpoons\sum_i\beta_i^a A_i,
  \qquad
  \nu_i^a:=\beta_i^a-\alpha_i^a,
  \qquad a=1,\ldots,N_R.
\label{eq:reaction-network}
\end{equation}
Each path comprises simultaneous forward and backward processes: its net rate may vanish while both directional rates are positive. Thermodynamic equilibrium additionally requires zero production in every selected dissipative mechanism (\cref{def:equilibrium-reversibility}).

{The corresponding mass productions are}
\begin{equation}
  r_i=M_i\sum_{a=1}^{N_R}\nu_i^a R_a,
  \qquad
  R_a=R_a^f-R_a^b,
\label{eq:mass-production}
\end{equation}
where $M_i$ is the molar mass and the superscripts $f$ and $b$ denote the forward and backward unidirectional rates. Throughout, ${\sf T}$ as a superscript denotes transposition, while the italic superscript $T$ used below labels thermal contributions. {Collect the mass stoichiometry in the matrix}
\begin{equation}
  \Stoich_{ia}:=M_i\nu_i^a,
  \qquad
  \boldsymbol r=\Stoich\boldsymbol R,
  \qquad
  \boldsymbol R:=(R_1,\ldots,R_{N_R})^{\sf T}.
\label{eq:stoichiometric-production-matrix}
\end{equation}
Multiplying the species balances by $\ell_i$ and summing eliminates the chemical source whenever $\boldsymbol\ell\in\ker\Stoich^{\sf T}$, giving a linear reaction invariant. Mass conservation,
\begin{equation}
  \sum_iM_i\nu_i^a=0 \qquad (a=1,\ldots,N_R),
\label{eq:stoich-mass}
\end{equation}
states $\bone\in\ker\Stoich^{\sf T}$ and implies the first identity in \cref{eq:source-conservation}. Other conserved combinations, such as elements or charge, occupy the same left nullspace when preserved by the network. These balance invariants differ from Wegscheider conditions on kinetic data for simultaneous detailed equilibrium. For mass-action kinetics in dimensionless activities, the latter require $\prod_a K_a^{\omega_a}=1$ for $\Stoich\boldsymbol\omega=0$, with $K_a$ the forward/backward rate-constant ratio at fixed temperature and pressure in the chosen activity convention. The right nullspace $\ker\Stoich$ contains pathway combinations  that leave aggregate mass production unchanged and are orthogonal to the affinity vector introduced below. Appendix~\ref{app:full-coupling} uses it to distinguish labelled pathways from independent thermodynamic reaction coordinates.

The total mass density, barycentric velocity, diffusion velocities, and diffusion mass fluxes are
\begin{equation}
  \rho:=\sum_i\rho_i,
  \qquad
  \rho\bv:=\sum_i\rho_i\bv_i,
  \qquad
  \bu_i:=\bv_i-\bv,
  \qquad
  \bj_i:=\rho_i\bu_i,
\label{eq:barycentric-definitions}
\end{equation}
with
\begin{equation}
    \sum_i\rho_i\bu_i=\sum_i\bj_i=\boldsymbol 0.
\label{eq:barycentric-constraint}
\end{equation}
We also use the mass fractions
\begin{equation}
  y_i:=\frac{\rho_i}{\rho},
  \qquad
  \boldsymbol y:=(y_1,\ldots,y_N)^{\sf T},
  \qquad\text{satisfying}\qquad
  \sum_i y_i=1.
\label{eq:mass-fractions}
\end{equation}

\subsection{Thermal energy and exact mixture identities}

The thermal energy density is
\begin{equation}
  \rho e:=\sum_i\rho_i e_i.
\label{eq:thermal-energy-density}
\end{equation}
Here $e$ is the thermal energy per unit mass of the mixture. The resolved diffusion motion carries the kinetic energy
\begin{equation}
  \half\sum_i\rho_i\bu_i^2.
\label{eq:diffusion-kinetic-energy}
\end{equation}
{This distinction is required by the selected entropy state: temperature is defined by differentiation with respect to thermal energy, whereas diffusion kinetic energy belongs to the resolved mechanical degrees of freedom.}

Summation of the partial momentum balances gives
\begin{equation}
  \partial_t(\rho\bv)+\Div(\rho\bv\otimes\bv-\bS)=\rho\bb,
\label{eq:mixture-momentum}
\end{equation}
where
\begin{equation}
  \rho\bb:=\sum_i\rho_i\bb_i,
  \qquad
  \bS:=\sum_i\bigl(\bS_i-\rho_i\bu_i\otimes\bu_i\bigr).
\label{eq:mixture-stress}
\end{equation}
Thus the mixture stress contains the diffusion kinetic stress. The total-energy density satisfies the exact decomposition
\begin{equation}
  \sum_i\bigl(\rho_i e_i+\half\rho_i\bv_i^2\bigr)
  =\rho e+\half\rho\bv^2+\half\sum_i\rho_i\bu_i^2.
\label{eq:energy-decomposition}
\end{equation}
{The corresponding single-continuum energy identity is obtained by setting}
\begin{align}
  \rho E&:=\rho e+\half\rho\bv^2
       +\half\sum_i\rho_i\bu_i^2,
  \label{eq:mixture-total-energy-density}
  \\
  \bq^{\mathrm{tot}}&:=\sum_i\Bigl[
     \bq_i-\bS_i\bu_i
     +\rho_i\bigl(e_i+\half|\bu_i|^2\bigr)\bu_i
  \Bigr],
  \label{eq:mixture-total-energy-flux}
  \\
  \rho\mathcal{P}&:=\sum_i\rho_i\bb_i\cdot\bu_i.
  \label{eq:diffusive-body-force-power}
\end{align}
Then summation of the constituent total-energy balances gives exactly
\begin{equation}
  \partial_t(\rho E)
  +\Div\bigl(\rho E\bv-\bS\bv+\bq^{\mathrm{tot}}\bigr)
  =\rho\bb\cdot\bv+\rho\mathcal{P}.
\label{eq:mixture-total-energy}
\end{equation}
The diffusive body-force power is also $\rho\mathcal{P}=\sum_i\rho_i(\bb_i-\bb)\cdot\bu_i$ by \cref{eq:barycentric-constraint}. It vanishes for a common body force per unit mass. Subtracting barycentric kinetic energy defines the objective single-continuum internal-energy density
\begin{equation}
  \rho e_{\mathrm{int}}
  :=\rho E-\half\rho\bv^2
  =\rho e+\half\sum_i\rho_i\bu_i^2.
\label{eq:mixture-internal-energy}
\end{equation}
Thus $\rho e_{\mathrm{int}}\ne\rho e$ in Classes~II and~III when relative motion is present. Class~III retains the separate $\rho_i e_i$ as primitive thermal state variables, while the additional diffusion kinetic energy remains mechanical.

To derive the partial thermal-energy balance, first use the mass balance:
\begin{align}
\bv_i\cdot\Bigl[\partial_t(\rho_i\bv_i)
  +\Div(\rho_i\bv_i\otimes\bv_i)\Bigr]
={}&\partial_t\!\bigl(\half\rho_i\bv_i^2\bigr)
 +\Div\!\bigl(\half\rho_i|\bv_i|^2\bv_i\bigr)
 +\half r_i\bv_i^2.
\label{eq:kinetic-identity}
\end{align}
Furthermore,
\begin{equation}
  \bv_i\cdot\Div\bS_i
  =\Div(\bS_i\bv_i)-\bS_i:\grad\bv_i.
\label{eq:stress-power-identity}
\end{equation}
Taking the scalar product of \cref{eq:momentum-balance} with $\bv_i$, using \cref{eq:kinetic-identity,eq:stress-power-identity}, and subtracting the resulting kinetic-energy equation from \cref{eq:energy-balance}, we obtain
\begin{equation}
  \partial_t(\rho_i e_i)+\Div(\rho_i e_i\bv_i+\bq_i)
  =\bS_i:\grad\bv_i
   +k_i-\bv_i\cdot\bfv_i+\half r_i\bv_i^2.
\label{eq:partial-internal-energy-raw}
\end{equation}

\subsection{Galilean-invariant exchange variables}

If mass is transferred, the conservative productions $\bfv_i$ and $k_i$ are not objective separately. Galilean covariance of the mixture balances \cite{RuggeriSimic2007,ArimaRuggeriSugiyamaTaniguchi2017} leads to the internal momentum and energy productions relative to the barycentric frame,
\begin{equation}
  \bmom_i:=\bfv_i-r_i\bv,
  \qquad
  \varepsilon_i:=k_i-\bv\cdot\bfv_i+\half r_i\bv^2.
\label{eq:objective-sources}
\end{equation}
They are objective and satisfy
\begin{equation}
  \sum_i\bmom_i=\boldsymbol 0,
  \qquad
  \sum_i\varepsilon_i=0.
\label{eq:objective-source-conservation}
\end{equation}
Conversely,
\begin{equation}
  \bfv_i=\bmom_i+r_i\bv,
  \qquad
  k_i=\varepsilon_i+\bv\cdot\bfv_i-\half r_i\bv^2.
\label{eq:source-reconstruction-raw}
\end{equation}
Inserting these relations and $\bv_i=\bv+\bu_i$ into \cref{eq:partial-internal-energy-raw} yields
\begin{equation}
  \partial_t(\rho_i e_i)+\Div(\rho_i e_i\bv_i+\bq_i)
  =\bS_i:\grad\bv_i
  +\varepsilon_i-\bu_i\cdot\bmom_i+\half r_i\bu_i^2.
\label{eq:partial-internal-energy-objective}
\end{equation}
The last three terms describe objective exchange of thermal and relative mechanical energy. Together with the mass balance they determine the $r_i$, $\bmom_i$, and $\varepsilon_i$ contributions to entropy production.

\begin{remark}[Covariance under admissible shifts of constituent energy zeros]
\label[remark]{rem:energy-reference-covariance}
\begingroup
Let $\Delta e_i$ be a constant energy-reference offset per unit mass of $A_i$, and set $e_i'=e_i+\Delta e_i$. These offsets change the energy convention, not the physical state. The partial mass balance shows that the same partial internal-energy balance then requires
\begin{equation}
  \varepsilon_i'=\varepsilon_i+r_i\Delta e_i.
\label{eq:energy-reference-source-shift}
\end{equation}
Writing the same physical entropy in the shifted energy variables gives $T_i'=T_i$ and $\mu_i'=\mu_i+\Delta e_i$. Hence the combined contribution of energy and chemical mass production to the entropy balance is unchanged for each constituent,
\begin{equation}
  \frac1{T_i'}\,\varepsilon_i'-\frac{\mu_i'}{T_i'}r_i
  =\frac1{T_i}\,\varepsilon_i-\frac{\mu_i}{T_i}r_i.
\label{eq:energy-reference-entropy-invariance}
\end{equation}
The local Euler relation is invariant for the same reason: $e_i-\mu_i$ is unchanged and the partial pressures need not be shifted.

The additional source in the summed energy balance is $\sum_ir_i\Delta e_i$. Thus the closed-system condition $\sum_i\varepsilon_i'=0$ is preserved for every independently admissible reaction-progress vector precisely when
\begin{equation}
  \sum_i r_i\Delta e_i=0,
\qquad\text{equivalently}\qquad
  \sum_i \Delta e_iM_i\nu_i^a=0
  \quad\text{for every reaction path }a.
\label{eq:admissible-energy-reference-gauge}
\end{equation}
With $\Delta\boldsymbol e:=(\Delta e_1,\ldots,\Delta e_N)^{\sf T}$, this condition reads
\begin{equation}
  \Delta\boldsymbol e\in\ker\Stoich^{\sf T},
\label{eq:energy-reference-gauge-nullspace}
\end{equation}
where $\Stoich$ is defined in \cref{eq:stoichiometric-production-matrix}.

An admissible offset adds the reaction-conserved density $\sum_i\rho_i\Delta e_i$ to the energy without changing the net energy increment of any reaction. A common offset is admissible since $\bone\in\ker\Stoich^{\sf T}$. Cancellation for one realized rate vector is insufficient. Offsets outside this nullspace are still changes of variables, but introduce $\sum_ir_i\Delta e_i$ into the summed energy balance. This conversion term records the changed convention, not external heating.

Constituent energy increments must transform together with the chemical potentials. At unequal temperatures, even an admissible offset can change the chemical affinity alone. Preserving the physical reaction relation requires the transformed energy increments (\cref{rem:event-reference-covariance}).
\endgroup
\end{remark}

Summation gives the thermal-energy identity
\begin{equation}
  \partial_t(\rho e)
  +\Div\bigl(\rho e\bv+\widetilde{\bq}\bigr)
  =\sum_i\bS_i:\grad\bv_i
  -\sum_i\bu_i\cdot\bmom_i
  +\half\sum_i r_i\bu_i^2,
\label{eq:thermal-energy-balance}
\end{equation}
where
\begin{equation}
  \widetilde{\bq}:=\sum_i(\bq_i+\rho_i e_i\bu_i).
\label{eq:preliminary-heat-flux}
\end{equation}
In Class~III, in contrast to Classes~I and~II, the partial energy balances are retained, and \cref{eq:thermal-energy-balance} provides their summed consistency identity.

\section{Entropy principle and model-class specification}
\label{sec:entropy-principle}

\subsection{Entropy postulates and constitutive mechanism principles}
\label{subsec:entropy-mechanisms}

The same entropy principle applies to Classes~I, II, and III: objectivity of $(\rho s,\bPhi)$, the universal local entropy inequality, and mechanism-wise sign conditions. Their balances, resolved states, and entropy-production representations differ.

A \emph{thermodynamic process} is a sufficiently smooth collection of fields satisfying the partial balances near an arbitrary point $(t_0,\boldsymbol x_0)$ for a prescribed constitutive class; it need not solve a complete initial--boundary-value problem. Resolved variables, constitutive state, dependencies, and dissipative mechanisms are fixed before entropy exploitation.
Local tests vary state values and spatial derivatives subject to the exact constraints, with partial time derivatives chosen to satisfy the balances, e.g.\ $\partial_t\rho_i=r_i-\Div(\rho_i\bv_i)$~\cite{BotheDreyer2015}. The entropy inequality restricts the constitutive equations for all such processes; it does not define the process class.

Principles~(I)--(IV) state the entropy, dimensional-parity, and mechanism requirements of Bothe and Dreyer~\cite{BotheDreyer2015}; Principle~(V) adds the balance anchoring of~\cite{BothePolymer2026}. Related sharp-interface extensions appear in~\cite{BotheSharpInterface2022,BotheSharpClassII2025}. We define equilibrium by zero production and specify the material class in~(M1)--(M3).

\begin{enumerate}[label=\textnormal{(\Roman*)},leftmargin=3.2em]
\item \textit{Entropy/entropy-flux pair.} There exists a material-dependent entropy/entropy-flux pair $(\rho s,\bPhi)$, where $s$ is the entropy per unit mass of the mixture and $\bPhi$ is the nonconvective entropy flux. The entropy density $\rho s$ is an objective scalar and $\bPhi$ is an objective vector.

\item \textit{Entropy balance and universal entropy inequality.} At every point $(t,\boldsymbol x)$ and for every thermodynamic process,
\begin{equation}
  \partial_t(\rho s)+\Div(\rho s\bv+\bPhi)=\zeta
  ,\qquad \zeta\ge0.
\label{eq:entropy-balance}
\end{equation}

\item \textit{Dimensional parity.} Every physical quantity is assigned a parity according to its physical dimension in SI base units. For the quantities of the present model class, write
\[
  [a]=\mathrm{kg}^{\alpha}\,\mathrm{m}^{\beta}\,
       \mathrm{s}^{\gamma}\,\mathrm{K}^{\delta}\,
       \mathrm{mol}^{\epsilon},
  \qquad \pi(a)=(-1)^{\gamma}.
\]
Even and odd time-unit exponents give positive and negative parity, respectively. Time differentiation reverses parity and spatial differentiation preserves it. Density, internal-energy density, temperature, chemical potential, and stress have positive parity; velocity, heat flux, reaction rate, and deformation rate have negative parity. Specific entropy has dimension
\[
  [s]=\mathrm{J}\,\mathrm{kg}^{-1}\mathrm{K}^{-1}
     =\mathrm{m}^{2}\mathrm{s}^{-2}\mathrm{K}^{-1},
  \qquad
  [\rho s]=\mathrm{J}\,\mathrm{m}^{-3}\mathrm{K}^{-1}.
\]
Thus $\rho s$ has positive parity; its time derivative, $\bPhi$, and $\zeta$ have negative parity. Each dimensionally consistent binary product representing $\zeta$ therefore has opposite-parity factors. This convention is motivated by time reversal of the unclosed balances~\cite{BotheDreyer2015}, without asserting time-reversal invariance of the closed dissipative equations.

\item \textit{Admissible entropy fluxes and mechanism-wise nonnegativity.}  A chosen entropy flux is admissible if the \emph{reduced entropy production}, obtained after inserting that flux and using the balance equations and exact constraints, can be represented as
\begin{equation}
  \zeta=\sum_{m=1}^{M}N_mP_m,
\label{eq:binary-product-axiom}
\end{equation}
where $N_m$ and $P_m$ have negative and positive parity, respectively.
The product notation includes scalar products of vectors, contractions of tensors, and duality pairings of specified coupled blocks.  Each binary product describes a dissipative mechanism, or a coupled block of mechanisms, introduced before constitutive closure, and
\[
  N_mP_m\ge0
  \qquad\text{for every }m\text{ and every thermodynamic process}.
\]
Parity ordering does not select the constitutive factor~\cite{BotheDreyer2015}: heat flux has negative parity and its reciprocal-temperature gradient positive parity, whereas viscous stress has positive parity and deformation rate negative parity.

\begingroup
\item \textit{Balance anchoring.} Fix a balance representation and its primitive state variables. A \emph{balance anchor} is a constitutive quantity, not fixed kinematically or thermodynamically before closure, that enters an additive contribution to a balance flux or source, directly or through a preassigned positive-parity coefficient. Every selected binary dissipative product $N_mP_m$ contains such an anchor. The anchor's parity is inherited from the fixed balance representation.
\endgroup
\end{enumerate}

\begingroup
\begin{definition}[Thermodynamic equilibrium]
\label[definition]{def:equilibrium-reversibility}
A thermodynamic process is in thermodynamic equilibrium at a point $(t,\boldsymbol x)$ if $\zeta(t,\boldsymbol x)=0$. It is in thermodynamic equilibrium on a space--time domain if this condition holds at every point of that domain. Equilibrium at one point need not imply equilibrium elsewhere.
\end{definition}
\endgroup

The entropy flux is constitutive. For fixed entropy density and balances, $\bPhi\mapsto\bPhi+\boldsymbol K$ changes production by $\Div\boldsymbol K$. The resulting expression must again satisfy~(IV). If this changes the volume-deformation coefficient, the Euler restriction derived below must also be reconsidered. In contrast, an invertible co-factor change preserving $\sum_mN_mP_m$ leaves both the flux and production unchanged.

Together with the entropy balance in Principle~(II), Principle~(IV) gives the local second law of thermodynamics,
\[
  \partial_t(\rho s)+\Div(\rho s\bv+\bPhi)
  =\zeta=\sum_{m=1}^{M}N_mP_m\ge0.
\]
It holds pointwise for every thermodynamic process. The stronger mechanism-wise requirement demands a physical decomposition on the constrained local process space~\cite{BotheDreyer2015}. Independent-variation arguments use independent dual coordinates. Selected redundant pathwise or pairwise products must nevertheless satisfy the individual sign condition~(IV). Zero total production then forces zero production in every selected mechanism or block. This strengthened detailed-balance requirement differs from chemical-network detailed balance.

\paragraph{Thermodynamic equilibrium and stationarity.}
After closure, $\zeta=0$ specifies local states, velocities, and derivatives with zero production in every selected mechanism or block. Whether these conditions form a stationary equilibrium configuration additionally depends on the stationary balances, frame, and boundary conditions.

\paragraph{Linear cross-effects and reciprocity.}
For a linear closure written in a fixed conjugate co-factor basis as $\boldsymbol{\mathcal J}=\mathsf L\boldsymbol{\mathcal X}$, with fixed coefficients tested over the independent force space\footnote{Varying a reciprocal-temperature difference or a reaction affinity changes the thermodynamic state and may also change the coupling coefficients. Keeping the coefficients fixed lets the sign test isolate the effect of varying the forces.}, the total entropy inequality constrains the symmetric dissipative part
\begin{equation}
  \mathsf L^{\mathrm s}:=\half(\mathsf L+\mathsf L^{\sf T})\succeq0,
  \qquad
  \boldsymbol{\mathcal X}\mathbin{\boldsymbol\cdot}
  \mathsf L^{\mathrm a}\boldsymbol{\mathcal X}=0,
  \qquad
  \mathsf L^{\mathrm a}:=\half(\mathsf L-\mathsf L^{\sf T}).
\label{eq:entropy-linear-symmetric-skew}
\end{equation}
This scalar criterion alone establishes neither flux admissibility nor the physical decomposition and its individual sign conditions. Before closure, specify \emph{cross-coupled mechanism groups} with balance-anchored factors. For a group $\mathcal G$, the dual block factors $\boldsymbol N_{\mathcal G}$ and $\boldsymbol P_{\mathcal G}$ carry mechanism and spatial indices. Their full pairing is \emph{one} product in~(IV):
\[
  \zeta_{\mathcal G}
  =\boldsymbol N_{\mathcal G}\mathbin{\boldsymbol\cdot}\boldsymbol P_{\mathcal G}
  =\sum_{a\in\mathcal G}N_aP_a\ge0.
\]
Individual summands may then be negative, but the pairing must be nonnegative for every process \emph{separately in each selected block}. Different blocks cannot compensate one another.

At the coarsest grouping, treating the entire reduced production as one block and imposing no separate sign conditions reduces the nonnegativity requirement exactly to the standard entropy inequality $\zeta\ge0$. For linear closures on the independent force spaces in the stated isotropic class, tensorial symmetry still separates scalar, vector, and deviatoric sectors, and this inequality constrains only the symmetric part of each permitted block, not the signs of its subprocess contributions or reciprocity. This weakens only the mechanism-wise sign requirement, not the entropy-flux, parity, or balance-anchoring requirements.

To couple specified elementary dissipative mechanisms, we use entropy-neutral mixing before closure. Suppressing the group index, use dual bases on the independent constrained spaces and invertible, dimensionally consistent transformations preserving tensorial character and each parity class:
\[
  \widetilde{\boldsymbol N}=\mathsf A\boldsymbol N,
  \qquad
  \widetilde{\boldsymbol P}=\mathsf B\boldsymbol P,
  \qquad
  \mathsf A^{\sf T}\mathsf B=\mathsf I.
\]
These preserve the entropy pairing. The selected elementary mechanisms in the mixed representation must each satisfy the sign condition in~(IV). For a fixed linear law tested on a neighbourhood of zero in the independent force space, independent variation and separate nonnegativity exclude linear couplings between distinct transformed mechanisms: in $L_{mm}\widetilde P_m^2+L_{mn}\widetilde P_m\widetilde P_n$, a nonzero $L_{mn}$ permits fixed $\widetilde P_n\ne0$ and sufficiently small $\widetilde P_m$ opposite in sign to $L_{mn}\widetilde P_n$ to give a negative product, arbitrarily near zero. The same argument applies componentwise to vectors and tensors. Thus under this test the closure is block-diagonal in the selected transformed mechanisms. The thermal and chemical forces are fixed at a given thermodynamic state. For these directions, the full-space test is a constitutive requirement or applies to a linearization admitting the independent force variations. For our isotropic scalar, vector, and deviatoric sectors,
\[
  \widetilde N_m=\lambda_m\widetilde P_m,
  \qquad \lambda_m\ge0,
  \qquad
  \widetilde N_m\widetilde P_m
  =\lambda_m|\widetilde P_m|^2\ge0.
\]
Here $\lambda_m$ multiplies the identity on the spatial components of a vector or deviatoric mechanism. With $\widetilde{\mathsf\Lambda}=\operatorname{diag}(\lambda_m)$, understood in this block sense,
\[
  \boldsymbol N=\mathsf L\boldsymbol P,
  \qquad
  \mathsf L=\mathsf B^{\sf T}\widetilde{\mathsf\Lambda}\mathsf B
  =\mathsf L^{\sf T}\succeq0.
\]
This entropy-neutral coupling of elementary dissipative mechanisms yields Onsager symmetry, generally with cross-couplings, in the original parity-ordered co-factors~\cite{BotheDreyer2015}. Conversely, every self-adjoint positive-semidefinite operator with these tensorial restrictions admits this factorization, used in \cref{app:full-coupling}.

The construction retains the individual sign conditions of the elementary mechanisms in the selected mixed representation. A different modelling choice is to regard a larger block as a single mechanism without prescribing the entropy productions of its subprocesses. Nonnegativity then constrains only the symmetric part of its coefficient operator, and internal reciprocity need not follow. Further invertible entropy-neutral mixing preserves this distinction: for $\mathsf L'=\mathsf A\mathsf L\mathsf A^{\sf T}$,
\[
 (\mathsf L')^{\mathrm a}
 =\mathsf A\mathsf L^{\mathrm a}\mathsf A^{\sf T}.
\]
It therefore preserves symmetry but cannot eliminate a nonzero skew-adjoint part, including when the block is combined with additional reciprocal mechanisms.

For the coupling of constituent $i$'s volume deformation to reaction $a$ of the mixture, the negative-parity variables $(\Div\bv_i,R_a)$ are conjugate to the positive-parity variables $(-\pi_i/T_i,-\Acal_a)$. With $\lambda_i^{\mathrm v}>0$, $L_a\ge0$, and coupling coefficient $\ell_{\theta R}$, the symmetric Onsager relation reads
\[
 \begin{pmatrix}\Div\bv_i\\ R_a\end{pmatrix}
 =\begin{pmatrix}
  \dfrac{T_i}{\lambda_i^{\mathrm v}} & \dfrac{T_i\ell_{\theta R}}{\lambda_i^{\mathrm v}}\\[1ex]
  \dfrac{T_i\ell_{\theta R}}{\lambda_i^{\mathrm v}} & L_a+\dfrac{T_i\ell_{\theta R}^{\,2}}{\lambda_i^{\mathrm v}}
 \end{pmatrix}
 \begin{pmatrix}-\pi_i/T_i\\ -\Acal_a\end{pmatrix}.
\]
Solving instead for the mixed-parity pair $(-\pi_i/T_i,R_a)$ gives
\[
 \begin{pmatrix}-\pi_i/T_i\\ R_a\end{pmatrix}
 =\begin{pmatrix}
  \lambda_i^{\mathrm v}/T_i & -\ell_{\theta R}\\
  \ell_{\theta R} & L_a
 \end{pmatrix}
 \begin{pmatrix}\Div\bv_i\\ -\Acal_a\end{pmatrix}.
\]
The opposite cross-coefficients $-\ell_{\theta R}$ and $+\ell_{\theta R}$ exhibit the Onsager--Casimir form: exchanging one conjugate pair converts Onsager symmetry into cross-antisymmetry, without introducing a new coupling~\cite[Section~6]{BotheDreyer2015}. A microscopic interpretation additionally requires appropriate time-reversal variables and symmetries~\cite{Onsager1931}.

\paragraph{Constitutive specification and balance anchoring.}
Mechanisms and dependencies for each constitutive quantity are chosen using material information, experiments, scale analysis, or the intended application. Conditions under which each mechanism ceases to dissipate guide factor selection and closure. The same principles apply to more extensively coupled classes specified consistently.

For a bulk balance
\[
  \partial_t a+\Div\boldsymbol J=r,
\]
anchoring refers to the constitutive quantities entering specified additive contributions to the flux $\boldsymbol J$ or source $r$, rather than necessarily to either complete quantity.\footnote{Relative to the selected balance representation, the anchored constitutive factor may be called the \emph{generalized flux} and its conjugate co-factor the \emph{driving force}. Here ``flux'' also includes constitutive production rates. This terminology specifies their constitutive roles, not their parities: a stress has positive parity, whereas a heat flux or reaction rate has negative parity.}
For chemical reactions, $R_a$ is the constitutive co-factor in $-R_a\Acal_a$, with conjugate $-\Acal_a$. The temperature-weighted affinity $\Acal_a$ is defined in~\cref{eq:classIII-affinity}. The rate is anchored through its occurrence in the partial mass sources
\[
  r_i=\sum_{a=1}^{N_R}M_i\nu_i^aR_a.
\]
The coefficients $M_i\nu_i^a$ have positive dimensional parity, irrespective of their signs. The coefficient-parity condition applies to an anchoring occurrence, not to every occurrence of the same rate in other balances or to its kinetic closure coefficients.
Contributions determined by kinematics, thermodynamics, or other selected constitutive quantities require no separate closure, but remain included in the balance and entropy identities. Prescribed external supplies are not constitutive anchors.

For an individual mechanism, at least one of the co-factors $N_m$ and $P_m$ is the anchor, possibly in an admissible representation as described next. For a coupled block, anchoring is understood in the full dual pairing. Its constitutive anchors need not all belong to the same parity class. An explicitly specified even-parity thermodynamic normalization or parity-preserving dual change of co-factors is admissible provided the original anchors remain explicitly recoverable from the transformed co-factors and the entropy production is unchanged. Such changes do not, by themselves, create new physical balance anchors.

Principle~(V) supplements, without restricting, the dimensional parity rule~(III). Its anchors refer to the fixed physical balance quantities and the specified physical normalization of reaction rates. For example, inserting an artificial time-dimensional constant $C_\tau$ with $[C_\tau]=\mathrm{s}$ gives
\begin{equation}
  N_mP_m=(C_\tau N_m)\Bigl(\frac{P_m}{C_\tau}\Bigr).
\label{eq:artificial-parity-refactorization}
\end{equation}
This reverses both displayed parities. Reconstructing an original anchoring contribution from its rescaled factor requires a coefficient of negative dimensional parity, which~(V) excludes. A consistent change of representation must transform the balance contributions, their entropy conjugates, all related products, and the constitutive relations. The available anchored co-factors of a given tensorial character are then unchanged up to an invertible, dimensionally consistent representation change. An auxiliary dimensional constant therefore supplies no new physical anchor and cannot generate a physical cross-effect.

Liu's calculation in \cref{app:liu-consistency} recovers the entropy conjugates and residual inequality after the state, balances,~(M1)--(M3), mechanisms~(IV), and anchoring~(V) have been specified. Its balance multipliers differ from those enforcing kinematic or algebraic constraints below.

\subsection{Specification of the fluid-mixture class}
\label{sec:model-class-specification}
\label{subsec:pressure-status-preview}

We specify the primitive thermodynamic variables and mechanisms for the material class through the following modelling assumptions before evaluating the entropy principle.

\begin{enumerate}[label=\textnormal{(M\arabic*)},leftmargin=3.2em]
\item \textit{Selected dissipative mechanisms.} The  baseline model of Class~III contains constituent shear and volume deformation, constituent heat conduction, relative momentum exchange, interspecies thermal relaxation, and chemical reaction. The baseline model includes the thermo-diffusive mixing of Section~\ref{subsec:thermodiffusion-mixing}. Further cross-effects require a specified entropy-invariant mixing or a separate rational argument.

\item \textit{Thermodynamic state and positive-temperature domain.} The entropy density has the local constitutive form
\begin{equation}
  \rho s
  =\widehat{\rho s}(\rho_1e_1,\ldots,\rho_Ne_N,\rho_1,\ldots,\rho_N),
\label{eq:entropy-state}
\end{equation}
where $\widehat{\rho s}$ is objective and twice continuously differentiable, and its Hessian with respect to the $2N$ thermodynamic variables in \cref{eq:entropy-state} is negative definite at every admissible state,
\begin{equation}
  D^2\widehat{\rho s}\prec0.
\label{eq:negative-definite-entropy-hessian}
\end{equation}
Thus the entropy is strictly concave along every line segment contained in the admissible state domain. We do not assume a state-independent strong-concavity modulus. In addition, the state domain and entropy function are assumed to satisfy
\begin{equation}
  \frac{\partial\widehat{\rho s}}{\partial(\rho_i e_i)}>0,
  \qquad i=1,\ldots,N,
\label{eq:positive-temperature-domain}
\end{equation}
at every admissible thermodynamic state. This monotonicity condition is an independent constitutive assumption. It is not implied by the negative-definite entropy Hessian.

\item \textit{Mechanical structure and partial thermodynamic pressures.} We assume vanishing constituent couple stresses and vanishing constituent body-couple productions. The constituent angular-momentum balances then imply that every partial Cauchy stress $\bS_i$ is symmetric. For the  volume-deformation mechanism of the baseline model, fix the thermodynamic state and let all constituents undergo the same dilatation rate $\alpha$. We assume that the mechanical pressure response $P_i(\alpha)$ is continuous at $\alpha=0$ and define $p_i:=P_i(0)$. The $p_i=p_i(\rho_1e_1,\ldots,\rho_Ne_N,\rho_1,\ldots,\rho_N)$ are differentiable state functions, called \emph{partial thermodynamic pressures}.  The constitutive laws of the baseline model are isotropic in the absence of additional material structure. Prescribed body forces are allowed, but no electromagnetic or radiation field balance is included.
\end{enumerate}

Specification~(M1) retains the stated mechanisms and thermo-diffusive coupling with few coefficients. Further couplings require physical or modelling grounds. Bedeaux's three-phase contact-line theory~\cite{Bedeaux2004} illustrates the rapid coefficient growth with unrestricted admissible couplings. Specification~(M2) excludes nonlocal capillarity, phase-field effects, and diffusion kinetic energy from the thermal state. The last exclusion preserves the mechanical interpretation of momentum-exchange co-factors. Its temperature positivity is independent of concavity. The Euler relation is derived, not assumed in~(M3), and constrains only $\sum_i p_i/T_i$.  The definition $p_i=P_i(0)$ in (M3) refers to the  isolated volume-deformation response of the baseline model.  The coupled model in Appendix~\ref{app:full-coupling} retains these thermodynamic state functions while replacing that scalar response by a coupled law. Consequently, $P_i\neq p_i$ may occur at zero dilatation when other scalar co-factors remain active.

 The fully coupled model variant considered in Appendix~\ref{app:full-coupling} replaces~(M1) by deviatoric, vector, and scalar groups of balance-anchored co-factors with separately nonnegative closure in the selected mixed representation.  The unmixed binary products of the baseline model need not remain nonnegative. Adding arbitrary cross-coefficients to  independent constitutive relations of the baseline model would neither specify nor verify this alternative decomposition.

 The state description in~(M2) assumes equilibration of unresolved degrees of freedom conditional on the retained mass and thermal-energy densities, justified when it is fast on the time scales considered. It does not impose mutual thermal or chemical equilibrium: one local composition can coexist with unequal temperatures and nonzero affinities. Existence of the entropy function is therefore distinct from zero-production equilibrium (\cref{def:equilibrium-reversibility}).

\subsection{Temperature and chemical potentials}
\label{subsec:temperature-potentials}

The constituent absolute temperatures $T_i$, through their reciprocals, and the chemical potentials per unit mass of $A_i$ are defined directly from \cref{eq:entropy-state} by
\begin{equation}
  \frac1{T_i}:=\frac{\partial\widehat{\rho s}}{\partial(\rho_i e_i)},
  \qquad
  -\frac{\mu_i}{T_i}:=\frac{\partial\widehat{\rho s}}{\partial\rho_i}.
\label{eq:entropy-conjugates}
\end{equation}
Assumption~\eqref{eq:positive-temperature-domain} independently ensures $T_i>0$. We retain the explicit conjugates $1/T_i$, $-\mu_i/T_i$ and balanced thermal variables $\rho_i e_i$.

The common potential imposes integrability on thermodynamic material data: tables or submodels for $T_i$ and $\mu_i$ must yield conjugates $1/T_i$ and $-\mu_i/T_i$ compatible with derivatives of the same $\widehat{\rho s}$. This fixes conjugates and their state derivatives, but not the individual pressure partition.

Independently of the positivity assumption, the pointwise negative-definite Hessian \cref{eq:negative-definite-entropy-hessian} gives the differential thermodynamic stability used here and, by the inverse-function theorem, local invertibility of the map from the variables $((\rho_i e_i,\rho_i))_i$ to their entropy derivatives. For fixed densities, the matrix
\begin{equation}
  -\Bigl(
  \frac{\partial(1/T_i)}{\partial(\rho_j e_j)}
  \Bigr)_{i,j=1}^N
\label{eq:thermal-hessian}
\end{equation}
is positive definite. At common temperature, its inverse reduces for an additive simple mixture to $T^2$ times the diagonal matrix of volumetric heat capacities.

For fixed mass densities, write $\boldsymbol z=(\rho_1e_1,\ldots,\rho_Ne_N)^{\sf T}$ and let $H_{zz}$ be the energy--energy block of $D^2\widehat{\rho s}$. Differentiating \cref{eq:entropy-conjugates} gives
\[
  \Bigl(\frac{\partial T_i}{\partial z_j}\Bigr)_{i,j=1}^N
  =-\operatorname{diag}(T_1^2,\ldots,T_N^2)\,H_{zz}.
\]
The principal block $H_{zz}\prec0$ and $T_i>0$ make this Jacobian nonsingular, permitting local temperature coordinates at fixed densities and the corresponding partial Legendre transformation with conjugates $1/T_i$. Neither global invertibility nor separate caloric relations $e_i=e_i(\rho_i,T_i)$ follow: each energy may depend on all temperatures and densities.

An equivalent thermodynamic potential can be used instead of the entropy density. Put $\beta_i=1/T_i$ and $\boldsymbol\rho=(\rho_i)_i$. The local partial Legendre transform in the energy variables is the Massieu density
\[
 \mathcal M(\boldsymbol\beta,\boldsymbol\rho)
 :=\widehat{\rho s}(\boldsymbol z,\boldsymbol\rho)
      -\sum_i\beta_i z_i,
 \qquad \boldsymbol z=\boldsymbol z(\boldsymbol\beta,\boldsymbol\rho).
\]
It satisfies
\[
 \frac{\partial\mathcal M}{\partial\beta_i}=-\rho_i e_i,
 \qquad
 \frac{\partial\mathcal M}{\partial\rho_i}=-\frac{\mu_i}{T_i}.
\]
Thus the entropy density and $\mathcal M$ contain the same local thermodynamic information, with the same integrability restrictions. At common temperature, the Helmholtz free-energy density is $f(T,\boldsymbol\rho)=-T\mathcal M(T^{-1}\bone,\boldsymbol\rho)$, and
\[
 \mu_i=\left.\frac{\partial f}{\partial\rho_i}\right|_{T,\rho_{j\ne i}},
 \qquad
 p=\sum_i\rho_i\mu_i-f.
\]
The Euler relation derived below gives the second identity and total pressure $p=\sum_i p_i$; at unequal temperatures it fixes only $\sum_i p_i/T_i$. An invertible Legendre transformation adds no pressure-partition information. Separate constituent free-energy formulas for $p_i$ require an additional constitutive decomposition. Neither potential determines kinetic coefficients.

Thus $p_i=p_i(T_1,\ldots,T_N,\rho_1,\ldots,\rho_N)$ remain individual thermal equations of state subject to the weighted Euler relation, rather than separate derivatives fixed by the joint potential.

\section{Detailed exploitation of the entropy principle}
\label{sec:entropy-exploitation}

\subsection{\texorpdfstring{ Non-reduced representation of the entropy production}{Non-reduced representation of the entropy production}}
\label{subsec:non-reduced-entropy-production}

Let
\begin{equation}
  \frac{\mathrm D}{\mathrm Dt}:=\partial_t+\bv\cdot\grad
\label{eq:barycentric-material-derivative}
\end{equation}
denote the material derivative associated with the barycentric velocity. Starting from \cref{eq:entropy-balance},
\begin{equation}
  \zeta
  =\frac{\mathrm D(\rho s)}{\mathrm Dt}
   +\rho s\Div\bv+\Div\bPhi.
\label{eq:entropy-start}
\end{equation}
The chain rule and \cref{eq:entropy-conjugates} give
\begin{equation}
  \frac{\mathrm D(\rho s)}{\mathrm Dt}
  =\sum_i\frac1{T_i}\frac{\mathrm D(\rho_i e_i)}{\mathrm Dt}
   -\sum_i\frac{\mu_i}{T_i}\frac{\mathrm D\rho_i}{\mathrm Dt}.
\label{eq:entropy-chain-rule}
\end{equation}

The partial mass balance, written with $\bv_i=\bv+\bu_i$, gives

\begin{equation}
  \frac{\mathrm D\rho_i}{\mathrm Dt}
  =r_i-\rho_i\Div\bv-\Div(\rho_i\bu_i).
\label{eq:mass-derivative-elimination}
\end{equation}
Similarly, \cref{eq:partial-internal-energy-objective} gives

\begin{align}
  \frac{\mathrm D(\rho_i e_i)}{\mathrm Dt}
  ={}&\bS_i:\grad\bv_i
  +\varepsilon_i-\bu_i\cdot\bmom_i
  +\half r_i\bu_i^2
  -\rho_i e_i\Div\bv
  -\Div(\rho_i e_i\bu_i+\bq_i).
\label{eq:energy-derivative-elimination}
\end{align}

Insertion of \cref{eq:mass-derivative-elimination,eq:energy-derivative-elimination} into \cref{eq:entropy-start,eq:entropy-chain-rule} yields,
\begin{align}
\zeta={}&\Div\bPhi
+\Bigl(\rho s-\sum_i\frac{\rho_i e_i}{T_i}
       +\sum_i\frac{\rho_i\mu_i}{T_i}\Bigr)\Div\bv
+\sum_i\frac1{T_i}\bS_i:\grad\bv_i
\label{eq:entropy-before-product-rules}
\\
&-\sum_i\frac1{T_i}\Div(\rho_i e_i\bu_i+\bq_i)
+\sum_i\frac{\mu_i}{T_i}\Div(\rho_i\bu_i)
\nonumber\\
&+\sum_i\Bigl[
  \frac1{T_i}\,\varepsilon_i
  -\frac1{T_i}\bu_i\cdot\bmom_i
  +\frac{r_i}{2T_i}\bu_i^2
  -\frac{\mu_i}{T_i}r_i
  \Bigr].
\nonumber
\end{align}
We use the product-rule identities
\begin{align}
-\frac1{T_i}\Div(\rho_i e_i\bu_i+\bq_i)
={}&-\Div\!\Bigl(\frac{\rho_i e_i\bu_i+\bq_i}{T_i}\Bigr)
 +(\rho_i e_i\bu_i+\bq_i)\cdot\grad\frac1{T_i},
\label{eq:product-rule-energy}
\\
\frac{\mu_i}{T_i}\Div(\rho_i\bu_i)
={}&\Div\!\Bigl(\frac{\rho_i\mu_i}{T_i}\bu_i\Bigr)
 -\rho_i\bu_i\cdot\grad\frac{\mu_i}{T_i}.
\label{eq:product-rule-mass}
\end{align}
Consequently,
\begin{align}
\zeta={}&\Div\!\Bigl[
\bPhi-\sum_i\frac{\bq_i+\rho_i e_i\bu_i}{T_i}
+\sum_i\frac{\rho_i\mu_i}{T_i}\bu_i
\Bigr]
\label{eq:entropy-first-representation}
\\
&+\Bigl(\rho s-\sum_i\frac{\rho_i e_i}{T_i}
       +\sum_i\frac{\rho_i\mu_i}{T_i}\Bigr)\Div\bv
+\sum_i\frac1{T_i}\bS_i:\grad\bv_i
+\sum_i(\bq_i+\rho_i e_i\bu_i)\cdot\grad\frac1{T_i}
\nonumber\\
&-\sum_i\rho_i\bu_i\cdot\grad\frac{\mu_i}{T_i}
+\sum_i\frac1{T_i}\Bigl[
  \varepsilon_i-\bu_i\cdot\bmom_i
  +\half r_i\bu_i^2-\mu_i r_i
  \Bigr].
\nonumber
\end{align}
This is the non-reduced Class-III entropy-production representation before the entropy flux and final co-factors have been selected.

\subsection{Velocity constraints and temporary Lagrange multipliers}

The diffusion velocities obey
\begin{equation}
  \Div\bu_i=\Div\bv_i-\Div\bv.
\label{eq:divergence-constraint}
\end{equation}
To evaluate the entropy inequality on the kinematically constrained local process space without eliminating the divergence constraint explicitly, we add the identically vanishing expression
\begin{equation}
  0
  =\sum_i\frac{\varpi_i}{T_i}
  \bigl(\Div\bv_i-\Div\bv-\Div\bu_i\bigr),
\label{eq:lm-zero}
\end{equation}

\noindent where $\varpi_i$ are temporary state-dependent multipliers of the \emph{kinematic identity} \cref{eq:divergence-constraint}. They retain explicit scalar co-factors and disappear when mechanically anchored thermodynamic pressures enter the scalar restriction.

The last term is rewritten as
\begin{equation}
  -\frac{\varpi_i}{T_i}\Div\bu_i
  =-\Div\!\Bigl(\frac{\varpi_i}{T_i}\bu_i\Bigr)
   +\bu_i\cdot\grad\frac{\varpi_i}{T_i}.
\label{eq:lm-product-rule}
\end{equation}
Using the stress decomposition \cref{eq:stress-decomposition} and the symmetry of $\bS_i$, we have
\begin{equation}
  \frac1{T_i}\bS_i:\grad\bv_i
  =-\frac{P_i}{T_i}\Div\bv_i
   +\frac1{T_i}\bS_i^{\circ}:\bD_i^{\circ},
\label{eq:stress-entropy-split}
\end{equation}
where
\begin{equation}
  \bD_i:=\half\bigl(\grad\bv_i+(\grad\bv_i)^{\sf T}\bigr),
  \qquad
  \bD_i^{\circ}:=\bD_i-\frac13(\Div\bv_i)\bI.
\label{eq:strain-rates}
\end{equation}
After adding \cref{eq:lm-zero} and using \cref{eq:lm-product-rule,eq:stress-entropy-split}, the entropy production becomes
\begin{align}
\zeta={}&\Div\!\Bigl[
\bPhi-\sum_i\frac{\bq_i+\rho_i e_i\bu_i}{T_i}
+\sum_i\frac{\rho_i\mu_i-\varpi_i}{T_i}\bu_i
\Bigr]
+\sum_i\frac1{T_i}\bS_i^{\circ}:\bD_i^{\circ}
+\sum_i\bq_i\cdot\grad\frac1{T_i}
\label{eq:entropy-with-multipliers}
\\
&+\Bigl[
\rho s-\sum_i\frac{\rho_i e_i}{T_i}
+\sum_i\frac{\rho_i\mu_i-\varpi_i}{T_i}
\Bigr]\Div\bv
-\sum_i\frac{P_i-\varpi_i}{T_i}\Div\bv_i
\nonumber\\
&+\sum_i\bu_i\cdot\Bigl[
\rho_i e_i\grad\frac1{T_i}
-\rho_i\grad\frac{\mu_i}{T_i}
+\grad\frac{\varpi_i}{T_i}
-\frac{\bmom_i}{T_i}
+\frac{r_i}{2T_i}\bu_i
\Bigr]
+\sum_i\frac1{T_i}\varepsilon_i
-\sum_i r_i\frac{\mu_i}{T_i}.
\nonumber
\end{align}
Equation~\eqref{eq:entropy-with-multipliers} is the unclosed entropy-production identity obtained directly from the balance equations.

\begin{remark}[{Multiplier representation of the divergence constraint}]
The divergence constraint can alternatively be eliminated directly by means of
\begin{equation}
  \Div\bv=\sum_i y_i\Div\bv_i+\sum_i\bu_i\cdot\grad y_i.
\label{eq:direct-divergence-elimination}
\end{equation}

Direct elimination transfers the coefficient of $\Div\bv$ into the vector diffusion block through $\grad y_i$ before scalar closure. Multiplier augmentation retains the common and constituent dilatations explicitly. Both yield the same result after constitutive restrictions, and only the final balance-anchored co-factors receive physical interpretation; cf.~\cite[Section~4]{BotheDreyer2015}.

\end{remark}

\subsection{Entropy-flux selection and multiplier representation}
\label{subsec:entropy-flux-pressure}

We first consider the following provisional family of candidate entropy fluxes:
\begin{equation}
  \bPhi_{\varpi}
  :=\sum_i\Bigl[
  \frac{\bq_i}{T_i}
  +\frac{\rho_i e_i+\varpi_i-\rho_i\mu_i}{T_i}\bu_i
  \Bigr].
\label{eq:entropy-flux-lambda}
\end{equation}

Each candidate cancels the explicit divergence in \cref{eq:entropy-with-multipliers}. Its admissibility depends on the mechanism decomposition and closure under~(IV). We select the final member with the pressure representation below. With
\begin{equation}
  C_{\varpi}
  :=\rho s-\sum_i\frac{\rho_i e_i}{T_i}
  +\sum_i\frac{\rho_i\mu_i-\varpi_i}{T_i},
\label{eq:scalar-euler-remainder}
\end{equation}
the entropy production becomes
\begin{align}
\zeta={}&
C_{\varpi}\Div\bv
-\sum_i\frac{P_i-\varpi_i}{T_i}\Div\bv_i
+\sum_i\frac1{T_i}\bS_i^{\circ}:\bD_i^{\circ}
+\sum_i\bq_i\cdot\grad\frac1{T_i}
\label{eq:entropy-after-flux-selection}
\\
&+\sum_i\bu_i\cdot\Bigl[
\rho_i e_i\grad\frac1{T_i}
-\rho_i\grad\frac{\mu_i}{T_i}
+\grad\frac{\varpi_i}{T_i}
-\frac{\bmom_i}{T_i}
+\frac{r_i}{2T_i}\bu_i
\Bigr]
+\sum_i\frac1{T_i}\varepsilon_i
-\sum_ir_i\frac{\mu_i}{T_i}.
\nonumber
\end{align}

Before selecting bulk viscosity, we restrict the mechanical pressure $P_i=-\tr\bS_i/3$. Its zero-rate response defines, by~(M3), differentiable state functions
\begin{equation}
  p_i=p_i(\rho_1e_1,\ldots,\rho_Ne_N,\rho_1,\ldots,\rho_N),
  \qquad i=1,\ldots,N.
\label{eq:partial-pressure-state-functions}
\end{equation}
These are defined by zero dilatation, not thermodynamic equilibrium, and are not individually determined by the entropy potential.

At a fixed thermodynamic state, consider the balance-compatible common-dilatation family
\begin{equation}
  \bu_i=\boldsymbol 0,
  \qquad
  \grad\bv_i=\frac{\alpha}{3}\bI,
  \qquad i=1,\ldots,N,
\label{eq:common-dilatation-test-family}
\end{equation}
so that
\begin{equation}
  \Div\bv_i=\Div\bv=\alpha,
  \qquad
  \bD_i^{\circ}=\boldsymbol 0.
\label{eq:common-dilatation-test-rates}
\end{equation}
Volume deformation is separately dissipative in the baseline model, so this test does not require vanishing thermal or chemical co-factors. The local process variation admits both signs of $\alpha$, with storage terms determined by the balances and $P_i(\alpha)\to p_i$ as $\alpha\to0$. This is a local constitutive-process premise. No initial--boundary-value existence result is used. The scalar contribution in \cref{eq:entropy-after-flux-selection} becomes
\begin{equation}
  \zeta_{\mathrm{sc}}(\alpha)
  =
  \alpha\Bigl[
    C_{\varpi}
    -\sum_i\frac{P_i(\alpha)-\varpi_i}{T_i}
  \Bigr].
\label{eq:common-dilatation-scalar-production}
\end{equation}
Continuity and nonnegativity of this scalar production for both signs of $\alpha$ imply
\begin{equation}
  C_{\varpi}
  =\sum_i\frac{p_i-\varpi_i}{T_i}.
\label{eq:scalar-euler-remainder-relation}
\end{equation}
Inserting \cref{eq:scalar-euler-remainder} cancels all multiplier terms and gives the local class-III Euler relation

\begin{equation}
\begin{gathered}
  \rho s=\sum_i\frac{\rho_i e_i+p_i-\rho_i\mu_i}{T_i}.
\end{gathered}
\label{eq:weighted-euler}
\end{equation}
Only $\sum_i p_i/T_i$ is constrained; the individual zero-rate functions remain material data.

The scalar restriction does not select an individual entropy flux from this family. For the baseline model we now select its member $\bPhi_p$ by putting
\begin{equation}
  \varpi_i\equiv p_i,
  \qquad
  \pi_i:=P_i-p_i.
\label{eq:lm-pressure}
\end{equation}

The quantity $\pi_i=P_i-p_i$ is commonly called the \emph{non-equilibrium pressure} or \emph{dynamical pressure}.

This selection gives $C_{\varpi}=0$ by \cref{eq:scalar-euler-remainder-relation}. It is not a consequence determining the individual $p_i$. Other flux choices are discussed in \cref{rem:entropy-flux-nonuniqueness}. Section~\ref{sec:closures} closes the anchored difference $P_i-p_i$. For  bulk viscosity in the baseline model, zero scalar dissipation implies $P_i=p_i$.

Combining \cref{eq:weighted-euler} with
\begin{equation}
  \dd(\rho s)
  =\sum_i\frac1{T_i}\,\dd(\rho_i e_i)
   +\sum_i\Bigl(-\frac{\mu_i}{T_i}\Bigr)\dd\rho_i
\label{eq:classIII-gibbs-differential}
\end{equation}
gives the differential compatibility relation
\begin{equation}
  \sum_i\Bigl[
    \frac1{T_i}\,\dd p_i
    +(\rho_i e_i+p_i)\,\dd\!\bigl(\frac1{T_i}\bigr)
    +\rho_i\,\dd\!\bigl(-\frac{\mu_i}{T_i}\bigr)
  \Bigr]=0,
\label{eq:classIII-gibbs-duhem}
\end{equation}
or, equivalently,
\begin{equation}
  \sum_i\Bigl[
    \frac1{T_i}(\dd p_i-\rho_i\,\dd\mu_i)
    +(\rho_i e_i+p_i-\rho_i\mu_i)
      \dd\!\bigl(\frac1{T_i}\bigr)
  \Bigr]=0.
\label{eq:classIII-gibbs-duhem-mu}
\end{equation}
We call \cref{eq:weighted-euler} the local Euler relation and \cref{eq:classIII-gibbs-duhem,eq:classIII-gibbs-duhem-mu} the class-III Gibbs--Duhem compatibility relation.

Equation~\eqref{eq:classIII-gibbs-duhem} requires only the sum of the bracketed one-forms to vanish, not each term separately. Under the hypotheses of~\cite{BotheRuggeriPartII}, their separate vanishing is necessary and sufficient for generating potentials of the prescribed constituent Euler fluxes. Additive entropy alone does not ensure this integrability without the corresponding constituent pressure relations.

The selected entropy flux is
\begin{equation}
  \bPhi
  =\sum_i\Bigl[
  \frac{\bq_i}{T_i}
  +\frac{\rho_i e_i+p_i-\rho_i\mu_i}{T_i}\bu_i
  \Bigr].
\label{eq:entropy-flux}
\end{equation}
Define the algebraic contributions
\begin{equation}
  \rho_i s_i^{\ast}
  :=\frac{\rho_i e_i+p_i-\rho_i\mu_i}{T_i}.
\label{eq:algebraic-partial-entropies}
\end{equation}
Then
\begin{equation}
  \rho s=\sum_i\rho_i s_i^{\ast},
  \qquad
  \rho s\bv+\bPhi
  =\sum_i\Bigl(\rho_i s_i^{\ast}\bv_i+\frac{\bq_i}{T_i}\Bigr).
\label{eq:constituent-entropy-transport-form}
\end{equation}
These algebraic contributions have the Euler form of additive constituent thermodynamics (cf.~\cite[Section~15]{BotheDreyer2015}), but neither additivity nor separate entropy balances are assumed. They coincide with constituent entropy densities when the thermodynamics and pressure partition satisfy individual Euler relations.

\begin{remark}[Nonuniqueness of entropy flux and Euler representation]
\label[remark]{rem:entropy-flux-nonuniqueness}
The divergence-cancelling choice \cref{eq:entropy-flux-lambda} is not unique. A constitutive addition $\boldsymbol K$ changes production by $\Div\boldsymbol K$ and is admissible only if reduction with the same balances and constraints gives the selected physical binary products, each satisfying~(IV) and anchoring~(V). A zero addition may motivate a different flux with different co-factors and mechanisms, but does not itself change flux or production. Flux, pressure split, Euler representation, and closure must be specified consistently.

\end{remark}

\subsection{Final reduced entropy production and co-factor selection}

With $P_i=p_i+\pi_i$, use
\begin{equation}
  \grad\frac{p_i}{T_i}
  =\frac1{T_i}\grad p_i+p_i\grad\frac1{T_i}
\label{eq:partial-pressure-product-rule}
\end{equation}
to obtain
\begin{align}
\zeta={}&
\sum_i\frac1{T_i}
\bigl(\bS_i^{\circ}:\bD_i^{\circ}-\pi_i\Div\bv_i\bigr)
+\sum_i\bq_i\cdot\grad\frac1{T_i}
+\sum_i\frac1{T_i}\varepsilon_i
-\sum_i r_i\frac{\mu_i}{T_i}
\label{eq:entropy-production-expanded}
\\
&-\sum_i\bu_i\cdot\Bigl[
\rho_i\grad\frac{\mu_i}{T_i}
-\frac1{T_i}\grad p_i
-(\rho_i e_i+p_i)\grad\frac1{T_i}
+\frac1{T_i}\bigl(\bmom_i-\half r_i\bu_i\bigr)
\Bigr].
\nonumber
\end{align}
Define the noninteraction part of the selected diffusion co-factor by
\begin{equation}
  \bB_i
  :=\rho_i\grad\frac{\mu_i}{T_i}
  -\frac1{T_i}\grad p_i
  -(\rho_i e_i+p_i)\grad\frac1{T_i}.
\label{eq:raw-diffusion-cofactor}
\end{equation}
The spatial form of \cref{eq:classIII-gibbs-duhem} gives
\begin{equation}
  \sum_i\bB_i=\boldsymbol 0.
\label{eq:raw-cofactor-zero-sum}
\end{equation}
Thus the thermodynamic co-factor already lies in the zero-sum dual space. Projection is needed only after adding thermo-diffusion and temperature-weighted momentum and mass-production terms.

The kinetic constitutive quantities for chemical transformation are the molar pathway rates $R_a$. The species mass productions $r_i$ are their stoichiometric combinations. Rewriting the chemical contribution with \cref{eq:mass-production} gives
\begin{equation}
  \sum_i r_i\frac{\mu_i}{T_i}
  =\sum_{a=1}^{N_R}R_a\Acal_a,
\label{eq:reaction-source-identity}
\end{equation}
where the conjugate, temperature-weighted reaction affinities are
\begin{equation}
  \Acal_a:=\sum_iM_i\nu_i^a\frac{\mu_i}{T_i}.
\label{eq:classIII-affinity}
\end{equation}
At common temperature (Classes~I and~II), $\Acal_a=T^{-1}\sum_iM_i\nu_i^a\mu_i$; at unequal constituent temperatures no common factor can be extracted. The entropy-conjugate force is $-\Acal_a=\sum_iM_i\nu_i^a(-\mu_i/T_i)$, so the reaction contribution can equivalently be written as $+\sum_aR_a(-\Acal_a)$.

The entropy production therefore takes the form
\begin{align}
\zeta={}&
\sum_i\frac1{T_i}
\bigl(\bS_i^{\circ}:\bD_i^{\circ}-\pi_i\Div\bv_i\bigr)
+\sum_i\bq_i\cdot\grad\frac1{T_i}
\label{eq:entropy-production}
\\
&-\sum_i\bu_i\cdot\Bigl[
\bB_i+\frac1{T_i}\bigl(\bmom_i-\half r_i\bu_i\bigr)
\Bigr]
+\sum_i\frac1{T_i}\varepsilon_i
-\sum_{a=1}^{N_R}R_a\Acal_a.
\nonumber
\end{align}

The source-dependent part of \cref{eq:entropy-production} can equivalently be left in the unrearranged Galilean-invariant form,
\begin{equation}
  \sum_i\Bigl[
  -\frac{\mu_i-\half\bu_i^2}{T_i}r_i
  -\frac1{T_i}\bu_i\cdot\bmom_i
  +\frac1{T_i}\,\varepsilon_i
  \Bigr].
\label{eq:ruggeri-source-pairing}
\end{equation}
No kinetic term is discarded: $r_i\bu_i^2/(2T_i)$ is assigned to relative momentum, while chemical rates pair with the thermochemical affinity \cref{eq:classIII-affinity}. This entropy-invariant regrouping extends the common-temperature reacting Class-II choice.

The selected representation contains the combined flux
\begin{equation}
  \bq_i+(\rho_i e_i+p_i)\bu_i.
\label{eq:enthalpy-carrying-flux}
\end{equation}
Here $\rho_i e_i+p_i$ is the partial enthalpy density. We retain \cref{eq:entropy-production} to distinguish constituent heat conduction $\bq_i$ from relative partial-enthalpy transport $(\rho_i e_i+p_i)\bu_i$. Combining them into one primitive heat flux would require closing the combined contribution. Sharp-interface thermodynamics similarly distinguishes intrinsic bulk/surface heat fluxes from enthalpy transported by one-sided mass transfer~\cite{BotheSharpInterface2022,BotheSharpClassII2025}.

\subsection{Levels of transport and closure representation}
\label{subsec:representation-levels}

Several distinct transformations are used in mixture thermodynamics. Some are purely kinematic. Others act on constitutive quantities. A closure that is diagonal in one representation need not remain diagonal in another.

\paragraph{Change of transport reference.}
The kinematic identity $\bv_i=\bv+\bu_i$ separates constituent convection into barycentric and diffusive transport, as in \cref{eq:constituent-entropy-transport-form}, without changing balances or constitutive quantities.

\paragraph{Repartition between energy and entropy fluxes.}
At common temperature, chemical-potential transport can be shifted between an energy-dimensional thermal flux and the entropy flux's explicit diffusion term. M\"uller- and Gurtin-type representations must be compared with their balances and force bases~\cite{Mueller1968,Gurtin1971,Prahs2026}. In multi-velocity models, the enthalpy-shifted thermal flux lacks the stress-work and diffusion-kinetic contributions of the physical total-energy flux. At unequal temperatures no common thermal factor exists, so we retain separate constituent entropy-flux contributions. Redistributing partial energy transport requires specifying the changed fluxes and deriving their entropy contributions. The explicit common-temperature M\"uller--Gurtin forms are given in \cref{rem:muller-gurtin-fluxes}.

\begin{remark}[Bulk equivalence and boundary fluxes]
\label[remark]{rem:boundary-port-repartition}

The bulk identities hold pointwise. Equivalence of boundary-value problems additionally requires expressing the boundary data in the transformed flux variables. On the common-temperature manifold define
\begin{equation}
  \bq^{\mathrm h}:=\sum_i[\bq_i+(\rho_i e_i+p_i)\bu_i],
  \qquad
  \boldsymbol\tau_i:=\bS_i+p_i\bI.
\label{eq:thermal-flux-and-irreversible-stress}
\end{equation}
Here $\boldsymbol\tau_i$ is the irreversible partial stress.
With $p:=\sum_i p_i$, rearranging the pressure work in \cref{eq:thermal-energy-balance} gives the exact thermal balance

\[
 \partial_t(\rho e)+\Div(\rho e\bv+\bq^{\mathrm h})
 =-p\Div\bv+\sum_i\bigl[\boldsymbol\tau_i:\grad\bv_i
   +\bu_i\cdot(\grad p_i-\bmom_i)+\half r_i\bu_i^2\bigr].
\]
The exact nonconvective energy-flux term in \cref{eq:mixture-total-energy-flux} satisfies
\begin{equation}
  \bq^{\mathrm{tot}}
  =\bq^{\mathrm h}-\sum_i\boldsymbol\tau_i\bu_i
   +\half\sum_i\rho_i|\bu_i|^2\bu_i.
\label{eq:physical-versus-thermal-energy-flux}
\end{equation}
Thus the selected entropy flux is $\bPhi=\bq^{\mathrm h}/T-\sum_i\mu_i\bj_i/T$. If this nonconvective energy-flux term is shifted instead, set
\begin{equation}
  \widehat{\bq}:=\bq^{\mathrm{tot}}-\sum_i\mu_i\bj_i.
\label{eq:shifted-energy-flux-boundary}
\end{equation}
The associated entropy representation is consequently
\begin{equation}
  \bPhi=\frac{\widehat{\bq}}T
  +\frac1T\sum_i\boldsymbol\tau_i\bu_i
  -\frac1{2T}\sum_i\rho_i|\bu_i|^2\bu_i,
\label{eq:physical-shifted-entropy-flux}
\end{equation}
not $\widehat{\bq}/T$ without additional mechanical restrictions. At a boundary with unit normal $\boldsymbol n$, the normal nonconvective flux component satisfies
\begin{equation}
  \bq^{\mathrm{tot}}\cdot\boldsymbol n
  =\widehat{\bq}\cdot\boldsymbol n
   +\sum_i\mu_i\bj_i\cdot\boldsymbol n.
\label{eq:boundary-energy-port-transform}
\end{equation}
The complete energy flux through a fixed spatial boundary is $(\rho E\bv-\bS\bv+\bq^{\mathrm{tot}})\cdot\boldsymbol n$, including convection and traction power as in \cref{eq:mixture-total-energy}.
The two normal nonconvective flux components coincide when every $\bj_i\cdot\boldsymbol n=0$, but need not coincide at a permeable boundary. At an impermeable wall, zero shifted normal nonconvective energy flux therefore gives zero physical normal nonconvective energy flux. The normal entropy flux can still contain partial stress power if tangential relative motion remains.

Thus the nonconvective energy flux and the entropy flux remain distinct: identifying them up to $1/T$ omits chemical diffusion or, after its shift, the displayed mechanical corrections. Writing $\bPhi=\bq/T$ is consistent only when $\bq$ denotes the corresponding entropy-weighted flux and the energy balance and boundary data are transformed accordingly.

\end{remark}

\paragraph{Change of bilinear co-factor basis.}
Let $\boldsymbol{\mathcal J}$ and $\boldsymbol{\mathcal X}$ collect selected co-factors of one tensorial rank so that their contribution to entropy production is $\boldsymbol{\mathcal J}\mathbin{\boldsymbol\cdot}\boldsymbol{\mathcal X}$. For an invertible transformation $\mathsf A$ acting on the mechanism index, the simultaneous change
\begin{equation}
  \boldsymbol{\mathcal J}'=\mathsf A\boldsymbol{\mathcal J},
  \qquad
  \boldsymbol{\mathcal X}'=(\mathsf A^{-1})^{\sf T}\boldsymbol{\mathcal X}
\label{eq:entropy-invariant-basis-change}
\end{equation}
preserves the pairing,
\begin{equation}
  \boldsymbol{\mathcal J}'\mathbin{\boldsymbol\cdot}\boldsymbol{\mathcal X}'
  =\boldsymbol{\mathcal J}\mathbin{\boldsymbol\cdot}\boldsymbol{\mathcal X}.
\label{eq:entropy-pairing-invariant}
\end{equation}
If an already selected linear relation is $\boldsymbol{\mathcal J}=\mathsf L\boldsymbol{\mathcal X}$, the same material relation in the transformed basis is
\begin{equation}
  \boldsymbol{\mathcal J}'
  =\mathsf L'\boldsymbol{\mathcal X}',
  \qquad
  \mathsf L'=\mathsf A\mathsf L\mathsf A^{\sf T}.
\label{eq:onsager-congruence}
\end{equation}
Congruence preserves positive semidefiniteness and reciprocal symmetry, but generally not diagonality. The same cross-coefficient reciprocity is expressed in the transformed co-factors~\cite{Prahs2026}. Alternatively, selecting mixed co-factors first and imposing nonnegative diagonal relations constructs the generally cross-coupled reciprocal matrix of Section~\ref{sec:entropy-principle}. The distinction between coupling elementary mechanisms and selecting an unresolved block is explained in Section~\ref{subsec:entropy-mechanisms}\ above.

\section{\texorpdfstring{{Constrained co-factor spaces and closure selection}}{Constrained co-factor spaces and closure selection}}
\label{sec:closure-philosophy}

\subsection{Dual constrained spaces}

Write $\boldsymbol y=(y_1,\ldots,y_N)^{\sf T}$ for the mass-fraction vector, where $y_i=\rho_i/\rho$ and $\sum_i y_i=1$, and $\bone=(1,\ldots,1)^{\sf T}$. At each spatial point, the diffusion velocities form the barycentrically constrained space
\begin{equation}
  \Ucal_{\boldsymbol y}
  :=\bigl\{
  \boldsymbol U=(\boldsymbol U_1,\ldots,\boldsymbol U_N):
  \sum_i y_i\boldsymbol U_i=\boldsymbol 0
  \bigr\}.
\label{eq:velocity-space}
\end{equation}
The entropy pairing is unchanged if an array $(\rho_i\boldsymbol w)_i$ is added to a diffusion co-factor, because
\begin{equation}
  \sum_i\bu_i\cdot\rho_i\boldsymbol w
  =\boldsymbol w\cdot\sum_i\rho_i\bu_i=0.
\label{eq:annihilator-pairing}
\end{equation}
Here $\boldsymbol w$ is an arbitrary common spatial vector with the units of a diffusion co-factor divided by mass density. It is not a velocity.
The dual zero-sum force space is
\begin{equation}
  \Fcal
  :=\bigl\{
  \boldsymbol F=(\boldsymbol F_1,\ldots,\boldsymbol F_N):
  \sum_i\boldsymbol F_i=\boldsymbol 0
  \bigr\}.
\label{eq:force-space}
\end{equation}
The pairing $\langle\boldsymbol U,\boldsymbol F\rangle=\sum_i\boldsymbol U_i\cdot\boldsymbol F_i$ makes these spaces dual: arrays differing by a multiple of $\boldsymbol y$ act identically on $\Ucal_{\boldsymbol y}$, and exactly one representative has zero sum. Thus every linear functional on $\Ucal_{\boldsymbol y}$ has a unique representative in $\Fcal$.
We introduce the oblique projection pair
\begin{equation}
  \bP_{\boldsymbol y}:=\boldsymbol I_N-\bone\otimes\boldsymbol y,
  \qquad
  \bP_{\boldsymbol y}^{\sf T}:=\boldsymbol I_N-\boldsymbol y\otimes\bone,
\label{eq:projection-pair}
\end{equation}
acting on the constituent index while leaving the spatial vector components unchanged. For an arbitrary array $\boldsymbol Z=(\boldsymbol Z_1,\ldots,\boldsymbol Z_N)$ of spatial co-factor vectors, these projections satisfy
\begin{equation}
  \bP_{\boldsymbol y}\boldsymbol U=\boldsymbol U
  \quad(\boldsymbol U\in\Ucal_{\boldsymbol y}),
  \qquad
  \bP_{\boldsymbol y}^{\sf T}\boldsymbol Z\in\Fcal,
\label{eq:projection-properties1}
\end{equation}
and
\begin{equation}
  \sum_i\boldsymbol U_i\cdot\boldsymbol Z_i
  =\sum_i\boldsymbol U_i\cdot
  (\bP_{\boldsymbol y}^{\sf T}\boldsymbol Z)_i
  \quad(\boldsymbol U\in\Ucal_{\boldsymbol y}).
\label{eq:projection-properties2}
\end{equation}
This is the permutation-symmetric counterpart of eliminating one distinguished constituent. The dual-projection structure and the associated distinction between constraint-induced coupling and additional constitutive cross-effects are developed for Class-I diffusion in Bothe and Druet~\cite[Sections~4--5 and~8]{BotheDruet2023}.  Projection onto the $N-1$ independent relative directions can produce off-diagonal component entries even from a diagonal core. These are constraint-induced couplings, as in the core-diagonal closure of Bothe and Druet~\cite[Section~8]{BotheDruet2023}. An additional constitutive cross-effect acts on the already constrained mechanism space. Likewise, eliminating resolved Class-III momenta can turn pairwise resistance into a dense Class-I mobility without adding microscopic mechanisms.

\paragraph{Resistance, mobility, and primitive-variable diffusion matrices.}
Resistance maps prescribed relative velocities to opposing temperature-weighted force co-factors, $\boldsymbol{\mathfrak b}=-\bM\bu$; mobility maps prescribed thermodynamic forces to velocities or fluxes. Solving a balance-compatible quasistationary relative-momentum relation for the relative velocities requires invertibility of the resulting resistance on the constrained space. In reacting mixtures, this operator includes the mass-production contributions retained by the reduction's scaling; see~\cite[Section~4.5]{BotheRuggeriPartII}. Multiplication by the constituent densities gives $\bj_i=\rho_i\bu_i$, with conjugate forces transformed consistently. Expressing the thermodynamic conjugates in density, energy, temperature, or composition gradients then yields the primitive-variable PDE diffusion matrix:
\begin{equation}
  \text{resistance}
  \xrightarrow{\text{constrained inverse}}
  \text{mobility}
  \xrightarrow{\text{thermodynamic chain rule}}
  \text{PDE diffusion matrix}.
\label{eq:resistance-mobility-diffusion-spine}
\end{equation}
These objects differ in dimensions, state dependence, null spaces, and experimental interpretation. Constrained inversion, projections, and thermodynamic derivatives alone can make the reduced matrices dense. Conversely, interpreting a fitted Fickian matrix as a Class-III pair resistance requires reconstructing the reduction's thermodynamic and mechanical relations.

Coefficient identification must specify the measured object and reference velocity. Inferring resistance from nonideal-mixture Fick data requires consistent frame conversion and independently justified thermodynamic derivatives.  Reproducing closure-generated coefficients verifies a calculation, whereas independent collective current-correlation data retaining interspecies correlations can test the closure. Fits at isolated states do not identify constitutive state dependence.

\subsection{Thermodynamic derivatives versus transport kinetics}
\label{subsec:susceptibility-versus-kinetics}

To separate thermodynamic derivatives from transport kinetics, let
\[
  \boldsymbol z
  :=(\rho_1e_1,\ldots,\rho_Ne_N,\rho_1,\ldots,\rho_N)
\]
and let $\boldsymbol w=D_{\boldsymbol z}\widehat{\rho s}$ collect the entropy conjugates. On the stable interior state domain, the negative entropy Hessian is positive definite:
\begin{equation}
  \mathsf G(\boldsymbol z)
  :=-D_{\boldsymbol z}^2\widehat{\rho s}(\boldsymbol z)
  \succ0.
\label{eq:thermodynamic-susceptibility}
\end{equation}
It gives $\delta\boldsymbol w=-\mathsf G\,\delta\boldsymbol z$, with $[G_{\alpha\beta}]=[\rho s]/([z_\alpha][z_\beta])$. The energy--energy, energy--mass, and mass--mass blocks have SI units $\mathrm{m^3/(J\,K)}$, $\mathrm{m^3/(kg\,K)}$, and $\mathrm{J\,m^3/(kg^2\,K)}$, respectively: they are thermodynamic derivatives, not kinetic coefficients. A linear transport relation in conjugate gradients,
\begin{equation}
  \boldsymbol{\mathcal J}
  =\mathsf L(\boldsymbol z)\,\grad\boldsymbol w,
\label{eq:kinetic-conjugate-law}
\end{equation}
becomes, in primitive thermodynamic gradients,
\begin{equation}
  \boldsymbol{\mathcal J}
  =-\mathsf L(\boldsymbol z)\,\mathsf G(\boldsymbol z)\,
   \grad\boldsymbol z,
\label{eq:kinetic-times-susceptibility}
\end{equation}
up to the mechanism's projections, pressure-partition terms, and sign conventions. In a reduced scalar isothermal relation, the ideal-dilute term $RT\log(c_i/c_i^\circ)$ in $\mu_i^{\mathrm m}$, for molar concentration $c_i>0$ and fixed reference $c_i^\circ>0$, contributes $RT/c_i$ to $\partial_{c_i}\mu_i^{\mathrm m}$. A mobility proportional to $c_i$ can therefore yield a finite concentration-gradient coefficient. This neither determines that mobility nor supplies a Class-III composition-boundary closure.

Thus PDE-level Fickian or temperature-gradient coefficients combine kinetics with thermodynamic restoring forces. Their matrix need not be symmetric even when both factors are self-adjoint and positive in their respective pairings: reciprocity concerns conjugate fluxes and forces, not arbitrary nonlinear thermodynamic coordinates. The analogous Class-I mobility/free-energy-Hessian factorization and distinction between Onsager and Fickian symmetry are discussed in Bothe and Druet~\cite{BotheDruet2023}.

Similarly, a block-diagonal kinetic operator in entropy-conjugate heat and diffusion variables can acquire temperature-gradient terms through the chain rule for quantities such as $\mu_i/T_i$. These coordinate-induced terms are not additional Soret--Dufour coefficients. An explicit thermo-diffusive cross-effect is an off-diagonal kinetic operator between the identified heat and relative-momentum co-factors, as in the  entropy-invariant coupling in the baseline model below.

{The closure of the baseline model parametrizes thermodynamics through $\widehat{\rho s}$ and the compatible pressure partition separately from viscosity, heat conduction, momentum resistance, thermal relaxation, and reaction kinetics.}

\subsection{\texorpdfstring{{Selection of dissipative mechanisms}}{Selection of dissipative mechanisms}}

For the baseline model, we organize the selected dissipative mechanisms in \cref{eq:entropy-production} as
\begin{equation}
  \zeta
  =\zeta_{\mathrm{visc}}
   +\zeta_{\mathrm{heat/diff}}
   +\zeta_{\mathrm{relax}}
   +\zeta_{\mathrm{chem}}.
\label{eq:mechanism-blocks}
\end{equation}
The baseline model makes each selected binary product nonnegative for every thermodynamic process. As in Class-II theory, an entropy-invariant zero addition first introduces thermo-diffusion and its reciprocal effect. Further cross-effects mix compatible co-factors within their parity classes before imposing nonnegative closure on each resulting mechanism. Appendix~\ref{app:full-coupling} gives the reciprocal linear\ (in the co-factors) construction. This requirement concerns the mixed representation: unmixed summands need not remain nonnegative. Physical mechanisms and coefficients still require microscopic, constitutive, or experimental input.

\section{Minimal admissible closure relations}
\label{sec:closures}

\subsection{Partial viscous stresses}

 The viscous closure of the baseline model combines the  dynamical pressure with an uncoupled deviatoric relation:
\begin{equation}
  \bS_i^{\circ}=2\eta_i^{\mathrm v}\bD_i^{\circ},
  \qquad
  \pi_i=-\lambda_i^{\mathrm v}\Div\bv_i,
  \qquad
  \eta_i^{\mathrm v}\ge0,
  \quad
  \lambda_i^{\mathrm v}\ge0.
\label{eq:viscous-closure}
\end{equation}
Hence
\begin{equation}
  \zeta_{\mathrm{visc}}
  =\sum_i\frac{2\eta_i^{\mathrm v}}{T_i}|\bD_i^{\circ}|^2
  +\sum_i\frac{\lambda_i^{\mathrm v}}{T_i}(\Div\bv_i)^2
  \ge0.
\label{eq:viscous-dissipation}
\end{equation}
Entropy-invariant mixing of the deviatoric mechanisms followed by nonnegative linear closure gives the isotropic cross-viscosity relation $\bS_i^{\circ}/T_i=2\sum_j\eta_{ij}\bD_j^{\circ}$ with $[\eta_{ij}]$ symmetric positive semidefinite; see \cref{app:full-deviatoric}. Symmetry pairs deviatoric rates with temperature-weighted stresses. Imposing it on unweighted stresses generally defines a different relation.

\subsection{Entropy-invariant thermo-diffusion mixing}
\label{subsec:thermodiffusion-mixing}

Let
\begin{equation}
  d_i^T=d_i^T(\rho_1e_1,\ldots,\rho_Ne_N,\rho_1,\ldots,\rho_N)
\label{eq:thermal-diffusivity-state}
\end{equation}
be scalar thermo-diffusion coefficients. They may have either sign. To retain the common-temperature convention with $N-1$ coefficient functions~\cite[Section~8]{BotheDreyer2015}, the baseline model assumes
\begin{equation}
  \sum_i d_i^T=0.
\label{eq:thermal-diffusivity-normalization}
\end{equation}

Bird, Stewart and Lightfoot~\cite[Section~24.2, p.~767; Appendix~D]{BirdStewartLightfoot2002} give a zero sum for thermal-diffusion \emph{mass-flux} coefficients. In dilute nonreacting monatomic gases, the common-temperature Chapman--Enskog derivation enforces it through barycentric momentum matching, $\sum_i\bj_i=0$~\cite[Equations~(15b), (34b)]{ArnaultGuisset2022}. These coefficients are distinct from the present $d_i^T$ in the partial heat-flux law.

In Class~II, the replacement $d_i^T\mapsto d_i^T-y_i\sum_jd_j^T$ preserves the summed heat flux and projected diffusion force because $\sum_i y_i\bu_i=0$ and the reciprocal-temperature gradient is common. In Class~III, applying this replacement in \cref{eq:heat-closure} with the other coefficients fixed generally changes the individual heat fluxes. Thus \cref{eq:thermal-diffusivity-normalization} is an admissible simplifying assumption of the baseline model, not a consequence of the entropy principle or a normalization without loss of generality. The identity below holds for unrestricted $d_i^T$; the general vector coupling in Appendix~\ref{app:full-vector} allows this additional freedom.

Starting from the heat and diffusion terms in \cref{eq:entropy-production}, we use the identity
\begin{align}
&\sum_i\bq_i\cdot\grad\frac1{T_i}
-\sum_i\bu_i\cdot\Bigl[
\bB_i+\frac1{T_i}\bigl(\bmom_i-\half r_i\bu_i\bigr)
\Bigr]
\label{eq:entropy-invariant-thermodiffusion}
\\
&\quad=
\sum_i(\bq_i-d_i^T\bu_i)\cdot\grad\frac1{T_i}
-\sum_i\bu_i\cdot\Bigl[
\bB_i-d_i^T\grad\frac1{T_i}
+\frac1{T_i}\bigl(\bmom_i-\half r_i\bu_i\bigr)
\Bigr].
\nonumber
\end{align}

\begingroup
The same $d_i^T$ couples both products. At common temperature they share the reciprocal-temperature gradient, while $\bB_i$ contains the enthalpy contribution. We require the selected heat and relative-momentum blocks to be separately nonnegative. The latter pairs co-factors on the barycentrically constrained space and may have signed constituent summands. The Fourier law below gives $(\bq_i-d_i^T\bu_i)\cdot\grad(1/T_i)=\kappa_i|\grad(1/T_i)|^2\ge0$ for each constituent, and the pairwise entropy resistance gives $f_{ij}\rho_i\rho_j(\bu_i-\bu_j)^2\ge0$ for each $i<j$.
\endgroup
At Class III the $N$ gradients $\grad(1/T_i)$ are distinct. We retain only the displayed same-constituent reciprocal heat/diffusion coupling. Coupling every $\bq_i$ and $\bu_i$ to every reciprocal-temperature gradient is formally admissible but requires microscopic or experimental justification.

We choose the constituent Fourier laws
\begin{equation}
  \bq_i-d_i^T\bu_i
  =\kappa_i\grad\frac1{T_i},
  \qquad \kappa_i\ge0.
\label{eq:heat-closure}
\end{equation}
Equivalently,
\begin{equation}
  \bq_i=-\lambda_i^q\grad T_i+d_i^T\bu_i,
  \qquad
  \lambda_i^q:=\frac{\kappa_i}{T_i^2}\ge0.
\label{eq:fourier-T}
\end{equation}
The terms represent intrinsic conduction and reciprocal diffusion-thermo transport. Diagonality refers to the mixed co-factor basis of \cref{eq:entropy-invariant-thermodiffusion}: transforming the same operator according to \cref{eq:onsager-congruence} generally creates off-diagonal blocks. Reimposing diagonality in another basis instead changes the material model~\cite{Prahs2026}.

A nonreacting multitemperature Boltzmann expansion with modified collision scaling yields pairwise relative-velocity contributions to partial heat fluxes~\cite[Equation~(21)]{SavenkoPetersKerkhof2008}. Establishing the sparse form $d_i^T\bu_i$ and \cref{eq:thermal-diffusivity-normalization} for a particular kinetic model requires matching its partial heat fluxes and exchange moments.

\subsection{Projected Maxwell--Stefan resistance}

Maxwell--Stefan resistance relates diffusion forces to pairwise relative velocities; see Bothe and Dreyer~\cite{BotheDreyer2015} for the common-temperature derivation and Bothe and Druet~\cite{BotheDruet2023} for comparison with other closures. Projection~\eqref{eq:projection-pair} subtracts each constituent's mass fraction times the summed co-factor, selecting its zero-sum representative without changing the entropy pairing.

Define the modified diffusion co-factor
\begin{equation}
  \widetilde{\bB}_i
  :=\bB_i-d_i^T\grad\frac1{T_i}
  +\frac1{T_i}\bigl(\bmom_i-\half r_i\bu_i\bigr).
\label{eq:mixed-diffusion-cofactor}
\end{equation}
Its canonical zero-sum representative is
\begin{equation}
  \boldsymbol{\mathfrak b}
  :=\bP_{\boldsymbol y}^{\sf T}\widetilde{\boldsymbol B},
  \qquad
  \sum_i\boldsymbol{\mathfrak b}_i=\boldsymbol 0.
\label{eq:canonical-force}
\end{equation}
The resistance closure is
\begin{equation}
  \boldsymbol{\mathfrak b}=-\bM\bu.
\label{eq:resistance-closure}
\end{equation}
Here $\bM$ is self-adjoint and positive semidefinite on the constituent index space, with
\begin{equation}
  \bM\bone=\boldsymbol 0,
  \qquad
  \ker\bM=\operatorname{span}\{\bone\}
\label{eq:resistance-properties}
\end{equation}
for strict relaxation. The sign in \cref{eq:resistance-closure} is fixed by the minus sign in the diffusion product of \cref{eq:entropy-production}.

For the baseline model we select binary-type interactions,
\begin{equation}
  (\bM\bu)_i
  =\sum_{j\ne i}f_{ij}\rho_i\rho_j(\bu_i-\bu_j),
  \qquad
  f_{ij}=f_{ji}\ge0.
\label{eq:pairwise-resistance}
\end{equation}
For $N\ge3$, nonnegative pair coefficients impose a stronger restriction than positivity of the constrained resistance alone~\cite{BotheDruet2023}. The composition boundary of $\{\boldsymbol y:y_i\ge0,\ \sum_i y_i=1\}$ means disappearance of a constituent. The density prefactor gives vanishing interaction only if $\rho_i\rho_j f_{ij}\to0$ along the boundary approach, as when $f_{ij}$ has a finite limit. Thermodynamics neither requires $f_{ij}\to0$ nor excludes singular dilute limits. Its extension is constitutive input. The strict kernel in \cref{eq:resistance-properties} therefore concerns interior active species. At disappearance, the interaction graph and unreduced $N$-species nullspace change, requiring an active-set model or compatible extension as in \cref{sec:closed-model}.
For these pairwise coefficients,
\begin{equation}
  \sum_i\bu_i\cdot(\bM\bu)_i
  =\sum_{i<j}f_{ij}\rho_i\rho_j(\bu_i-\bu_j)^2\ge0.
\label{eq:resistance-dissipation}
\end{equation}
With the thermodynamic state and pair coefficients held fixed, $\mathcal{R}_{\mathrm s}(\bu):=\tfrac12\sum_{i<j}f_{ij}\rho_i\rho_j(\bu_i-\bu_j)^2$ is a Rayleigh-type potential with $D_{\bu_i}\mathcal{R}_{\mathrm s}=(\bM\bu)_i$ and twice its value equal to \cref{eq:resistance-dissipation}~\cite{RayleighDissipation1873}. It has entropy-production units; $T\mathcal{R}_{\mathrm s}$ is the mechanical Rayleigh function at common temperature.
For the strict kernel statement, the graph with active constituents as vertices and edges $f_{ij}\rho_i\rho_j>0$ must be connected: every two vertices must be joined by a path of active edges. Zero production then makes all $\bu_i$ equal, hence zero by the barycentric constraint.

To compare the coefficients in \cref{eq:pairwise-resistance} with conventional Maxwell--Stefan diffusivities $\mathcal D_{ij}$, restrict to the common-temperature, nonreactive mechanical block. Let $c=\sum_i\rho_i/M_i$ and $x_i=\rho_i/(M_i c)$, where $M_i$ is the molar mass. Matching \cref{eq:resistance-dissipation} with the molar entropy production $cR\sum_{i<j}x_ix_j(\bu_i-\bu_j)^2/\mathcal D_{ij}$ gives
\[
 f_{ij}=\frac{R}{cM_iM_j\mathcal D_{ij}}.
\]
The mechanical pair-drag coefficient is $Tf_{ij}\rho_i\rho_j$. Pairwise velocity differences are unchanged by a common reference-velocity shift. The corresponding individual diffusion fluxes and constraints must nevertheless be transformed consistently.

The projected relation determines the physical objective momentum productions uniquely. Write
\begin{equation}
  \boldsymbol G_i
  :=\bB_i-d_i^T\grad\frac1{T_i}
   -\frac{r_i}{2T_i}\bu_i.
\label{eq:noninteraction-cofactor}
\end{equation}
Then
\begin{equation}
  \widetilde{\bB}_i=\boldsymbol G_i+\frac{\bmom_i}{T_i}.
\label{eq:cofactor-source-split}
\end{equation}
Since $\bP_{\boldsymbol y}^{\sf T}\widetilde{\boldsymbol B}=-\bM\bu$, there exists one spatial vector $\blambda$ such that
\begin{equation}
  \widetilde{\bB}_i=-(\bM\bu)_i+y_i\blambda.
\label{eq:gauge-representation}
\end{equation}
Consequently,
\begin{equation}
  \bmom_i
  =T_i\bigl[-(\bM\bu)_i+y_i\blambda-\boldsymbol G_i\bigr].
\label{eq:momentum-reconstruction}
\end{equation}
The constraint $\sum_i\bmom_i=\boldsymbol 0$ determines
\begin{equation}
  \blambda
  =\Bigl(\sum_jT_j\bigl[(\bM\bu)_j+\boldsymbol G_j\bigr]\Bigr)
    \Big/\Bigl(\sum_jy_jT_j\Bigr).
\label{eq:momentum-multiplier}
\end{equation}
The denominator is positive, and $\blambda$ is the annihilator component enforcing total momentum conservation. Its temperature weighting reduces to the common-temperature form when all $T_i$ coincide. At unequal temperatures, however, this reconstruction need not preserve the pair graph of $\bM$: the conserving $y_iT_i\blambda$ term can contribute to a constituent with no incident resistance edge. Thus the binary structure in \cref{eq:pairwise-resistance} concerns entropy resistance. Identifying it with pairwise physical momentum exchange requires additional constitutive justification.

Combining \cref{eq:heat-closure,eq:resistance-closure} gives
\begin{equation}
  \zeta_{\mathrm{heat/diff}}
  =\sum_i\kappa_i\Bigl|\grad\frac1{T_i}\Bigr|^2
  +\sum_i\bu_i\cdot(\bM\bu)_i\ge0.
\label{eq:heat-diff-dissipation}
\end{equation}

Freezing coefficients and linearizing the homogeneous, nonreactive mechanical subsystem without differential body forces about a rest state with $\bv_i=0$ and $T_i=T$ for all $i$ gives $\operatorname{diag}(\rho_i)\dot{\bu}=-T\bM\bu$. The nonzero decay rates therefore solve
\[
 T\bM\boldsymbol a=\lambda\operatorname{diag}(\rho_i)\boldsymbol a,
 \qquad \sum_i\rho_i\boldsymbol a_i=\boldsymbol0.
\]
For $\mathsf D_{\rho}:=\operatorname{diag}(\rho_i)$, the generator is similar to the symmetric negative-semidefinite matrix $-T\mathsf D_{\rho}^{-1/2}\bM\mathsf D_{\rho}^{-1/2}$. Connectedness and the barycentric constraint exclude its common-velocity null mode, leaving real positive relative-momentum decay rates. These frozen homogeneous rates neither establish full spatial stability nor equal the diffusion eigenvalues after quasistationary elimination and the thermodynamic chain rule.

\subsection{Dissipative thermal relaxation and general energy redistribution}
\label{subsec:thermal-relaxation}

\begingroup
Closing $r_i$, $\bmom_i$, and $\varepsilon_i$ determines the conservative sources through \cref{eq:source-reconstruction-raw}. No independent $k_i$ closure is needed. Objective energy exchange contributes $\sum_i\varepsilon_i/T_i$ with $\sum_i\varepsilon_i=0$, whereas the actual nonkinetic-energy source $\varepsilon_i-\bu_i\cdot\bmom_i+\half r_i\bu_i^2$ need not sum to zero; see \cref{eq:partial-internal-energy-objective}. The thermal co-factors are consequently reciprocal-temperature differences, but admissible energy redistribution need not depend only on them.
\endgroup

For the baseline model, we sum pairwise thermal exchanges:
\begin{equation}
  \varepsilon_i^{T}
  =\sum_{j=1}^{N}b_{ij}\Bigl(\frac1{T_i}-\frac1{T_j}\Bigr),
  \qquad
  b_{ij}=b_{ji}\ge0,
  \qquad
  b_{ii}=0.
\label{eq:thermal-relaxation}
\end{equation}
In matrix form, $\boldsymbol\varepsilon^T=\mathsf L_b\boldsymbol\beta$, with $\beta_i=1/T_i$, $(\mathsf L_b)_{ii}=\sum_{j\ne i}b_{ij}$ and $(\mathsf L_b)_{ij}=-b_{ij}$ for $i\ne j$. This weighted graph Laplacian has constituents as vertices and edges $b_{ij}>0$: it has the structure of \cref{eq:pairwise-resistance}, with thermal weights $b_{ij}$ and reciprocal temperatures replacing $f_{ij}\rho_i\rho_j$ and velocities. Hence
\begin{equation}
  \sum_i\varepsilon_i^{T}=0,
\label{eq:thermal-relaxation-zero-sum}
\end{equation}
and
\begin{equation}
  \zeta_{\mathrm{relax}}
  =\sum_i\frac1{T_i}\,\varepsilon_i^{T}
  =\sum_{i<j}\Bigl(\frac1{T_i}-\frac1{T_j}\Bigr)
   \Bigl[b_{ij}\Bigl(\frac1{T_i}-\frac1{T_j}\Bigr)\Bigr]\ge0.
\label{eq:thermal-relaxation-dissipation}
\end{equation}

Pair contributions are nonnegative, although $\varepsilon_i^T/T_i$ may have either sign. Zero production forces equal temperatures within every connected component of the thermal graph, hence a common temperature if it is connected. The entropy product fixes the reciprocal-temperature co-factors; kinetic theory or experiment must determine the coefficients in the applicable regime.

\par\smallskip\noindent\emph{Remark (optional process-resolved thermal relaxation).} A purely thermal forward process transferring $E_{ij}>0$ per mole from constituent $j$ to $i$, without composition or momentum change, has entropy increment $\chi_{ij}=(E_{ij}/R)(1/T_i-1/T_j)$, with molar gas constant $R$. Positive molar rates satisfying $\log(W_{ij}^f/W_{ij}^b)=\chi_{ij}$ and pair source $E_{ij}(W_{ij}^f-W_{ij}^b)$ yield $R(W_{ij}^f-W_{ij}^b)\chi_{ij}\ge0$. Linearization about a state with $T_i=T_j$ and \mbox{$W_{ij}^f=W_{ij}^b$} gives \cref{eq:thermal-relaxation} with $b_{ij}^{\mathrm{eq}}=(E_{ij}^{\mathrm{eq}})^2W_{ij}^{\mathrm{eq}}/R$, where $W_{ij}^{\mathrm{eq}}:=W_{ij}^{f,\mathrm{eq}}=W_{ij}^{b,\mathrm{eq}}$. Outside this regime, both the kinetic prefactor and energy increment need constitutive justification. The forward and reverse transfers must be physically specified, not merely an algebraic split of a net source. This is the thermal specialization of \cref{subsec:general-event-channels}. Any composition or momentum changes must also enter the complete entropy increment. The second law selects neither linear nor logarithmic closure.

 The coefficients of the baseline model $f_{ij}$ and $b_{ij}$ are independent. The entropy inequality does not identify their relaxation times. A common microscopic collision model requires consistent calculation of its first velocity and energy moments, which determine different coefficients and need not give equal relaxation times. Momentum-transport agreement alone therefore does not validate thermal exchange, especially for unequal masses with different momentum and thermal equilibration efficiencies per encounter.

More generally, an objective zero-sum energy redistribution may be decomposed, as part of a chosen constitutive representation, as
\begin{equation}
  \varepsilon_i=\varepsilon_i^{T}+\varepsilon_i^{0},
  \qquad
  \sum_i\varepsilon_i^{0}=0.
\label{eq:energy-redistribution-decomposition}
\end{equation}
If the additional redistribution is entropy neutral in the selected representation, it must also satisfy
\begin{equation}
  \sum_i\frac1{T_i}\,\varepsilon_i^{0}=0.
\label{eq:entropy-neutral-energy-redistribution}
\end{equation}

At a fixed common-temperature state the second condition follows from the first. Otherwise the two constraints are independent: they force zero redistribution for a binary mixture and leave dimension $N-2$ for $N\ge3$. The split~\eqref{eq:energy-redistribution-decomposition} need not be unique and distinguishes the graph-Laplacian model from general objective redistribution. Energy transfer associated with relative motion or reaction must be evaluated jointly with its momentum or reaction contribution in the full entropy pairing.

\begingroup
This pointwise freedom is restricted by regularity: any continuous $\boldsymbol\varepsilon^0$ satisfying both conditions on an open full Class-III domain vanishes at common temperature. Indeed, use the locally independent coordinates $\boldsymbol\beta=(1/T_i)_i$ at fixed densities and set $\boldsymbol\beta_0=\alpha\bone$. For every $\boldsymbol d\perp\bone$ and sufficiently small nonzero $t$,
\[
  0=(\boldsymbol\beta_0+t\boldsymbol d)\cdot
       \boldsymbol\varepsilon^0(\boldsymbol\beta_0+t\boldsymbol d)
    =t\boldsymbol d\cdot
       \boldsymbol\varepsilon^0(\boldsymbol\beta_0+t\boldsymbol d).
\]
Continuity gives $\boldsymbol d\cdot\boldsymbol\varepsilon^0(\boldsymbol\beta_0)=0$ for all such $\boldsymbol d$; zero sum then forces $\boldsymbol\varepsilon^0(\boldsymbol\beta_0)=0$. A zero-sum redistribution within an equal-temperature group is neutral on that restricted manifold, but this does not specify its extension when group temperatures separate. A continuous, everywhere neutral full-domain extension must satisfy both conditions and vanish at full common temperature. Any prescribed nonzero common-temperature redistribution therefore requires an entropy-active contribution in such an extension. This statement does not exclude finite exchanges required to maintain a common-temperature constraint, or arising formally in a singular fast-relaxation limit. Such exchanges are distinct from a continuous, everywhere entropy-neutral Class-III source law; see~\cite[Section~4.4]{BotheRuggeriPartII}.
\endgroup

The baseline model sets $\varepsilon^0=0$. The resolved-process formulation below specifies each constituent's energy increments. Only their component satisfying \cref{eq:entropy-neutral-energy-redistribution} is neutral. Entropy-active reaction--energy and dilatation--energy couplings require a different specification, as in \cref{app:full-coupling}.

\subsection{Chemical reaction rates}

{In this section and \cref{subsec:general-event-channels}, $R=N_{\mathrm A}k_{\mathrm B}$ is the molar gas constant. Reaction and resolved-process progress are measured in moles, and their rates in moles per unit volume and time.}
\label{subsec:reaction-rates}

Under specification~(M1) for the baseline model, chemical conversion with affinity~\eqref{eq:classIII-affinity} is separate from zero-sum thermal redistribution. Its driving co-factor can remain finite at a locally well-defined thermodynamic state. We therefore retain positive forward and backward rates and relate their ratio logarithmically to the resolved process's entropy change. This local-detailed-balance interpretation concerns the relative statistical weights of a process and its reverse. It does not require an activation barrier, which, when present, affects the magnitudes of the one-way rates. Write
\begin{equation}
  R_a=R_a^f-R_a^b,
  \qquad R_a^f>0,\quad R_a^b>0,
\label{eq:one-way-rates}
\end{equation}
on the interior thermodynamic state domain, where all active constituent densities are positive. The logarithmic forward--backward closure
\begin{equation}
  \log\frac{R_a^f}{R_a^b}
  =-\frac{\gamma_a}{R}\Acal_a,
  \qquad
  \gamma_a>0,
\label{eq:reaction-closure}
\end{equation}
goes back to Marcelin, Kohnstamm, and De Donder~\cite{Marcelin1910,Lengyel1989,Pekar2005}. Its balance-anchored application to elementary sorption and one-sided transfer was developed in multi-velocity sharp-interface theory~\cite{BotheSharpClassII2025}. For unmodified local detailed balance, the logarithmic ratio equals the selected mechanism's dimensionless entropy change. This statistical interpretation requires sufficient forward--reverse resolution beyond entropy nonnegativity~\cite{Maes2021}. Algebraically splitting a net rate does not identify physical one-way processes or establish local detailed balance. Logarithmic monotonicity gives
\begin{equation}
  -R_a\Acal_a
  =\frac{R}{\gamma_a}(R_a^f-R_a^b)
  \log\!\Bigl(\frac{R_a^f}{R_a^b}\Bigr)\ge0.
\label{eq:reaction-dissipation}
\end{equation}
Hence the reactive entropy production vanishes for path $a$ exactly when
\begin{equation}
  R_a^f=R_a^b
  \quad\Longleftrightarrow\quad
  \Acal_a=0.
\label{eq:path-detailed-equilibrium}
\end{equation}

\begingroup
This is detailed balance of one reaction mechanism; thermodynamic equilibrium in \cref{def:equilibrium-reversibility} requires zero production of every selected mechanism. For bidirectional mass action at fixed common temperature, pressure, and standard-state convention, simultaneous detailed equilibrium requires the classical Wegscheider cycle conditions on rate constants. General positive one-way rate relations need not admit such constants. In either case, existence of an interior state solving all affinity equations in the chosen compatibility class and thermodynamic domain is separate from both cycle compatibility and equality of one rate pair.
\endgroup

For the separated reaction mechanism, \cref{eq:reaction-closure} fixes the rate ratio, while one rate requires kinetic, transition-state, or experimental input. The baseline model uses unmodified local detailed balance, $\gamma_a=1$, unless stated otherwise. Other positive values preserve nonnegativity and the pathwise detailed-balance set but phenomenologically rescale the entropy term, generally changing its relation to conventional mass action. Thus the ratio follows from the selected decomposition and logarithmic constitutive ansatz.

The value $\gamma_a=1$ refers to a fixed physical process, complete entropy increment, and molar normalization. The second law alone does not fix it. Renormalizing progress changes stoichiometric and source increments: $\nu_i^a\mapsto\lambda\nu_i^a$ and $R_a^f,R_a^b\mapsto R_a^f/\lambda,R_a^b/\lambda$ for $\lambda>0$ preserve species productions and the rate ratio, but multiply $\Acal_a$ by $\lambda$, requiring $\gamma_a\mapsto\gamma_a/\lambda$. Changing $\gamma_a$ at fixed data instead changes the relation, independently of the common kinetic scale.

For this closure define
\[
 X_a:=-\frac{\gamma_a\Acal_a}{R},
 \qquad q_a:=\sqrt{R_a^fR_a^b}>0.
\]
Then
\[
 R_a=2q_a\sinh(X_a/2),
 \qquad
 \frac{R_a^f-R_a^b}{R_a^f+R_a^b}=\tanh(X_a/2).
\]
Thus $|X_a|\ll1$ means that the forward and backward rates differ little compared with their sum, whether both are large or small. Linearization at regular detailed balance gives $\delta R_a=q_a^{\mathrm{eq}}\delta X_a$. Small absolute entropy production does not imply near balance: both rates may be small at finite $X_a$. Since $q_a$ may depend on state, the hyperbolic-sine form alone does not prescribe the full nonlinear response.

On $T_i=T$, the weighted Euler relation~\eqref{eq:weighted-euler} and $\rho e=\sum_i\rho_i e_i$ give
\begin{equation*}
  T\rho s=\rho e+p-\sum_i\rho_i\mu_i,
  \qquad
  p:=\sum_i p_i.
\end{equation*}
Here $p$ is the thermodynamic mixture pressure. For comparison with mass action, introduce
\begin{equation}
  \mu_i^{\mathrm m}:=M_i\mu_i,
  \qquad
  \mu_i^{\mathrm m}=g_i^{\mathrm m}(T,p)+RT\log a_i,
\label{eq:activity-chemical-potential}
\end{equation}
where $a_i$ is the activity relative to the chosen standard state and $g_i^{\mathrm m}(T,p)$ is the corresponding standard-state molar Gibbs energy at the common temperature $T$ and mixture pressure $p$. Then, for $\gamma_a=1$,
\begin{equation}
  \frac{R_a^f}{R_a^b}
  =\exp\!\Bigl[-\frac1{RT}\sum_i\nu_i^a g_i^{\mathrm m}(T,p)\Bigr]
   \prod_i a_i^{-\nu_i^a}.
\label{eq:reaction-ratio-activities}
\end{equation}
If path $a$ is elementary and one models
\begin{equation}
  R_a^f=k_a^f\prod_i a_i^{\alpha_i^a},
\label{eq:forward-mass-action}
\end{equation}
then \cref{eq:reaction-ratio-activities} determines the reverse rate as
\begin{equation}
  R_a^b=k_a^b\prod_i a_i^{\beta_i^a},
  \qquad
  \frac{k_a^b}{k_a^f}
  =\exp\!\Bigl[\frac1{RT}\sum_i\nu_i^a g_i^{\mathrm m}(T,p)\Bigr].
\label{eq:thermodynamic-rate-constant-ratio}
\end{equation}
Either one-way mass-action law determines the other. The ideal-mixture pure-component standard state at $(T,p)$ gives $a_i=x_i$; another ideal-gas convention uses the corresponding pressure-based activity. Nonideal activities depend on the full state. Under the additional ansatz $\mu_i^{\mathrm m}=g_i^{\mathrm m}(T,p)+\varphi_i(T,x_i)$ on a connected open subset of the interior mole-fraction simplex with $N\ge3$, Gibbs--Duhem compatibility requires $\varphi_i(T,x_i)=C(T)\log x_i$, up to composition-independent terms, with common $C(T)$~\cite[Appendix~B]{BotheDreyerDruet2023}. Separability neither fixes $C(T)=RT$ nor follows from \cref{eq:entropy-state}. Activities and the kinetic prefactor remain constitutive input.

At unequal constituent temperatures,  the separated-mechanism closure of the baseline model uses \cref{eq:reaction-closure} with the affinity defined in \cref{eq:classIII-affinity}.

Equation~\eqref{eq:reaction-closure} does not apply unchanged to every elementary process. Specified zero-sum objective-energy increments $h_i$ per mole add $R^{-1}\sum_i h_i/T_i$ to its dimensionless entropy change, using the same energy zeros as the internal energies and chemical potentials. The rate ratio must then use the full thermochemical increment of \cref{subsec:general-event-channels} below, not $-\Acal_a/R$ alone. Extending the common-temperature activity-based rate ratio to a resolved process at unequal constituent temperatures requires specifying the process and its constituent objective-energy increments, as in \cref{sec:saha}. A model that also resolves reactive momentum inheritance must satisfy the full source-pairing condition in \cref{sec:chemical-momentum}. The baseline model excludes reaction--reaction and reaction--thermal cross-couplings.

At common temperature,
\begin{equation}
  \Acal_a=\frac1T\sum_iM_i\nu_i^a\mu_i,
\label{eq:common-temp-affinity}
\end{equation}
so \cref{eq:reaction-closure} reduces to the established common-temperature closure.

\subsection{Joint closure of reaction and energy exchange}
\label{subsec:general-event-channels}

For simultaneous composition change and energy redistribution, we close the combined entropy product. A \emph{resolved process} specifies a forward transformation, its reverse, and constituent mass and objective-energy increments. Equal net stoichiometry need not imply equal energy transfers. For $m$ processes indexed by $\ell$, let $h_i^\ell$ be the energy gained (negative for loss) by constituent $i$ per mole of forward progress, measured in the barycentric frame, and let the dimensionless $\omega_a^\ell$ give moles of reaction-coordinate progress $a$ per mole of process progress. Collect them as
\begin{equation}
  \boldsymbol h^{\ell}=(h_i^{\ell})_{i=1}^{N},
  \qquad
  \boldsymbol\omega^{\ell}=(\omega_a^{\ell})_{a=1}^{N_R}\text{.}
\label{eq:event-columns}
\end{equation}

The product $h_i^{\ell}W_\ell$ contributes to the objective energy production $\varepsilon_i$ in \cref{eq:objective-sources}. The nonkinetic-energy source additionally contains $-\bu_i\cdot\bmom_i+\half r_i\bu_i^2$. One mole of reaction $a$ corresponds to $\omega_a^\ell=1$ with all other coordinate increments zero, possibly for several processes. Objective energy conservation requires
\begin{equation}
  \sum_i h_i^{\ell}=0.
\label{eq:event-energy-conservation}
\end{equation}

The dimensionless thermochemical entropy change of the process in this source decomposition is
\begin{equation}
  \chi_{\ell}
  :=\frac1R\Bigl[
  \sum_i\frac{h_i^{\ell}}{T_i}
  -\sum_a \omega_a^{\ell}\Acal_a
  \Bigr].
\label{eq:event-entropy}
\end{equation}
For strictly positive one-way rates $W_{\ell}^{f}$ and $W_{\ell}^{b}$, impose local detailed balance in the form
\begin{equation}
  \log\frac{W_{\ell}^{f}}{W_{\ell}^{b}}=\chi_{\ell},
  \qquad
  W_{\ell}:=W_{\ell}^{f}-W_{\ell}^{b}.
\label{eq:event-local-detailed-balance}
\end{equation}

\begingroup
One column is assumed to describe opposite energy increments in the forward and reverse directions. Bidirectionality alone does not ensure this: collision-energy-averaged directions may have different conditional mean transfers. They must then be resolved further by energy transfer or supplemented by a separate, nonoverlapping energy-exchange mechanism. Species-resolved rate coefficients alone do not determine this construction.
\endgroup

The aggregate productions are reconstructed as
\begin{equation}
  \varepsilon_i^{\mathrm{proc}}=\sum_{\ell=1}^{m}h_i^{\ell}W_{\ell},
  \qquad
  R_a=\sum_{\ell=1}^{m}\omega_a^{\ell}W_{\ell}.
\label{eq:event-production-map}
\end{equation}
The coefficients $h_i^\ell$ and $M_i\sum_a\nu_i^a\omega_a^\ell$ have positive dimensional parity, so a nonzero energy or mass increment anchors $W_\ell$. Purely thermal processes are anchored through their energy increments.
Then
\begin{equation}
  \sum_i\frac1{T_i}\,\varepsilon_i^{\mathrm{proc}}
  -\sum_aR_a\Acal_a
  =R\sum_{\ell=1}^{m}W_{\ell}\chi_{\ell}
\label{eq:event-dissipation}
  =R\sum_{\ell=1}^{m}
  (W_{\ell}^{f}-W_{\ell}^{b})
  \log\frac{W_{\ell}^{f}}{W_{\ell}^{b}}
  \ge0.
\end{equation}
Purely thermal processes have $\boldsymbol\omega^{\ell}=\boldsymbol0$; purely chemical paths with no assigned objective energy have $\boldsymbol h^{\ell}=\boldsymbol0$. General columns specify physical threshold, excess-energy, or cross-reaction transfers. A dense positive-matrix factorization only reparameterizes a closure; it does not identify these physical increments.

\begin{remark}[Covariance of resolved-process data under energy-zero shifts]
\label[remark]{rem:event-reference-covariance}
For a general resolved-process column $\boldsymbol\omega^\ell=(\omega_a^\ell)_a$, define its net species-mass production per mole of process progress by
\begin{equation}
  g_i^\ell:=M_i\sum_{a=1}^{N_R}\nu_i^a \omega_a^\ell,
  \qquad
  r_i=\sum_{\ell=1}^{m}g_i^\ell W_\ell.
\label{eq:event-species-production}
\end{equation}
\begingroup
Use the same constant offsets $\Delta e_i$ as in \cref{rem:energy-reference-covariance}. Under $e_i'=e_i+\Delta e_i$ and $\mu_i'=\mu_i+\Delta e_i$, the same physical process is represented by
\begin{equation}
  h_i^{\ell\prime}=h_i^\ell+\Delta e_i g_i^\ell.
\label{eq:event-reference-transform}
\end{equation}
The additional term in $\sum_i h_i^{\ell\prime}/T_i$ is exactly
\[
  \sum_i\frac{\Delta e_i g_i^\ell}{T_i}
  =
  \sum_a \omega_a^\ell\sum_iM_i\nu_i^a\frac{\Delta e_i}{T_i},
\]
which cancels the energy-reference shift of the affinity contribution in \cref{eq:event-entropy}. Thus $\chi_\ell$ and the forward/backward rate ratio are unchanged under the coordinated shifts. To preserve the zero-sum energy-source condition for the full independently admissible reaction network, impose
\begin{equation}
  \sum_i\Delta e_iM_i\nu_i^a=0
  \qquad\text{for every reaction coordinate }a,
\label{eq:event-reference-gauge-condition}
\end{equation}
equivalently $\Delta\boldsymbol e\in\ker\Stoich^{\sf T}$. Then $\sum_i\Delta e_i g_i^\ell=0$ for every resolved process, so $\sum_i h_i^\ell=0$ is preserved as well.

For a restricted realization, set $\mathsf C=(\omega_a^\ell)_{a,\ell}$ and $\mathsf G=\Stoich\mathsf C=(g_i^\ell)_{i,\ell}$. Preserving zero-sum energy increments of every independent process requires exactly $\mathsf G^{\sf T}\Delta\boldsymbol e=0$. The sufficient full-network condition~\eqref{eq:event-reference-gauge-condition} is equivalent when $\operatorname{im}(\Stoich\mathsf C)=\operatorname{im}\Stoich$. Otherwise the restricted realization admits more shifts. These conditions preserve the represented reaction energetics.

 The separated-mechanism baseline model is tied to fixed energy zeros: even for $\Delta\boldsymbol e\in\ker\Stoich^{\sf T}$,
\[
  \Acal_a'-\Acal_a=\sum_i\Stoich_{ia}\frac{\Delta e_i}{T_i},
\]
which need not vanish at unequal temperatures. Reimposing zero assigned reaction energy and the same thermal source then generally changes the model. The same material instead requires the transformed process increments and $\varepsilon_i'=\varepsilon_i+r_i\Delta e_i$, potentially leaving the separated closure class. At common temperature the affinity shift vanishes.
\endgroup
\end{remark}

Resolved processes replace the corresponding separated reaction/thermal block. If momentum inheritance is also resolved, the full remaining thermochemical--mechanical pairing must satisfy \eqref{eq:resolved-event-total-admissibility}. The same energy pairing cannot supply both thermochemical and separate kinetic dissipation independently.

\par\smallskip
Once $r_i$, $\bmom_i$, and $\varepsilon_i$ are known, the sources in the conservative balances are reconstructed by \cref{eq:source-reconstruction-raw}.

Conservation follows from stoichiometry, momentum reconstruction, and zero-sum objective energy, given by \cref{eq:thermal-relaxation-zero-sum} or \cref{eq:event-energy-conservation}.

\subsection{Common-temperature entropy/energy-flux representations}

The pressure and entropy-flux representation was fixed in \cref{subsec:entropy-flux-pressure} before phenomenological closure.  Restriction to $T_i=T$ gives the following flux identities. This algebraic specialization does not by itself define an autonomous Class-II model or establish invariance of the common-temperature manifold under the Class-III dynamics. The thermodynamic reconstruction and reduced balances are treated separately in~\cite{BotheRuggeriPartII}.

\begin{remark}[Molecular, M\"uller-type, and Gurtin-type entropy-flux representations]
\label[remark]{rem:muller-gurtin-fluxes}

At $T_i=T$, the entropy-transport identity~\eqref{eq:constituent-entropy-transport-form} gives the following M\"uller--Gurtin repartitions~\cite{Gurtin1971,Prahs2026}. With $\bq^{\mathrm h}$ from \cref{eq:thermal-flux-and-irreversible-stress}, set
\begin{equation}
\begin{gathered}
  \bq^{\mathrm M}:=\bq^{\mathrm h},
  \qquad
  \bq^{\mathrm G}:=\bq^{\mathrm M}-\sum_i\mu_i\bj_i.
\end{gathered}
\label{eq:repartitioned-energy-flux}
\end{equation}
The M\"uller-type form is
\begin{equation}
  {\text{thermal-balance flux: }\bq^{\mathrm M}},
  \qquad
  \bPhi=\frac{\bq^{\mathrm M}}{T}
  -\sum_i\frac{\mu_i}{T}\bj_i,
\label{eq:muller-type-reduced-representation}
\end{equation}
or, equivalently, in the Gurtin-type form
\begin{equation}
  {\text{thermal transport: }}
  \bq^{\mathrm G}+\sum_i\mu_i\bj_i=\bq^{\mathrm M},
  \qquad
  \bPhi=\frac{\bq^{\mathrm G}}{T}.
\label{eq:gurtin-type-reduced-representation}
\end{equation}

Since $\sum_i\bj_i=\boldsymbol0$, only $N-1$ relative chemical potentials are needed.
Balances, constitutive relations, and boundary data must be transformed together, as discussed in \cref{subsec:representation-levels,rem:boundary-port-repartition}. In particular, $\bq^{\mathrm M}$ is not the physical total-energy flux; their difference is given in \cref{eq:physical-versus-thermal-energy-flux}.

\end{remark}

\begin{remark}[ Invariance of the thermodynamic pressure functions under enlarged cross-couplings]
 The coupled construction in Appendix~\ref{app:full-coupling} mixes the selected scalar co-factors entropy neutrally while preserving the entropy flux, Euler/pressure representation, and mechanically anchored state functions $p_i$. Its cross-operators modify $P_i-p_i$ away from zero production. Vanishing scalar dissipation under the stated coercivity gives $P_i=p_i$. Changing the pressure representation would require a more general zero addition affecting the entropy flux and Euler relation, hence a different constitutive model.
\end{remark}

\section{Chemically induced momentum and energy redistribution}
\label{sec:chemical-momentum}

The aggregate closure in \cref{sec:closures} determines the total objective momentum production $\bmom_i$, without separating nonreactive friction from reactive momentum inheritance. Its graph-Laplacian energy closure describes minimal dissipative thermal relaxation. The resolved-process formulation in \cref{subsec:general-event-channels} couples conversion and energy redistribution, but its columns do not determine momentum inheritance. The latter requires additional constitutive data and evaluation of the complete entropy contribution for the chosen source split.

The class-II derivation assigns forward products the reactants' centre-of-mass velocity, and conversely for the backward path~\cite{BotheDreyer2015}:
\begin{equation}
  \bfv_i^{C}-r_i\bv_i
  =-\sum_jC_{ij}(\bv_i-\bv_j),
\label{eq:BD-chemical-force}
\end{equation}
where
\begin{equation}
  C_{ij}
  =\sum_{a=1}^{N_R}
  \frac{M_iM_j}{\sum_k\alpha_k^aM_k}
  \bigl(
  R_a^f\beta_i^a\alpha_j^a
  +R_a^b\alpha_i^a\beta_j^a
  \bigr).
\label{eq:chemical-matrix}
\end{equation}
This generally nonsymmetric matrix is symmetric at pathwise detailed chemical equilibrium. Stoichiometric mass conservation gives
\begin{equation}
  r_i=\sum_j(C_{ij}-C_{ji}).
\label{eq:chemical-matrix-row-column}
\end{equation}
Indeed, equal reactant and product masses make the second- and first-index sums in \cref{eq:chemical-matrix} equal to $M_i(R_a^f\beta_i^a+R_a^b\alpha_i^a)$ and $M_i(R_a^f\alpha_i^a+R_a^b\beta_i^a)$ per path. In barycentric objective variables,
\begin{equation}
  \bmom_i^C
  =-\sum_jC_{ij}(\bu_i-\bu_j)+r_i\bu_i.
\label{eq:objective-chemical-momentum}
\end{equation}
Equation~\eqref{eq:chemical-matrix-row-column} then implies
\begin{equation}
  \sum_i\bmom_i^C=\boldsymbol0,
\label{eq:chemical-objective-momentum-conservation}
\end{equation}
without an additional projection or multiplier.

For this reactive contribution, split the sources before closure:
\begin{equation}
  \bmom_i=\bmom_i^M+\bmom_i^C,
  \qquad
  \varepsilon_i=\varepsilon_i^T+\varepsilon_i^C.
\label{eq:chemical-source-split}
\end{equation}
Here $\bmom_i^M$ is the nonreactive thermo-mechanical interaction and $\varepsilon_i^T$ the thermal-relaxation production. The entropy production becomes
\begin{equation}
\zeta={}\zeta_{\mathrm{visc}}+\zeta_{\mathrm{heat/diff}}^M
+\sum_i\frac1{T_i}\,\varepsilon_i^T
-\sum_aR_a\Acal_a
\label{eq:resolved-chemical-zeta}
+\underbrace{\sum_i\frac1{T_i}
\bigl(
\varepsilon_i^C-\bu_i\cdot\bmom_i^C
+\half r_i\bu_i^2
\bigr)}_{\displaystyle \zeta_{\mathrm{chem,kin}}}.
\end{equation}
The thermo-mechanical co-factor is
\begin{equation}
  \widetilde{\bB}_i^M
  :=\bB_i-d_i^T\grad\frac1{T_i}+\frac{\bmom_i^M}{T_i},
\label{eq:resolved-mechanical-cofactor}
\end{equation}
The term $-r_i\bu_i/(2T_i)$ in \cref{eq:mixed-diffusion-cofactor} is absent: its entropy contribution $r_i\bu_i^2/(2T_i)$ now belongs entirely to $\zeta_{\mathrm{chem,kin}}$. Close the projected $\widetilde{\bB}_i^M$ by the same resistance relation and reconstruct $\bmom^M$ using $\sum_i\bmom_i^M=\boldsymbol0$.

At common temperature, every zero-sum $\varepsilon_i^C$ has zero direct thermal entropy pairing, and \cref{eq:chemical-matrix-row-column,eq:objective-chemical-momentum} give
\begin{align}
  \frac1T\sum_i\bigl[
  -\bu_i\cdot\bmom_i^C+\half r_i\bu_i^2
  \bigr]
  &=\frac1{2T}\sum_{i,j}C_{ij}(\bu_i-\bu_j)^2
  \ge0.
\label{eq:common-T-chemical-mechanical-diss}
\end{align}
This uses only $C_{ij}\ge0$ and the row--column identity. Unequal weights $1/T_i$ obstruct the symmetrization, requiring joint evaluation of momentum inheritance and reaction-energy redistribution. The uncompensated contribution can be negative: take three equal-density constituents, $C_{12}=C_{21}=C>0$ with all other entries zero, $\bu_1=U\boldsymbol e$, $\bu_2=2U\boldsymbol e$, and $\bu_3=-3U\boldsymbol e$, where $|\boldsymbol e|=1$. The barycentric constraint holds. Equal one-way rates for $A_1\rightleftharpoons A_2$, with equal molar masses and inert $A_3$, realize these data. With $T_1=T>0$, $T_2=4T$, and $\boldsymbol\varepsilon^C=0$, one obtains $r_i=0$ and $\zeta_{\mathrm{chem,kin}}=-CU^2/(2T)<0$ for $U\ne0$. The conservative kinetic-energy completion below removes this obstruction.

\begin{proposition}[Joint entropy condition for reactive momentum inheritance]
\label[proposition]{prop:chemical-joint-condition}
For the source split \cref{eq:chemical-source-split}, with $\zeta_{\mathrm{chem,kin}}$ evaluated separately as in \cref{eq:resolved-chemical-zeta}, the rational momentum production \cref{eq:objective-chemical-momentum} is entropy admissible only together with a reaction-energy redistribution satisfying
\begin{equation}
  \sum_i\varepsilon_i^C=0
\label{eq:chemical-energy-zero-sum}
\end{equation}
and
\begin{equation}
  \zeta_{\mathrm{chem,kin}}
  =\sum_i\frac1{T_i}
  \bigl(
  \varepsilon_i^C-\bu_i\cdot\bmom_i^C
  +\half r_i\bu_i^2
  \bigr)\ge0.
\label{eq:chemical-joint-admissibility}
\end{equation}
\end{proposition}

These conservation and summed entropy restrictions must be supplemented by physical binary mechanisms satisfying Principles~(IV)--(V): each contribution must be nonnegative, possibly after entropy-invariant mixing, and anchored in the balances. Reassigning terms to mechanisms requires a consistent new binary-product representation.

Alternatively, let $\varepsilon^{\mathrm{proc}}$ be the complete objective-energy source of the resolved reactive processes, with $r_i$, $R_a$, and $\bmom_i^C$ referring to the same processes. After thermo-mechanical closure through \cref{eq:resolved-mechanical-cofactor}, require
\begin{equation}
  \zeta_{\mathrm{proc,mech}}
  :=\sum_i\frac1{T_i}\bigl(
       \varepsilon_i^{\mathrm{proc}}-\bu_i\cdot\bmom_i^C
       +\half r_i\bu_i^2\bigr)
       -\sum_aR_a\Acal_a\ge0.
\label{eq:resolved-event-total-admissibility}
\end{equation}
Adding \cref{eq:event-dissipation,eq:chemical-joint-admissibility} with the same energy source counts its reciprocal-temperature pairing twice and does not imply this condition. Either evaluate \cref{eq:resolved-event-total-admissibility} directly or assign disjoint energy-source contributions to thermochemical and reactive kinetic blocks, using their sum in the partial-energy balances. Each closure must also supply the mechanism representation and balance anchoring of Principles~(IV)--(V). The separated reaction relation with \cref{prop:chemical-joint-condition} remains a distinct sufficient specialization for the total entropy inequality.

\begingroup
\begin{example}[A conservative source for transferred kinetic energy]
\label[example]{ex:reactive-kinetic-allocation}
For the nonnegative matrix $C_{ij}$ in \cref{eq:chemical-matrix}, one explicit phenomenological completion is
\[
  \varepsilon_i^{C,\mathrm{kin}}
  :=\half\sum_j\bigl(C_{ij}\bu_j^2-C_{ji}\bu_i^2\bigr).
\]
Interchanging the indices gives $\sum_i\varepsilon_i^{C,\mathrm{kin}}=0$. Using \cref{eq:chemical-matrix-row-column,eq:objective-chemical-momentum}, one obtains the constituentwise identity
\[
 \varepsilon_i^{C,\mathrm{kin}}-\bu_i\cdot\bmom_i^C
       +\half r_i\bu_i^2
 =\half\sum_j C_{ij}(\bu_i-\bu_j)^2\ge0,
\]
and hence, at arbitrary positive constituent temperatures,
\[
 \zeta_{C,\mathrm{kin}}
 =\half\sum_{i,j}\frac{C_{ij}}{T_i}(\bu_i-\bu_j)^2\ge0.
\]
The balance-anchored mechanism is directed transfer from $j$ to $i$, with mass, momentum, and kinetic-energy contributions $C_{ij}$, $C_{ij}\bu_j$, and $C_{ij}\bu_j^2/2$ and corresponding donor losses, followed by equilibration with the receiver. Subtracting the constituent kinetic-energy balance gives the nonnegative receiver heating above. Once the one-way rates and $C_{ij}$ are specified, no additional energy-sharing coefficients are needed: this is a minimal closure and computational reference. Sharing reaction enthalpy, excitation energy, or excess collision energy requires additional energy-source data.

For a process-resolved thermochemical block, use disjoint energy-source contributions,
\[
 \varepsilon_i
 =\varepsilon_i^{\mathrm{proc,tc}}+\varepsilon_i^{C,\mathrm{kin}},
 \qquad
 \varepsilon_i^{\mathrm{proc,tc}}=\sum_\ell h_i^\ell W_\ell,
 \qquad \bmom_i=\bmom_i^M+\bmom_i^C,
\]
For each reactive process, construct its contribution to $C$ from \cref{eq:chemical-matrix} using its actual reactant and product coefficients, including participating colliders, and the corresponding one-way rates $W_\ell^f,W_\ell^b$. The net mass increments must match those in the resolved-process block. Close $\bmom^M$ using \cref{eq:resolved-mechanical-cofactor}. The remaining production is then
\[
 R\sum_\ell W_\ell\chi_\ell
 +\half\sum_{i,j}\frac{C_{ij}}{T_i}(\bu_i-\bu_j)^2\ge0.
\]
The partial-energy balances use both energy sources; the thermochemical entropy contribution uses only $\varepsilon^{\mathrm{proc,tc}}$. The kinetic source can depend on both one-way rates even at zero net rate and need not be a single energy-increment column times that net rate. This proves continuum entropy admissibility for the stated split. Microscopic local detailed balance for the complete momentum-resolved process additionally requires a forward--reverse description of its mechanical degrees of freedom.
\end{example}
\endgroup

Mean product velocities fixed by the centre-of-mass argument do not specify reaction-energy sharing among nonkinetic-energy balances. The second law also leaves entropy-neutral redistribution undetermined. These require a reactive kinetic model, a transition-state model~\cite{LaidlerKing1983}, or a specified phenomenological partition. Adding $\bmom^C$ to the closed aggregate source introduces an entropy term $-\sum_i\bu_i\cdot\bmom_i^C/T_i$ of indefinite sign. Counting the full reactive kinetic block in addition to the aggregate vector block would also count the mass-production term twice. The source split above assigns it once.

\section{Constitutively closed class-III model}
\label{sec:closed-model}

The primitive fields are
\begin{equation}
  (\rho_i,\bv_i,\rho_i e_i)_{i=1}^{N}.
\label{eq:primitive-fields}
\end{equation}
The fields satisfy the partial balances \cref{eq:mass-balance,eq:momentum-balance,eq:energy-balance}. The single entropy potential $\widehat{\rho s}$ generates $T_i$ and $\mu_i$ and their Maxwell-type integrability restrictions. The mechanically anchored zero-rate pressures $p_i$ require additional constitutive data subject to the weighted Euler/Gibbs--Duhem structure. Equivalent Massieu or common-temperature Helmholtz potentials (\cref{subsec:temperature-potentials}) leave this freedom unchanged. Viscosities, conductivities, resistance, thermal-relaxation parameters, and kinetic prefactors are separate transport/source data. Bulk closure requires consistent specification of both sets. Scalar transport coefficients and one-way rate data are objective constitutive functions, invariant under common Galilean boosts and spatial rotations.

The constitutive data of the baseline model are:
\begin{enumerate}[label=(\roman*)]
\item an objective $C^2$ entropy density $\widehat{\rho s}$ of the form \cref{eq:entropy-state}, with negative-definite thermodynamic Hessian \cref{eq:negative-definite-entropy-hessian} on a state domain satisfying the positive-temperature condition \cref{eq:positive-temperature-domain};
\item partial thermodynamic pressures $p_i$ of the form \cref{eq:partial-pressure-state-functions}, supplied by the scalar mechanical stress model (or an equivalent constitutive model for its zero-rate limit) and satisfying the weighted Euler relation \cref{eq:weighted-euler};
\item the entropy-flux and pressure representation \cref{eq:entropy-flux,eq:weighted-euler,eq:lm-pressure}, together with the scalar mechanical closure $P_i-p_i=-\lambda_i^{\mathrm v}\Div\bv_i$;
\item partial viscosities $\eta_i^{\mathrm v}$ and $\lambda_i^{\mathrm v}$;
\item heat conductivities $\kappa_i$ and thermo-diffusion functions $d_i^T$ satisfying \cref{eq:thermal-diffusivity-normalization};
\item the pairwise resistance \cref{eq:pairwise-resistance}, defining a positive-semidefinite map $\bM:\Ucal_{\boldsymbol y}\to\Fcal$ under the entropy pairing;
\item  thermal-relaxation coefficients $b_{ij}$, with the sparse baseline model setting any additional entropy-neutral redistribution $\varepsilon^0$ in \cref{eq:energy-redistribution-decomposition} to zero;
\item  one kinetic one-way rate for every bidirectional reaction path and the local-detailed-balance choice $\gamma_a=1$. A different positive $\gamma_a$ is retained only as an explicitly identified phenomenological rescaling.
\end{enumerate}

The closure and the theorem below apply to the interior active-species domain $\rho_i>0$, $T_i>0$. At the composition-simplex boundary, a vanishing constituent's velocity and temperature cease to be interior variables. Logarithmic chemical potentials and rate ratios can diverge, and active stress, heat-flux, resistance, thermal-relaxation, and reaction mechanisms can change. An active-species reduction or a specified extension of the surviving constitutive quantities is therefore required. This state-space boundary is distinct from the mixture's spatial boundary. Density prefactors in pair resistance alone do not control possibly singular material coefficients as a density vanishes.

Constitutive closure specifies all bulk material functions on the admissible domain, including one thermodynamic potential or its Legendre equivalent, not an additional independently chosen free energy. An initial--boundary-value problem also needs initial data and spatial boundary conditions compatible with the energy and entropy fluxes, including at walls and interfaces.

The PDE system has the following constitutive dependencies, not a prescribed numerical evaluation order:
\begin{enumerate}[label=(\arabic*)]
\item Define $1/T_i$ and $-\mu_i/T_i$ by \cref{eq:entropy-conjugates} and supply differentiable $p_i$ satisfying \cref{eq:weighted-euler}. Under (M3), these are the partial mechanical pressures at zero common dilatation and fixed thermodynamic state.
\item Choose separate reaction/thermal closure or joint resolved-process closure. The baseline model uses \cref{eq:classIII-affinity,eq:reaction-closure,eq:mass-production} for affinities, reaction rates, and mass productions, given one one-way rate per reaction. Joint closure uses \cref{eq:event-local-detailed-balance,eq:event-production-map} for the rates, $R_a$, and $\varepsilon_i^{\mathrm{proc}}$, given process increments and one one-way rate per process. \cref{eq:mass-production} gives $r_i$.
\item Specify $\pi_i$, $\bS_i^{\circ}$, and the mechanical stresses $\bS_i=-(p_i+\pi_i)\bI+\bS_i^{\circ}$ by \cref{eq:viscous-closure}.
\item Specify the heat fluxes by \cref{eq:heat-closure}.
\item Determine $\bmom_i$ from $\bB_i$, $\boldsymbol G_i$, and \cref{eq:resistance-closure,eq:momentum-reconstruction,eq:momentum-multiplier}, using the mass productions from item~2.
\item Set $\varepsilon_i=\varepsilon_i^T$ by \cref{eq:thermal-relaxation} in the baseline model. Joint closure replaces that reaction/thermal block by the energy production from item~2, with the same process rates as in $r_i$.
\item Recover the conservative productions $\bfv_i$ and $k_i$ from \cref{eq:source-reconstruction-raw}.
\end{enumerate}

\begin{theorem}[Entropy admissibility of the baseline model]
\label[theorem]{thm:entropy-admissibility}
Under Specifications~(M1)--(M3), assume that the thermodynamic states remain in the interior active-species, positive-temperature domain $\rho_i>0$, $T_i>0$ specified by \cref{eq:entropy-state,eq:positive-temperature-domain}, together with differentiable partial thermodynamic pressures satisfying \cref{eq:weighted-euler} and the entropy-flux and pressure representation of \cref{subsec:entropy-flux-pressure}.
Impose the viscous, heat-flux, pairwise resistance, thermal-relaxation, and reaction laws in \cref{eq:viscous-closure,eq:heat-closure,eq:resistance-closure,eq:pairwise-resistance,eq:thermal-relaxation,eq:reaction-closure}, with $\varepsilon_i=\varepsilon_i^T$. Assume nonnegative viscosity and heat-conduction coefficients, symmetric nonnegative graph coefficients $f_{ij}=f_{ji}\ge0$ and $b_{ij}=b_{ji}\ge0$, and strictly positive one-way reaction rates with $\gamma_a>0$. Then every smooth solution with objective momentum sources given by \cref{eq:momentum-reconstruction,eq:momentum-multiplier} and conservative sources reconstructed by \cref{eq:source-reconstruction-raw} satisfies \cref{eq:entropy-balance} with
\begin{align}
\zeta={}&
\sum_i\frac{2\eta_i^{\mathrm v}}{T_i}|\bD_i^{\circ}|^2
+\sum_i\frac{\lambda_i^{\mathrm v}}{T_i}(\Div\bv_i)^2
+\sum_i\kappa_i\Bigl|\grad\frac1{T_i}\Bigr|^2
\label{eq:closed-zeta}
\\
&+\sum_i\bu_i\cdot(\bM\bu)_i
+\half\sum_{i,j}b_{ij}
\Bigl(\frac1{T_i}-\frac1{T_j}\Bigr)^2
-\sum_aR_a\Acal_a
\ge0.
\nonumber
\end{align}
\end{theorem}

\begin{proof}
Insert the viscous, heat, resistance, thermal-relaxation, and reaction closures into \cref{eq:entropy-production,eq:entropy-invariant-thermodiffusion}. The first five terms are nonnegative by the positivity assumptions and \cref{eq:resistance-dissipation,eq:thermal-relaxation-dissipation}. The last term is nonnegative by \cref{eq:reaction-dissipation}.
\end{proof}
The theorem is conditional on smooth solutions remaining in the stated domain. Entropy admissibility alone establishes neither invariance of the positive-density and positive-temperature domain, nor PDE well-posedness or dynamical stability.

\par\smallskip\noindent Two variants follow. Replacing the separated thermal/reaction block by \cref{eq:event-production-map,eq:event-local-detailed-balance} gives \cref{eq:event-dissipation}, with nonnegative binary products $R\,W_\ell\chi_\ell$, and preserves Class-III entropy admissibility. The blocks must be replaced, not added.

For separated reactions, use \cref{eq:chemical-source-split,eq:resolved-mechanical-cofactor} and impose \cref{eq:chemical-energy-zero-sum,eq:chemical-joint-admissibility} on the reactive energy source. For resolved processes, impose instead \cref{eq:resolved-event-total-admissibility}, or use its disjoint energy-source construction. Local detailed balance of the resolved-process rates alone does not ensure admissibility after an independent reactive momentum source is added.

\section{\texorpdfstring{{Internal-state ionization as a thermochemical example}}{Internal-state ionization as a thermochemical example}}
\label{sec:saha-overview}

Consider dilute Maxwell--Boltzmann species in a homogeneous common-velocity state, with heavy constituents at $T_{\mathrm h}$ and electrons at $T_{\mathrm e}$. These temperatures are prescribed for the stationary population calculation. Electromagnetic evolution, radiation, and a self-consistent spatial plasma closure are outside this example of the energetic information beyond reaction stoichiometry.

Each electronic level $A^z_\alpha$ has number density $n_{z\alpha}$, degeneracy $g_{z\alpha}$, and molar excitation energy $\epsilon_{z\alpha}^{\mathrm m}$ above its charge-state ground level. Its transitions are Class-III reactions whose energy increments identify the constituent supplying the excitation or ionization threshold: electrons for electron impact, the heavy collider for heavy impact. That constituent's reciprocal temperature enters the entropy change and rate ratio. No additional excitation temperature is introduced.

If the bound--bound networks are connected and individually detailed-balanced by one collider bath at temperature $T_c$, their populations satisfy
\[
 \frac{n_{z\alpha}}{n_z}
 =\frac{g_{z\alpha}\exp[-\epsilon_{z\alpha}^{\mathrm m}/(RT_c)]}{U_z(T_c)},
 \qquad
 U_z(T_c)=\sum_{\alpha}g_{z\alpha}
             \exp[-\epsilon_{z\alpha}^{\mathrm m}/(RT_c)],
\]
where $n_z=\sum_\alpha n_{z\alpha}$ and sums run over the finite resolved level set. Detailed balance thus yields Boltzmann populations and the internal partition function at $T_c=T_{\mathrm e}$ for electron-controlled excitation or $T_c=T_{\mathrm h}$ for heavy-controlled excitation. Heavy species translate at $T_{\mathrm h}$ in both cases.

If an ionization process connecting charge states $z$ and $z+1$ is also detailed-balanced and obtains its threshold energy from the same bath, the aggregate populations obey
\[
 \frac{n_{z+1}n_{\mathrm e}}{n_z}
 =\frac{2}{\lambda_{\mathrm e}(T_{\mathrm e})^3}
   \frac{\lambda_z(T_{\mathrm h})^3}{\lambda_{z+1}(T_{\mathrm h})^3}
   \frac{U_{z+1}(T_c)}{U_z(T_c)}
   \exp\!\left[-\frac{I_z^{\mathrm m}}{RT_c}\right].
\]
Here $I_z^{\mathrm m}$ is the ground-state molar ionization energy, $\lambda_i(T)=h/\sqrt{2\pi m_i k_{\mathrm B}T}$ the thermal de Broglie wavelength, $h$ Planck's constant, and $m_i$ the particle mass. The electron degeneracy is two. Electron- and heavy-controlled excitation/ionization select $T_c=T_{\mathrm e}$ and $T_c=T_{\mathrm h}$, respectively, while the electron translational factor always uses $T_{\mathrm e}$. Both relations reduce to Saha--Eggert at $T_{\mathrm e}=T_{\mathrm h}=T$. Neglecting electron mass makes the heavy-species wavelength ratio unity.

These formulas use the threshold increments of \cref{sec:saha}, with zero reciprocal-temperature pairing for any additional redistribution. An extra molar transfer $q_{\mathrm e}$ to electrons, compensated by the heavy group, multiplies the level-resolved stationary relation by $\exp[q_{\mathrm e}(1/T_{\mathrm e}-1/T_{\mathrm h})/R]$. Thus the same balances yield equilibrium and conditional two-temperature relations, but the latter depend on physical energy exchange, not net stoichiometry alone.

Competing electron- and heavy-impact processes at $T_{\mathrm e}\ne T_{\mathrm h}$ generally cannot both be detailed-balanced. Opposite net excitation rates $W^{\mathrm e}=-W^B=J\ne0$ can nevertheless maintain stationary level populations while transferring energy from electrons to the heavy group at rate $\Delta E^{\mathrm m}J$ per unit volume, where $\Delta E^{\mathrm m}>0$ is the molar excitation energy. They produce
\[
 \zeta_{\mathrm{cyc}}
 =\Delta E^{\mathrm m}J
   \left(\frac1{T_{\mathrm h}}-\frac1{T_{\mathrm e}}\right)>0
\]
in the direction selected by the rate relations. Maintaining the temperatures requires compensating exchange with the surroundings or transport. Stationary populations are not full source equilibrium. See \cref{sec:saha} for the derivation and restrictions on additional collision-energy transfers.

The explicit two-level calculation in \cref{eq:explicit-two-bath-stationary-cycle} makes this distinction quantitative. For illustration, take equal level degeneracies, $\Delta E^{\mathrm m}/(RT_{\mathrm h})=2$, and equal de-excitation coefficients $k_{\mathrm e}^-=k_B^-=k_0$. \Cref{tab:two-bath-cycle} evaluates the exact stationary formulas. These are dimensionless model calculations, not fitted plasma data. Here $c_*$ is the total molar level concentration and positive $J$ transfers energy from electrons to the heavy bath.
\begin{table}[htbp]
\centering
\caption{Stationary populations and dissipative cycle currents for two prescribed baths.}
\label{tab:two-bath-cycle}
\begin{tabular}{rrrr}
\toprule
$T_{\mathrm e}/T_{\mathrm h}$ & $c_\beta^{\mathrm{st}}/c_*$ & $J/(c_*k_0)$ & $\zeta_{\mathrm{cyc}}/(Rc_*k_0)$\\
\midrule
0.5 & 0.0713 & -0.0543 & 0.1087\\
1 & 0.1192 & 0.0000 & 0.0000\\
2 & 0.2010 & 0.0929 & 0.0929\\
4 & 0.2706 & 0.1719 & 0.2578\\
\bottomrule
\end{tabular}
\end{table}
At equal temperatures the cycle stops. At unequal temperatures its direction reverses with the thermal bias, while its entropy production remains positive.

\section{Constitutive zero production and equilibrium}
\label{sec:equilibrium}

Thermodynamic equilibrium at $(t,\boldsymbol x)$ means $\zeta(t,\boldsymbol x)=0$ (\cref{def:equilibrium-reversibility}), restricting the local state, velocities, and spatial derivatives after closure. A \emph{stationary equilibrium configuration} additionally satisfies the stationary balances and frame/boundary conditions accounting for common rigid motion and external forcing. Bulk equilibrium requires zero production at every point; whole-system equilibrium also requires zero production in separately modelled boundary, interface, and reservoir mechanisms.

If the viscosity, conductivity, resistance, and relaxation coefficients are strictly positive on their mechanism spaces, the interaction and relaxation graphs are connected, and each active reaction has strictly positive one-way rates obeying \cref{eq:reaction-closure} with $\gamma_a>0$, constitutive zero production in the baseline model implies
\begin{align}
\bD_i^{\circ}&=\boldsymbol0,
&\Div\bv_i&=0,
&\grad T_i&=\boldsymbol0,
\label{eq:equilibrium1}\\
\bu_i&=\boldsymbol0,
&T_i&=T_j,
&R_a^f&=R_a^b,
&\boldsymbol{\mathfrak b}&=\boldsymbol0.
\label{eq:equilibrium2}
\end{align}
On a connected spatial domain, $\bD_i=0$ gives rigid constituent velocities and $\bu_i=0$ makes them identical:
\[
 \bv_i(t,\boldsymbol x)=\bv(t,\boldsymbol x)
 =\boldsymbol V(t)+\boldsymbol\Omega(t)\boldsymbol x,
 \qquad \boldsymbol\Omega^{\sf T}=-\boldsymbol\Omega.
\]
Zero temperature gradients and connected thermal exchange give one spatially uniform temperature $T_i=T$, also constant in time at stationary equilibrium. Stationary no-slip walls typically enforce $\bv_i=0$, and every active reaction is detailed-balanced. Constituent densities need not be uniform: composition and pressure profiles must satisfy the momentum balances with the prescribed body forces and boundary conditions.

The canonical projected co-factor $\boldsymbol{\mathfrak b}=\bP_{\boldsymbol y}^{\sf T}\widetilde{\bB}$ includes objective interaction production, so its zero-production condition does not imply constituentwise $\bB_i=\boldsymbol0$. The scalar closure gives $\pi_i=0$ and $P_i=p_i$. Stationary mechanical equilibrium must additionally satisfy the momentum balances, including body forces. No constituentwise condition $\bB_i=\boldsymbol0$ follows from this zero-production argument alone.

\section{Conclusions}
\label{sec:conclusions}

Net stoichiometry and chemical potentials do not determine multi-temperature reaction kinetics: identical species changes can have different rate ratios through different energy redistribution. This dependence disappears from the direct zero-sum energy pairing at common temperature, so agreement there does not fix the Class-III extension. Rates must be specified with constituent energy increments, their reference convention, and any separately represented momentum inheritance.

The thermodynamic potential supplies integrable state derivatives. The logarithmic rate relation fixes process bias but leaves one kinetic activity scale free. The normalized forward/backward difference tests the near-balanced approximation, whereas small net rates or entropy production can reflect weak activity at finite bias. A local entropy function describes unresolved equilibration without requiring equal retained temperatures or zero affinities. It does not justify kinetic linearization.

Measured primitive-gradient matrices combine kinetics, thermodynamic derivatives, and projections: off-diagonal entries need not be additional physical cross-effects, nor their symmetry Onsager reciprocity. Linear entropy-co-factor relations may be nonlinear in temperature, composition, and velocity. Representation changes preserve a material model only when fluxes, operators, and boundary conditions transform together. Equivalent free energies supply neither kinetics nor the partial-pressure partition left open by the weighted Euler relation.

Resolved-level detailed balance yields Boltzmann populations and partition functions. Compatible excitation/ionization energy transfers then give conditional two-temperature Saha relations and their common-temperature limit. Stationary populations alone do not justify eliminating levels: opposing processes can cancel species sources while transferring energy and producing entropy. Thermal reductions must preserve these quantities as well as composition, and slowly relaxing populations must remain resolved unless elimination is separately justified.

The reciprocal completion establishes constitutive admissibility but does not determine coefficient magnitudes. For $N=2,3,4$ and $N_R=r_{\Stoich}=N-1$, its algebraic counts of scalar coefficient functions are $19,49,93$, versus $10,19,30$ in the baseline model, even before thermodynamic-potential and pressure data. These counts fix $\gamma_a=1$ and use near-balanced reaction coefficients; see \cref{app:full-identifiability} for assumptions and block counts. The baseline model retains restricted mechanisms and its specified thermo-diffusive coupling. Additional responses require data that distinguish their coefficients. Neither positive-operator factorization nor artificial directional rate splitting identifies physical elementary processes. The conservative reactive-energy construction is a minimal constitutive choice, while microscopic energy sharing requires further information.

The connected nondegenerate baseline model gives one uniform temperature, common rigid velocity, and detailed balance at zero production throughout a connected domain. Stationary walls can remove rigid motion, while body forces sustain nonuniform pressure and composition. Degenerate coefficients can leave nonzero co-factors undetected. The results establish local bulk constitutive admissibility. State-domain extensions, spatial boundary models, and quantitative kinetic validation remain separate tasks.  The companion paper~\cite{BotheRuggeriPartII} characterizes the fixed-flux-compatible subclass by constituentwise Gibbs--Duhem identities, which imply locally additive entropy with compatible partial pressures. Under its further constitutive hypotheses, it constructs a degenerate symmetric main-field formulation and principal subsystems for that subclass. Its separate entropy-maximizing Class-III-to-Class-II reduction also applies locally to nonadditive parent entropies under the stated domain and attainment assumptions.

\appendix
\section{Fully coupled constitutive completion}
\label{app:full-coupling}

The main text selects a restricted set of identifiable couplings. This appendix gives its reciprocal completion, linear in the selected co-factors and subject to their parity, tensorial character, balance anchoring, and conservation constraints. It specifies alternative dissipative mechanisms: within each tensorial rank, parity-preserving entropy-invariant mixing followed by nonnegative diagonal closure yields individually nonnegative binary products under Principle~(IV). Re-expression in the original co-factors produces reciprocal cross-effects. The  unmixed binary products of the baseline model need not also be separately nonnegative.

We retain  the baseline model's isotropic, nonpolar, interior active-species domain, excluding composition-boundary changes of mechanism spaces and singular coefficient asymptotics. Coefficient operators depend on the scalar thermodynamic state and act linearly on the selected co-factors at fixed state, without relative-velocity directions. The constitutive relations may nevertheless be nonlinear in primitive variables. Near-equilibrium linearization is only one application. The Curie principle separates deviatoric tensors, vectors, and scalars in this specialization. Objective nonlinear couplings using relative-velocity directions, possible even in isotropic materials, and couplings involving directors or lattice tensors lie outside it. Completeness refers only to this co-factor representation and reciprocal linear closure class. The distinction between this construction from elementary dissipative mechanisms and an unresolved block closure is explained in Section~\ref{subsec:entropy-mechanisms}. Compensation between distinct preselected blocks remains excluded by Principle~(IV).

The construction follows the entropy-flux and pressure selection in \cref{subsec:entropy-flux-pressure}. Entropy-neutral regrouping changes the closure of $P_i-p_i$, while retaining the entropy flux, Euler relation, and thermodynamic-pressure functions $p_i$ with their original state dependence. Thus $p_i$ retains  its definition in the baseline model through the isolated zero-dilatation response. It need not equal the enlarged model's mechanical pressure when only the dilatational rate vanishes and other scalar co-factors remain active. For one dilatational rate $\theta$ and affinity $\Acal$, for example,
\[
 \sigma=v\theta+\ell_{\theta R}\Acal,
 \qquad R_a=\ell_{\theta R}\theta-k\Acal,
 \qquad v,k>0.
\]
Then $\theta\sigma-R_a\Acal=v\theta^2+k\Acal^2\ge0$, but at $\theta=0$ the pressure correction is $\pi=-T\sigma=-T\ell_{\theta R}\Acal$. Reaction can therefore produce isotropic stress without volume deformation, as in the one-reaction specialization of \cref{eq:app-scalar-resistance-pressure,eq:app-scalar-resistance-rates}.

\subsection{Entropy products and mechanism spaces}
\label{app:full-products}

We retain the entropy-invariant thermo-diffusion representation from \cref{eq:entropy-invariant-thermodiffusion} and introduce
\begin{equation}
  \widehat{\bq}_i:=\bq_i-d_i^T\bu_i,
  \qquad
  \boldsymbol\xi_i:=\grad\frac1{T_i},
\label{eq:app-qhat-xi}
\end{equation}
and
\begin{equation}
  \boldsymbol{\mathfrak b}
  :=\bP_{\boldsymbol y}^{\sf T}\widetilde{\boldsymbol B}.
\label{eq:app-vector-force}
\end{equation}
Because $\bu\in\Ucal_{\boldsymbol y}$, the vector part of the entropy production is
\begin{equation}
  \zeta_{\mathrm{vec}}
  =\sum_i\widehat{\bq}_i\cdot\boldsymbol\xi_i
   -\sum_i\bu_i\cdot\boldsymbol{\mathfrak b}_i.
\label{eq:app-zeta-vector}
\end{equation}
Under Principle~(III), $\widehat{\bq}$ and $\bu$ have negative parity; $\boldsymbol\xi$ and $-\boldsymbol{\mathfrak b}$ have positive parity. Their primitive balance anchors under~(V) are $\bq_i$ and $\bmom_i$. The triangular transformation \cref{eq:entropy-invariant-thermodiffusion} changes the heat and conjugate diffusion co-factors together, preserving their balance origin and parity.

For the scalar mechanisms, set
\begin{equation}
  \theta_i:=\Div\bv_i,
  \qquad
  \boldsymbol\beta:=\Bigl(\frac1{T_1},\ldots,\frac1{T_N}\Bigr),
  \qquad
  \boldsymbol Q:=\boldsymbol I_N-\frac1N\bone\otimes\bone,
  \qquad
  \boldsymbol\beta^{\perp}:=\boldsymbol Q\boldsymbol\beta.
\label{eq:app-scalar-forces}
\end{equation}
Thus $\beta_i^{\perp}=1/T_i-N^{-1}\sum_j1/T_j$ records deviations from the mean reciprocal temperature. Let
\begin{equation}
  \mathcal E:=\bigl\{\boldsymbol z\in\R^N:\bone\cdot\boldsymbol z=0\bigr\}.
\label{eq:app-energy-space}
\end{equation}
Since $\boldsymbol\varepsilon:=(\varepsilon_i)_i\in\mathcal E$, one has $\boldsymbol\varepsilon\cdot\boldsymbol\beta=\boldsymbol\varepsilon\cdot\boldsymbol\beta^{\perp}$. With
\begin{equation}
  \pi_i:=P_i-p_i,
  \qquad
  \boldsymbol\sigma:=\Bigl(-\frac{\pi_1}{T_1},\ldots,-\frac{\pi_N}{T_N}\Bigr),
  \qquad
  \boldsymbol R:=(R_a)_{a=1}^{N_R},
  \qquad
  \boldsymbol{\mathcal A}:=(\Acal_a)_{a=1}^{N_R},
\label{eq:app-scalar-arrays}
\end{equation}
the scalar contribution reads
\begin{equation}
  \zeta_{\mathrm{sc}}
  =\boldsymbol\theta\cdot\boldsymbol\sigma
   +\boldsymbol\varepsilon\cdot\boldsymbol\beta^{\perp}
   +\boldsymbol R\cdot(-\boldsymbol{\mathcal A}).
\label{eq:app-zeta-scalar}
\end{equation}
Under Principle~(III), $(\boldsymbol\theta,\boldsymbol\varepsilon,\boldsymbol R)$ has negative parity and $(\boldsymbol\sigma,\boldsymbol\beta^{\perp},-\boldsymbol{\mathcal A})$ positive parity. The mechanical partial stress anchors $\pi_i=P_i-p_i$ under~(V), although pressure and dilatation appear in the opposite order to conventional ``flux--force'' terminology. Solving for different pair members can reverse cross-block signs, as in the resistance forms below. This change of dependent variables is distinct from adding a skew-adjoint term to the fixed parity-ordered operator \cref{eq:app-general-linear-law}.

The reaction-force factor in \cref{eq:app-zeta-scalar} has an intrinsic stoichiometric reduction. Put
\begin{equation}
  r_{\Stoich}:=\operatorname{rank}\Stoich,
  \qquad
  \mathcal A_{\Stoich}:=\operatorname{im}\Stoich^{\sf T}
  \subseteq\R^{N_R}.
\label{eq:app-affinity-space}
\end{equation}
With $\boldsymbol\psi:=(\mu_i/T_i)_{i=1}^N$, \cref{eq:classIII-affinity,eq:stoichiometric-production-matrix} give
\begin{equation}
  \boldsymbol{\mathcal A}=\Stoich^{\sf T}\boldsymbol\psi
  \in\mathcal A_{\Stoich},
  \qquad
  \dim\mathcal A_{\Stoich}=r_{\Stoich}.
\label{eq:app-affinity-range}
\end{equation}
For every $\boldsymbol d\in\ker\Stoich$ one has both $\Stoich\boldsymbol d=0$ and
\begin{equation}
  \boldsymbol{\mathcal A}\cdot\boldsymbol d
  =\boldsymbol\psi\cdot\Stoich\boldsymbol d=0.
\label{eq:app-path-null-pairing}
\end{equation}
Thus $\boldsymbol R\mapsto\boldsymbol R+\boldsymbol d$, with $\Stoich\boldsymbol d=0$, can change individual path rates without changing the species productions or $-\boldsymbol R\cdot\boldsymbol{\mathcal A}$: its stoichiometric changes cancel. Moreover, mass conservation \cref{eq:stoich-mass} gives $\bone^{\sf T}\Stoich=0$, so
\begin{equation}
  r_{\Stoich}\le N-1.
\label{eq:app-stoich-rank-bound}
\end{equation}
All $N_R$ labels may be retained to distinguish kinetic pathways, while the independent thermodynamic force dimension is $r_{\Stoich}$. The coordinates are redundant when $r_{\Stoich}<N_R$.

Finally, the deviatoric contribution is
\begin{equation}
  \zeta_{\mathrm{dev}}
  =\sum_i\bD_i^{\circ}:\frac{\bS_i^{\circ}}{T_i}.
\label{eq:app-zeta-deviatoric}
\end{equation}
By Principle~(III), the deviatoric rates $\bD_i^{\circ}$ have negative parity and the temperature-weighted partial stresses $\bS_i^{\circ}/T_i$ have positive parity. The partial stresses supply their balance anchors under~(V).

For arrays of vectors or tensors we use the canonical Euclidean pairings, for example
\begin{equation}
  \langle\boldsymbol X,\boldsymbol Y\rangle
  :=\sum_i\boldsymbol X_i\cdot\boldsymbol Y_i,
  \qquad
  \langle\boldsymbol A,\boldsymbol B\rangle
  :=\sum_i\boldsymbol A_i:\boldsymbol B_i.
\label{eq:app-array-pairings}
\end{equation}
The $\Ucal_{\boldsymbol y}$--$\Fcal$ pairing is nondegenerate. Throughout, ${\sf T}$ denotes the adjoint for these pairings, represented by matrix transposition in dual bases. It does not identify $\Ucal_{\boldsymbol y}$ with $\Fcal$.

\subsection{\texorpdfstring{Total entropy inequality and reciprocal closure}{Total entropy inequality and reciprocal closure}}
\label{app:general-linear-criterion}

For negative-parity co-factors $\boldsymbol J$ and their positive-parity conjugates $\boldsymbol X$, write $\zeta_{\mathrm{block}}=\langle\boldsymbol J,\boldsymbol X\rangle$. With the coefficient operator fixed at a thermodynamic state or obtained from a specified linearization, the general linear relation is

\begin{equation}
  \boldsymbol J=\mathbb L\boldsymbol X,
  \qquad
  \mathbb L=\mathbb L^{\mathrm s}+\mathbb L^{\mathrm a},
  \qquad
  (\mathbb L^{\mathrm s})^{\sf T}=\mathbb L^{\mathrm s},
  \qquad
  (\mathbb L^{\mathrm a})^{\sf T}=-\mathbb L^{\mathrm a}.
\label{eq:app-general-linear-law}
\end{equation}
Nonnegativity of the total quadratic entropy production of this fixed block is equivalent to
\begin{equation}
  \langle\boldsymbol X,\mathbb L^{\mathrm s}\boldsymbol X\rangle\ge0
  \qquad\text{for all admissible }\boldsymbol X.
\label{eq:app-general-positivity}
\end{equation}

The frozen-coefficient test imposes the quadratic inequality on all independent force directions. With state-dependent coefficients, the process inequality tests the co-factors realized at each state. This need not give linear dependence on primitive fields. Arbitrary dependence of the coefficient matrix on nonequilibrium forces instead defines a nonlinear relation: matrix positivity remains sufficient, but need not be necessary in directions untested at that state. This quadratic criterion alone does not encode the elementary mechanism construction of Section~\ref{subsec:entropy-mechanisms}.
For the reciprocal closure used below, couple elementary dissipative mechanisms by entropy-neutral mixing on the independent conjugate spaces. In dual bases, choose parity-preserving, dimensionally consistent invertible maps with $\mathsf A^{\sf T}\mathsf B=\mathsf I$, put $\widetilde{\boldsymbol J}=\mathsf A\boldsymbol J$ and $\widetilde{\boldsymbol X}=\mathsf B\boldsymbol X$, and impose $\widetilde J_m=\lambda_m\widetilde X_m$ with $\lambda_m\ge0$. Then
\[
  \mathbb L=\mathsf B^{\sf T}\operatorname{diag}(\lambda_m)\mathsf B,
  \qquad
  \widetilde J_m\widetilde X_m=\lambda_m|\widetilde X_m|^2\ge0.
\]
Thus the mixed products are individually nonnegative and $\mathbb L$ is self-adjoint and positive semidefinite; conversely, every such finite-dimensional operator admits this factorization. This is the continuum reciprocity construction of Section~\ref{sec:entropy-principle} on independent paired spaces. That section also explains the Onsager--Casimir cross-signs induced by changing the target members of conjugate pairs.

When the relevant block is invertible, mobility and resistance forms describe the same constitutive class by solving for different members of the balance-anchored pairs. The resistance forms below determine stresses and interaction productions from primitive velocities and dilatational rates.

\subsection{Full deviatoric-stress coupling}
\label{app:full-deviatoric}

For an isotropic fluid, the most general reciprocal linear relation between the constituent deviatoric rates and stresses is
\begin{equation}
  \frac{\bS_i^{\circ}}{T_i}
  =2\sum_{j=1}^{N}\eta_{ij}\bD_j^{\circ},
  \qquad
  \eta_{ij}=\eta_{ji},
\label{eq:app-full-shear}
\end{equation}
where $\boldsymbol\eta=[\eta_{ij}]$ is positive semidefinite. The entropy product fixes these temperature weights. Symmetry for unweighted stresses would generally be a different assumption. Consequently,
\begin{equation}
  \zeta_{\mathrm{dev}}
  =2\sum_{i,j=1}^{N}\eta_{ij}\bD_i^{\circ}:\bD_j^{\circ}\ge0.
\label{eq:app-full-shear-dissipation}
\end{equation}
 The constitutive relation of the baseline model in \cref{eq:viscous-closure} is recovered by taking
\begin{equation}
  \eta_{ij}=\delta_{ij}\frac{\eta_i^{\mathrm v}}{T_i}.
\label{eq:app-shear-baseline}
\end{equation}
For anisotropic materials, each scalar $\eta_{ij}$ may be replaced by a fourth-rank operator on symmetric trace-free tensors. Positivity and self-adjointness are then imposed with respect to the tensor contraction in \cref{eq:app-array-pairings}.

\subsection{Full vector coupling of heat conduction and relative momentum exchange}
\label{app:full-vector}

Let
\begin{equation}
  \mathcal H_q:=(\R^3)^N,
  \qquad
  \boldsymbol\xi\in\mathcal H_q,
  \qquad
  \widehat{\bq}\in\mathcal H_q,
  \qquad
  \bu\in\Ucal_{\boldsymbol y},
  \qquad
  -\boldsymbol{\mathfrak b}\in\Fcal.
\label{eq:app-vector-spaces}
\end{equation}
The full reciprocal mobility closure is
\begin{equation}
  \begin{pmatrix}
    \widehat{\bq}\\[0.2em]
    \bu
  \end{pmatrix}
  =
  \begin{pmatrix}
    \mathbb A & \mathbb C\\
    \mathbb C^{\sf T} & \mathbb L
  \end{pmatrix}
  \begin{pmatrix}
    \boldsymbol\xi\\[0.2em]
    -\boldsymbol{\mathfrak b}
  \end{pmatrix}.
\label{eq:app-vector-mobility}
\end{equation}
Here
\begin{equation}
  \mathbb A:\mathcal H_q\to\mathcal H_q,
  \qquad
  \mathbb C:\Fcal\to\mathcal H_q,
  \qquad
  \mathbb L:\Fcal\to\Ucal_{\boldsymbol y},
\label{eq:app-vector-operators}
\end{equation}
where $\mathbb A$ is self-adjoint. As in \cref{sec:closure-philosophy}, $\Fcal$ satisfies $\sum_i\boldsymbol Z_i=0$ and $\Ucal_{\boldsymbol y}$ satisfies $\sum_i y_i\bu_i=0$. These generally distinct spaces are dual under the entropy pairing. Accordingly, $\mathbb L$ is pairing-symmetric when
\begin{equation}
  \langle\mathbb L\boldsymbol{\mathfrak b},\boldsymbol z\rangle
  =\langle\mathbb L\boldsymbol z,\boldsymbol{\mathfrak b}\rangle
  \qquad(\boldsymbol{\mathfrak b},\boldsymbol z\in\Fcal),
\label{eq:app-vector-pairing-symmetry}
\end{equation}
and positive semidefinite when $\langle\mathbb L\boldsymbol{\mathfrak b},\boldsymbol{\mathfrak b}\rangle\ge0$. The dual-pairing adjoint $\mathbb C^{\sf T}:\mathcal H_q\to\Ucal_{\boldsymbol y}$ is defined by
\begin{equation}
  \langle\boldsymbol\xi,\mathbb C\boldsymbol{\mathfrak b}\rangle
  =\langle\mathbb C^{\sf T}\boldsymbol\xi,\boldsymbol{\mathfrak b}\rangle.
\label{eq:app-vector-adjoint}
\end{equation}

Isotropy requires these operators and the resistance operators below to commute with common spatial rotations: their spatial blocks are scalar multiples of $\bI$. The block \cref{eq:app-vector-mobility} must be positive semidefinite on $\mathcal H_q\times\Fcal$. It contains all cross-conductivities, diffusion mobilities, and reciprocal Soret/Dufour-type couplings for the selected vector products.

If $\mathbb L$ is coercive on $\Fcal$, define
\begin{equation}
  \mathbb M:=\mathbb L^{-1}:\Ucal_{\boldsymbol y}\to\Fcal,
  \qquad
  \mathbb D:=\mathbb C\mathbb M:\Ucal_{\boldsymbol y}\to\mathcal H_q,
  \qquad
  \mathbb K:=\mathbb A-\mathbb C\mathbb M\mathbb C^{\sf T}.
\label{eq:app-vector-schur-data}
\end{equation}
Solving \cref{eq:app-vector-mobility} for the force $-\boldsymbol{\mathfrak b}$ gives the equivalent resistance representation
\begin{align}
  \widehat{\bq}&=\mathbb K\boldsymbol\xi+\mathbb D\bu,
\label{eq:app-vector-resistance-q}
\\
  \boldsymbol{\mathfrak b}
  &=-\mathbb M\bu+\mathbb D^{\sf T}\boldsymbol\xi.
\label{eq:app-vector-resistance-b}
\end{align}
The reciprocal cross-terms cancel in the entropy production:
\begin{equation}
  \zeta_{\mathrm{vec}}
  =\langle\boldsymbol\xi,\mathbb K\boldsymbol\xi\rangle
   +\langle\bu,\mathbb M\bu\rangle\ge0.
\label{eq:app-vector-diagonal-zeta}
\end{equation}
The dissipative conductivity and resistance are $\mathbb K$ and $\mathbb M$. The paired cross-map $\mathbb D$, $\mathbb D^{\sf T}$ cancels explicitly but is not a skew-adjoint addition as in \cref{eq:app-general-linear-law}. With positive Schur-diagonal blocks fixed, entropy positivity places no further magnitude bound on $\mathbb D$: any admissible cross-map reconstructs the positive reciprocal mobility block through
\begin{equation}
  \mathbb L=\mathbb M^{-1},\qquad
  \mathbb C=\mathbb D\mathbb M^{-1},\qquad
  \mathbb A=\mathbb K+\mathbb D\mathbb M^{-1}\mathbb D^{\sf T}.
\label{eq:app-vector-reconstruct-mobility}
\end{equation}
Its magnitude requires additional constitutive data. Conversely, every positive reciprocal block \cref{eq:app-vector-mobility} with coercive $\mathbb L$ has the unique resistance form \cref{eq:app-vector-resistance-q,eq:app-vector-resistance-b}, with positivity equivalent to
\begin{equation}
  \mathbb M>0\quad\text{on }\Ucal_{\boldsymbol y},
  \qquad
  \mathbb K\ge0\quad\text{on }\mathcal H_q.
\label{eq:app-vector-schur-positivity}
\end{equation}

In the semidefinite mobility case the block ordering in \cref{eq:app-vector-mobility} gives
\begin{equation}
  \mathbb C(\ker\mathbb L)=\{0\},
  \qquad
  \operatorname{im}\mathbb C^{\sf T}\subseteq\operatorname{im}\mathbb L,
  \qquad
  \mathbb A-\mathbb C\mathbb L^{+}\mathbb C^{\sf T}\succeq0.
\label{eq:app-vector-singular-compatibility}
\end{equation}
The first two conditions are equivalent. Here the generalized inverse is taken in dual coordinates for the $\Fcal$--$\Ucal_{\boldsymbol y}$ pairing, or as the inverse between the corresponding quotient and active range. Solving for $-\boldsymbol{\mathfrak b}$ requires $\bu\in\operatorname{im}\mathbb L$ and gives
\begin{equation}
  -\boldsymbol{\mathfrak b}=\mathbb L^{+}(\bu-\mathbb C^{\sf T}\boldsymbol\xi)
                 +\boldsymbol k,
  \qquad \boldsymbol k\in\ker\mathbb L.
\label{eq:app-vector-singular-reconstruction}
\end{equation}
Since $\mathbb C\boldsymbol k=0$, the heat response is unique, but physical source reconstruction requires a chosen force representative or null-force closure. Smooth generalized inverses require constant-rank state patches. Pointwise positivity gives no regularity across rank changes.

Alternatively, \cref{eq:app-vector-resistance-q,eq:app-vector-resistance-b} may be prescribed directly with a positive-semidefinite resistance $\mathbb M$. The entropy cancellation \cref{eq:app-vector-diagonal-zeta} still holds for every admissible relative velocity. This direct zero-resistance case is not an inversion of a finite zero-mobility block, which instead constrains the velocity to its active range.

The main-text choice is obtained from
\begin{equation}
  (\mathbb K\boldsymbol\xi)_i=\kappa_i\boldsymbol\xi_i,
  \qquad
  \mathbb D=0,
  \qquad
  \mathbb M=\bM.
\label{eq:app-vector-baseline}
\end{equation}
The main-text $d_i^T$ select an entropy-invariant representation without restricting this full class: their constituentwise map may be absorbed into $\mathbb D$ with $d_i^T=0$, or retained with $\mathbb D$ describing only additional cross-effects. The two descriptions differ by an entropy-neutral co-factor change.

For implementation, \cref{eq:app-vector-resistance-b} replaces \cref{eq:resistance-closure}. Since $\boldsymbol{\mathfrak b}=\bP_{\boldsymbol y}^{\sf T}\widetilde{\boldsymbol B}$, there is a vector $\blambda$ such that
\begin{equation}
  \widetilde{\bB}_i
  =-(\mathbb M\bu)_i+(\mathbb D^{\sf T}\boldsymbol\xi)_i+y_i\blambda.
\label{eq:app-full-gauge-representation}
\end{equation}
Writing $\widetilde{\bB}_i=\boldsymbol G_i+\bmom_i/T_i$ as in \cref{eq:cofactor-source-split}, the objective momentum productions are
\begin{equation}
  \bmom_i
  =T_i\Bigl[-(\mathbb M\bu)_i
  +(\mathbb D^{\sf T}\boldsymbol\xi)_i
  +y_i\blambda-\boldsymbol G_i\Bigr].
\label{eq:app-full-momentum-reconstruction}
\end{equation}
Total momentum conservation fixes
\begin{equation}
  \blambda
  =\frac{\displaystyle\sum_jT_j\Bigl[(\mathbb M\bu)_j
  -(\mathbb D^{\sf T}\boldsymbol\xi)_j+\boldsymbol G_j\Bigr]}
  {\displaystyle\sum_jy_jT_j}.
\label{eq:app-full-momentum-multiplier}
\end{equation}
Thus the full vector coupling retains the constrained geometry and the temperature-weighted annihilator reconstruction.

\subsection{Full scalar coupling of volume deformation, thermal exchange, and reaction}
\label{app:full-scalar}

Define the scalar force and rate spaces
\begin{equation}
  \mathcal X_{\mathrm{sc}}
  :=\R^N\oplus\mathcal E\oplus\R^{N_R},
  \qquad
  \mathcal J_{\mathrm{sc}}
  :=\R^N\oplus\mathcal E\oplus\R^{N_R}.
\label{eq:app-scalar-spaces}
\end{equation}

These spaces retain all labelled reaction paths. Their thermodynamically attainable force subspace is
\begin{equation}
  \mathcal X_{\mathrm{sc}}^{\mathrm{ind}}
  :=\R^N\oplus\mathcal E\oplus\mathcal A_{\Stoich},
  \qquad
  \dim\mathcal X_{\mathrm{sc}}^{\mathrm{ind}}
  =2N-1+r_{\Stoich}\le3N-2.
\label{eq:app-independent-scalar-space}
\end{equation}
The thermodynamic part of this dimension count is locally realized under Specification~(M2): the nonsingular entropy Hessian makes the map from $((\rho_i e_i)_i,\boldsymbol\rho)$ to $(\boldsymbol\beta,-\boldsymbol\psi)$ locally invertible. Its composition with $(\boldsymbol\beta,\boldsymbol\psi)\mapsto(\boldsymbol Q\boldsymbol\beta,\Stoich^{\sf T}\boldsymbol\psi)$ has rank $N-1+r_{\Stoich}$. This concerns local state variation. It does not make thermal or chemical forces independently variable while the thermodynamic state is held fixed.

Since $(\ker\Stoich)^{\perp}=\operatorname{im}\Stoich^{\sf T}$, $\mathcal A_{\Stoich}$ also represents $\R^{N_R}/\ker\Stoich$. The relation below retains the labelled-pathway extension to distinguish kinetics with dependent stoichiometric columns. Its compression to \cref{eq:app-independent-scalar-space} is the minimal independent thermodynamic block.

The fully coupled reciprocal mobility relation is
\begin{equation}
  \begin{pmatrix}
    \boldsymbol\theta\\[0.2em]
    \boldsymbol\varepsilon\\[0.2em]
    \boldsymbol R
  \end{pmatrix}
  =
  \begin{pmatrix}
    \mathbb L_{\theta\theta} & \mathbb L_{\theta T} & \mathbb L_{\theta R}\\
    \mathbb L_{\theta T}^{\sf T} & \mathbb L_{TT} & \mathbb L_{TR}\\
    \mathbb L_{\theta R}^{\sf T} & \mathbb L_{TR}^{\sf T} & \mathbb L_{RR}
  \end{pmatrix}
  \begin{pmatrix}
    \boldsymbol\sigma\\[0.2em]
    \boldsymbol\beta^{\perp}\\[0.2em]
    -\boldsymbol{\mathcal A}
  \end{pmatrix}.
\label{eq:app-full-scalar-mobility}
\end{equation}
The labelled-pathway extension used here is required to be self-adjoint and positive semidefinite on all of $\mathcal X_{\mathrm{sc}}$. The entropy pairing tests only its compression to $\mathcal X_{\mathrm{sc}}^{\mathrm{ind}}$. Its ranges are chosen so that the second row lies in $\mathcal E$, thereby enforcing $\sum_i\varepsilon_i=0$. Stoichiometric mass conservation then enforces $\sum_i r_i=0$ independently of the cross-coupling.

The individual blocks have a direct interpretation:
\begin{itemize}[leftmargin=2.2em]
\item {$\mathbb L_{\theta\theta}$ contains the dilatational mobilities. Their inverse gives the volume-viscosity resistance in the coercive case;}
\item $\mathbb L_{TT}$ contains the most general reciprocal linear thermal-relaxation operator on $\mathcal E$;
\item $\mathbb L_{RR}$ contains cross-effects between labelled reaction paths. Its compression to $\mathcal A_{\Stoich}$ represents the couplings between independent thermodynamic reaction forces when $r_{\Stoich}<N_R$;
\item $\mathbb L_{TR}$ couples reaction progress and the redistribution of thermal energy;
\item $\mathbb L_{\theta T}$ describes a direct constitutive coupling between volume deformation and thermal exchange. After elimination of the thermal modes it can contribute to an effective dynamical pressure, but such a pressure correction may already arise from the balance equations and
the temperature dependence of the partial thermodynamic pressures even when $\mathbb L_{\theta T}=0$;
\item $\mathbb L_{\theta R}$ describes reaction--volume coupling.
\end{itemize}
All these operators couple arrays of scalar co-factors, so rotational invariance does not exclude them. Their inclusion is a modelling decision, not a consequence of the entropy inequality alone.

Tensorial separation concerns entropy products, not source-evaluation order. The scalar reaction rates determine $r_i$ by \cref{eq:mass-production}, which enters the vector co-factor $\boldsymbol G_i$ through $-r_i\bu_i/(2T_i)$ in \cref{eq:noninteraction-cofactor}. Hence scalar evaluation must precede vector momentum reconstruction. This dependence comes from the kinetic mass-production term in the objective energy balance and its assignment to the vector co-factor.

For a constitutive implementation in which the nonequilibrium mechanical partial pressures are solved from the given dilatational rates, group
\begin{equation}
  \boldsymbol z:=(\boldsymbol\beta^{\perp},-\boldsymbol{\mathcal A})
  \in\mathcal E\oplus\R^{N_R},
  \qquad
  \boldsymbol w:=(\boldsymbol\varepsilon,\boldsymbol R)
  \in\mathcal E\oplus\R^{N_R}.
\label{eq:app-scalar-grouping}
\end{equation}
Writing \cref{eq:app-full-scalar-mobility} as
\begin{equation}
  \begin{pmatrix}\boldsymbol\theta\\ \boldsymbol w\end{pmatrix}
  =
  \begin{pmatrix}\mathbb L&\mathbb C\\ \mathbb C^{\sf T}&\mathbb A\end{pmatrix}
  \begin{pmatrix}\boldsymbol\sigma\\ \boldsymbol z\end{pmatrix},
\label{eq:app-scalar-block-short}
\end{equation}
assume that $\mathbb L$ is coercive and set
\begin{equation}
  \mathbb V:=\mathbb L^{-1},
  \qquad
  \mathbb C_{\mathrm{sc}}:=\mathbb V\mathbb C,
  \qquad
  \mathbb K_{\mathrm{sc}}
  :=\mathbb A-\mathbb C^{\sf T}\mathbb V\mathbb C.
\label{eq:app-scalar-schur-data}
\end{equation}
Then the equivalent resistance form is
\begin{align}
  \boldsymbol\sigma
  &=\mathbb V\boldsymbol\theta-\mathbb C_{\mathrm{sc}}\boldsymbol z,
\label{eq:app-scalar-resistance-pressure}
\\
  \boldsymbol w
  &=\mathbb C_{\mathrm{sc}}^{\sf T}\boldsymbol\theta
    +\mathbb K_{\mathrm{sc}}\boldsymbol z.
\label{eq:app-scalar-resistance-rates}
\end{align}
The cross-terms again cancel in the entropy production:
\begin{equation}
  \zeta_{\mathrm{sc}}
  =\boldsymbol\theta\cdot\mathbb V\boldsymbol\theta
   +\boldsymbol z\cdot\mathbb K_{\mathrm{sc}}\boldsymbol z\ge0.
\label{eq:app-scalar-diagonal-zeta}
\end{equation}
In components, \cref{eq:app-scalar-resistance-pressure} reconstructs the mechanical partial pressures as
\begin{equation}
  P_i=p_i-T_i(\mathbb V\boldsymbol\theta)_i
      +T_i(\mathbb C_{\mathrm{sc}}\boldsymbol z)_i.
\label{eq:app-mechanical-pressure-reconstruction}
\end{equation}
Thermal or reaction forces can therefore contribute to $P_i-p_i$ even at $\boldsymbol\theta=\boldsymbol0$. At zero total production, coercive diagonal dissipative operators instead give $\boldsymbol\theta=\boldsymbol0$ and $\boldsymbol z=\boldsymbol0$, hence $P_i=p_i$.

As in the vector block, the cancelling cross-map $\mathbb C_{\mathrm{sc}}$ is a reciprocal paired response, not a skew-adjoint addition. For fixed $\mathbb V>0$ and $\mathbb K_{\mathrm{sc}}\ge0$, its magnitude is unrestricted by entropy positivity: any admissible cross-map reconstructs a positive reciprocal mobility through
\begin{equation}
  \mathbb L=\mathbb V^{-1},\qquad
  \mathbb C=\mathbb V^{-1}\mathbb C_{\mathrm{sc}},\qquad
  \mathbb A=\mathbb K_{\mathrm{sc}}
  +\mathbb C_{\mathrm{sc}}^{\sf T}\mathbb V^{-1}\mathbb C_{\mathrm{sc}}.
\label{eq:app-scalar-reconstruct-mobility}
\end{equation}
Its magnitude requires constitutive data. The dissipative restrictions are
\begin{equation}
  \mathbb V>0\quad\text{on }\R^N,
  \qquad
  \mathbb K_{\mathrm{sc}}\ge0
  \quad\text{on }\mathcal E\oplus\R^{N_R}.
\label{eq:app-scalar-positivity}
\end{equation}
The frozen-coefficient quadratic test over all independent thermal and reaction forces requires positivity only of the compression to $\mathcal E\oplus\mathcal A_{\Stoich}$. For state-dependent coefficients, positivity of this compression suffices for the process inequality at the realized $\boldsymbol z$.

For the semidefinite scalar mobility, $\mathbb C:\mathcal E\oplus\R^{N_R}\to\R^N$. The block ordering in \cref{eq:app-scalar-block-short} requires the \emph{opposite adjoint placement} from \cref{eq:app-vector-singular-compatibility}:
\begin{equation}
  \mathbb C^{\sf T}(\ker\mathbb L)=\{0\}
  \quad\Longleftrightarrow\quad
  \operatorname{im}\mathbb C\subseteq\operatorname{im}\mathbb L.
\label{eq:app-scalar-singular-compatibility}
\end{equation}
Indeed, testing the full quadratic form on $(t\boldsymbol k,\boldsymbol z)$, with $\boldsymbol k\in\ker\mathbb L$ and arbitrary $t\in\R$, forces $\langle\boldsymbol k,\mathbb C\boldsymbol z\rangle=0$. Under this compatibility, completion of the square gives
\begin{equation}
 \boldsymbol\sigma\cdot\mathbb L\boldsymbol\sigma
 +2\boldsymbol\sigma\cdot\mathbb C\boldsymbol z
 +\boldsymbol z\cdot\mathbb A\boldsymbol z
 =(\boldsymbol\sigma+\mathbb L^{+}\mathbb C\boldsymbol z)
 \cdot\mathbb L(\boldsymbol\sigma+\mathbb L^{+}\mathbb C\boldsymbol z)
 +\boldsymbol z\cdot
 (\mathbb A-\mathbb C^{\sf T}\mathbb L^{+}\mathbb C)\boldsymbol z.
\label{eq:app-scalar-singular-square}
\end{equation}
Consequently the full block is positive semidefinite exactly when $\mathbb L\succeq0$, \cref{eq:app-scalar-singular-compatibility} holds, and $\mathbb A-\mathbb C^{\sf T}\mathbb L^{+}\mathbb C\succeq0$. Reconstruction from prescribed rates requires $\boldsymbol\theta\in\operatorname{im}\mathbb L$ and gives
\begin{equation}
  \boldsymbol\sigma=\mathbb L^{+}(\boldsymbol\theta-\mathbb C\boldsymbol z)
                         +\boldsymbol k,
  \qquad \boldsymbol k\in\ker\mathbb L.
\label{eq:app-scalar-singular-reconstruction}
\end{equation}
The response $\boldsymbol w$ is independent of $\boldsymbol k$, but a unique physical pressure requires a null-stress prescription. Rates outside $\operatorname{im}\mathbb L$ are constrained, not freely reconstructible. Generalized inverses again require constant-rank patches.

Direct semidefinite resistance is a distinct admissible case: \cref{eq:app-scalar-resistance-pressure,eq:app-scalar-resistance-rates} may be prescribed with $\mathbb V\succeq0$ and $\mathbb K_{\mathrm{sc}}\succeq0$, since cancellation requires no inverse. A dilatational rate in $\ker\mathbb V$ gives no viscous contribution; at $\boldsymbol z\ne0$, however, $-\mathbb C_{\mathrm{sc}}\boldsymbol z$ may still contribute to $\boldsymbol\sigma$. A null mobility mode instead constrains admissible rates by the reconstruction range condition. Substituting a Moore--Penrose inverse into coercive formulas does not establish equivalence.

{In the direct resistance formulation, including vanishing viscosity coefficients, the baseline model is the special case}
\begin{equation}
  \mathbb C_{\mathrm{sc}}=0,
  \qquad
  (\mathbb V\boldsymbol\theta)_i
  =\frac{\lambda_i^{\mathrm v}}{T_i}\theta_i,
\label{eq:app-scalar-baseline1}
\end{equation}
for which \cref{eq:app-scalar-resistance-pressure} is exactly $\pi_i=P_i-p_i=-\lambda_i^{\mathrm v}\theta_i$, together with a graph-Laplacian thermal block and diagonal pathwise reaction relations.
\begingroup
With coefficients fixed at equilibrium, the reaction block in \cref{eq:app-full-scalar-mobility} gives the usual linear approximation in the affinities. State-dependent coefficients can also represent nonlinear reaction relations. For  the logarithmic closure of the baseline model, for example, $R_a=-L_a\Acal_a$ with
\[
 L_a=\frac{2\gamma_a q_a}{R}\frac{\sinh(X_a/2)}{X_a}>0,
 \qquad L_a\big|_{X_a=0}=\frac{\gamma_a q_a}{R}.
\]
This exact rewriting supplies no independent kinetic derivation: one-way rates remain those of the reaction model, and validity beyond its calibrated regime needs justification. A general coupled scalar closure replaces separated path relations unless compatibility is established. For one scalar mechanism with strictly positive mobility, a linear net rate admits an algebraic split into positive terms with the required logarithmic ratio, but physical one-way rates require a specified event model. Nonlinear extensions must preserve nonnegativity for every selected binary mechanism, including mixed mechanisms. Local detailed balance with arbitrary state-dependent prefactors does not imply the stronger property of global monotonicity.
\endgroup

\subsection{Nonlinear completion of the thermal/reaction block by resolved processes}
\label{app:nonlinear-scalar}

\begingroup
The nonlinear resolved-process representation is given in \cref{subsec:general-event-channels}. Let $\mathbb G$ map the process-rate array to $\boldsymbol w=(\boldsymbol\varepsilon,\boldsymbol R)$, with columns $(\boldsymbol h^\ell,\boldsymbol\omega^\ell)\in\mathcal E\oplus\R^{N_R}$. Then $\boldsymbol\chi=R^{-1}\mathbb G^{\sf T}\boldsymbol z$, where $\boldsymbol z=(\boldsymbol\beta^{\perp},-\boldsymbol{\mathcal A})$. For fixed resolved-process columns, linearization about a detailed-balanced reference with $W_\ell^{f,\mathrm{eq}}=W_\ell^{b,\mathrm{eq}}=:W_\ell^{\mathrm{eq}}>0$ gives $\delta\boldsymbol w=\mathbb K_{\mathrm{sc}}^{\mathrm{eq}}\delta\boldsymbol z$, with
\begin{equation}
  \mathbb K_{\mathrm{sc}}^{\mathrm{eq}}
  =\frac1R\,
   \mathbb G_{\mathrm{eq}}\,
   \operatorname{diag}(W_1^{\mathrm{eq}},\ldots,W_m^{\mathrm{eq}})
   \mathbb G_{\mathrm{eq}}^{\sf T}.
\label{eq:app-linearized-event-operator}
\end{equation}
Indeed, differentiating $W_\ell=W_\ell^b(\exp\chi_\ell-1)$ at $\chi_\ell=0$ gives $\delta W_\ell=W_\ell^{\mathrm{eq}}\delta\chi_\ell$. The displayed operator is pairing-symmetric and positive semidefinite. The same formula applies to differentiable state-dependent columns at a reference with $\boldsymbol z^{\mathrm{eq}}=0$, in particular at common temperature with vanishing affinities. At a more general detailed-balanced reference, where only $\mathbb G_{\mathrm{eq}}^{\sf T}\boldsymbol z^{\mathrm{eq}}=0$, their state dependence must also be retained:
\begin{equation}
 \delta\boldsymbol w
 =\frac1R\mathbb G_{\mathrm{eq}}
 \operatorname{diag}(W_\ell^{\mathrm{eq}})
 \Bigl[\mathbb G_{\mathrm{eq}}^{\sf T}\delta\boldsymbol z
       +(\delta\mathbb G)^{\sf T}\boldsymbol z^{\mathrm{eq}}\Bigr].
\label{eq:app-state-dependent-event-linearization}
\end{equation}
Thus a fixed-column factorization cannot be assumed at every state of detailed balance for every resolved process. Conversely, an arbitrary dense positive factorization only reparameterizes a phenomenological matrix and does not identify physical resolved-process data.
\endgroup

\subsection{Admissibility theorem and reconstruction of the conservative sources}
\label{app:full-admissibility}

\begin{theorem}[Entropy admissibility of the fully coupled model]
\label[theorem]{thm:app-full-admissibility}
Retain the interior thermodynamic domain, entropy conjugates, weighted Euler relation, and selected entropy flux of the main text, and use the alternative mechanism specification of this appendix. Assume the reciprocal constitutive relations \cref{eq:app-full-shear,eq:app-vector-resistance-q,eq:app-vector-resistance-b,eq:app-scalar-resistance-pressure,eq:app-scalar-resistance-rates}. {Let $\boldsymbol\eta$, $\mathbb K$, $\mathbb M$, $\mathbb V$, and $\mathbb K_{\mathrm{sc}}$ be self-adjoint and positive semidefinite on their stated paired mechanism spaces. These hypotheses suffice when the displayed resistance relations are prescribed directly. If they are obtained by inversion of a semidefinite original mobility, restrict the admissible rates to its active range, impose \cref{eq:app-vector-singular-compatibility,eq:app-scalar-singular-compatibility}, and specify any null-force or null-stress representative needed for physical closure.
\begingroup
For such inverse constructions, the displayed resistance relations use the zero null representative. A general reconstruction retains the additional null-force or null-stress terms in \cref{eq:app-vector-singular-reconstruction,eq:app-scalar-singular-reconstruction}. These terms have zero pairing with the admissible active-range rates. Entropy admissibility does not by itself fix the representative or imply a regular evolution problem.
\endgroup
} Then every smooth solution satisfying these closures has entropy production
\begin{equation}
  \zeta={}
  2\sum_{i,j}\eta_{ij}\bD_i^{\circ}:\bD_j^{\circ}
  +\langle\boldsymbol\xi,\mathbb K\boldsymbol\xi\rangle
  +\langle\bu,\mathbb M\bu\rangle
\label{eq:app-full-zeta}
  +\boldsymbol\theta\cdot\mathbb V\boldsymbol\theta
  +\boldsymbol z\cdot\mathbb K_{\mathrm{sc}}\boldsymbol z
  \ge0.
\end{equation}
The explicit cross terms associated with $\mathbb D$ and $\mathbb C_{\mathrm{sc}}$ cancel in the chosen resistance variables. They are reciprocal paired cross-responses rather than skew-adjoint additions. Their magnitudes are not bounded by entropy positivity once the Schur-diagonal blocks are fixed. Physical identification requires additional constitutive input.
\end{theorem}

\begin{proof}
Insert \cref{eq:app-full-shear} into \cref{eq:app-zeta-deviatoric}, \cref{eq:app-vector-resistance-q,eq:app-vector-resistance-b} into \cref{eq:app-zeta-vector}, and \cref{eq:app-scalar-resistance-pressure,eq:app-scalar-resistance-rates} into \cref{eq:app-zeta-scalar}. The cross-terms cancel pairwise by the defining adjoint relations. The remaining terms are the quadratic forms in \cref{eq:app-full-zeta} and are nonnegative by assumption.
\end{proof}

\begingroup
\paragraph{Binary-product representation of the reciprocal resistance relations.}
The same constitutive relations also satisfy the mechanism-wise requirement of Principle~(IV). This can be seen directly, without inverting any operator. In the vector group, define the mixed co-factors
\begin{align*}
 \boldsymbol q^{\mathrm{mix}}
   &:=\widehat{\bq}-\mathbb D\bu,
 &
 \boldsymbol{\mathfrak b}^{\mathrm{mix}}
   &:=\boldsymbol{\mathfrak b}-\mathbb D^{\sf T}\boldsymbol\xi.
\end{align*}
Then
\begin{align*}
 \langle\widehat{\bq},\boldsymbol\xi\rangle
   -\langle\bu,\boldsymbol{\mathfrak b}\rangle
 &=
 \langle\boldsymbol q^{\mathrm{mix}},\boldsymbol\xi\rangle
   -\langle\bu,\boldsymbol{\mathfrak b}^{\mathrm{mix}}\rangle,
 \\
 \boldsymbol q^{\mathrm{mix}}&=\mathbb K\boldsymbol\xi,
 \qquad
 -\boldsymbol{\mathfrak b}^{\mathrm{mix}}=\mathbb M\bu.
\end{align*}
In the scalar group, set
\begin{align*}
 \boldsymbol\sigma^{\mathrm{mix}}
   &:=\boldsymbol\sigma+\mathbb C_{\mathrm{sc}}\boldsymbol z,
 &
 \boldsymbol w^{\mathrm{mix}}
   &:=\boldsymbol w-\mathbb C_{\mathrm{sc}}^{\sf T}\boldsymbol\theta.
\end{align*}
The corresponding identity and constitutive relations are
\begin{align*}
 \boldsymbol\theta\cdot\boldsymbol\sigma
   +\boldsymbol w\cdot\boldsymbol z
 &=
 \boldsymbol\theta\cdot\boldsymbol\sigma^{\mathrm{mix}}
   +\boldsymbol w^{\mathrm{mix}}\cdot\boldsymbol z,
 \\
 \boldsymbol\sigma^{\mathrm{mix}}&=\mathbb V\boldsymbol\theta,
 \qquad
 \boldsymbol w^{\mathrm{mix}}=\mathbb K_{\mathrm{sc}}\boldsymbol z.
\end{align*}
These triangular transformations preserve entropy production and parity. At fixed thermodynamic state, diagonalization of the self-adjoint positive-semidefinite operators on their paired spaces, including the deviatoric block, gives individually nonnegative mixed binary products under Principle~(IV). This also verifies direct semidefinite resistance without inversion or deletion of null modes. The stated reconstruction and representative conditions remain necessary.
\endgroup

After evaluating the scalar block, the mass productions are reconstructed from the reaction rates by \cref{eq:mass-production}. The vector block gives $\bmom_i$ through \cref{eq:app-full-momentum-reconstruction,eq:app-full-momentum-multiplier}. The conservative sources then follow from \cref{eq:source-reconstruction-raw}. Thus the enlarged closure preserves
\begin{equation}
  \sum_i r_i=0,
  \qquad
  \sum_i\bmom_i=\boldsymbol 0,
  \qquad
  \sum_i\varepsilon_i=0,
\label{eq:app-full-source-conservation}
\end{equation}
and hence the original conservation laws, with the triangular order scalar block $\to$ mass productions $\to$ vector block $\to$ objective momentum productions required by the kinetic term in the vector co-factor.

If the operators in \cref{eq:app-full-zeta} are coercive on the physical mechanism spaces, zero production forces vanishing deviatoric rates, reciprocal-temperature gradients, relative velocities, dilatational rates, and active thermal/reaction forces $\boldsymbol z$. Then \cref{eq:app-mechanical-pressure-reconstruction} gives $P_i=p_i$. In semidefinite models, additional zero-production modes are exactly the joint null spaces of the dissipative Schur-diagonal operators. Reciprocal cross-maps leave this quadratic zero set unchanged, although their contributions to physical responses on null modes need not vanish.

The theorem concerns aggregate sources. If the momentum inheritance in \cref{sec:chemical-momentum} is resolved separately and accounts for all reactive mass production, apply the vector closure to the co-factor in \cref{eq:resolved-mechanical-cofactor} and evaluate the remaining scalar and reactive kinetic terms jointly, or partition them into nonoverlapping energy sources with nonnegative entropy products. The resolved-process condition is \cref{eq:resolved-event-total-admissibility}, including its mechanism and balance-anchoring requirements. Reusing the scalar energy pairing in \cref{eq:chemical-joint-admissibility} counts it twice. Any separate reactive energy contribution must be added consistently to the physical source and not already included in the scalar block. Retain each scalar cross-response together with its reciprocal partner when splitting the source contributions.

On the common-temperature manifold, the reciprocal-temperature gradients coincide and the zero-sum thermal-exchange pairing vanishes, recovering the Class-II entropy-production structure. The restriction gives the separated Class-II closure when the retained diagonal laws agree and the additional couplings have vanishing net contributions to the retained laws. Exact nonlinear agreement also requires matching reaction-rate laws. Reaction--dilatation coupling can still shift $P_i$ away from $p_i$, while dilatation--thermal or reaction--thermal coupling can drive a zero-sum exchange in the constituent energy equations. Such an exchange still belongs to the reciprocal scalar law of the full Class-III model.

\subsection{Completeness versus identifiability}
\label{app:full-identifiability}

Under the stated isotropic reciprocal assumptions, \cref{eq:app-full-shear,eq:app-vector-mobility,eq:app-full-scalar-mobility} exhaust the finite linear graphs expressing deviatoric stresses in terms of rates and vector/scalar rates in terms of their co-factors. This algebraic classification does not identify material coefficients. Mobility/resistance equivalence requires invertibility: a singular deviatoric mobility constrains strain rates and leaves null stresses undetermined, so cannot be represented by the finite viscosity matrix \cref{eq:app-full-shear}. Direct singular resistances remain admissible but are not finite inverses of singular mobilities. Completeness is limited to these specified graph classes.

The isotropic deviatoric and vector conductivity blocks each contain $N(N+1)/2$ scalar functions; constrained mobility/resistance contributes $N(N-1)/2$, and its reciprocal cross-operator $N(N-1)$. These counts do not identify nonnegative binary weights: for $N\ge3$, the nonnegative Maxwell--Stefan graph is a proper subclass of positive constrained operators. For the scalar block, the full labelled-pathway representation \cref{eq:app-scalar-spaces} has dimension
\begin{equation}
  d_{\mathrm{path}}=2N-1+N_R,
\label{eq:app-scalar-path-dimension}
\end{equation}
and $d_{\mathrm{path}}(d_{\mathrm{path}}+1)/2$ formal reciprocal entries before structural simplification, possibly redundant. The independent thermodynamic space \cref{eq:app-independent-scalar-space} instead has dimension
\begin{equation}
  d_{\mathrm{ind}}=2N-1+r_{\Stoich}\le3N-2,
\label{eq:app-scalar-dimension}
\end{equation}
and $d_{\mathrm{ind}}(d_{\mathrm{ind}}+1)/2$ algebraically independent reciprocal entries before further reductions. The entropy pairing detects only the compression of the operator to the attainable force space. State variation alone need not identify even this compressed state-dependent operator, since the thermal and chemical force components are fixed at each thermodynamic state. All $N_R$ coordinates may still distinguish kinetics. In particular, resolved processes with equal mass-production columns can assign different constituent energy increments and must not be identified on mass columns alone. Identifying these coefficients requires material- and regime-specific measurements or kinetic calculations.

For an illustration with $N_R=r_{\Stoich}=N-1$, the algebraic counts are listed in Table~\ref{tab:coefficient-counts}.
\begin{table}[htbp]
\centering
\caption{Numbers of scalar constitutive functions for the  specified fully coupled and baseline models.}
\label{tab:coefficient-counts}
\begin{tabular}{c r r r r r}
\toprule
$N$ & Deviatoric & Vector & Scalar & Fully coupled & Baseline model\\
\midrule
2 & 3 & 6 & 10 & 19 & 10\\
3 & 6 & 15 & 28 & 49 & 19\\
4 & 10 & 28 & 55 & 93 & 30\\
\bottomrule
\end{tabular}
\end{table}
 The coefficient count for the fully coupled model is $N(N+1)/2+N(2N-1)+d_{\mathrm{ind}}(d_{\mathrm{ind}}+1)/2$. Here the vector block is counted in one fixed entropy-invariant representation, so the $d_i^T$ are not counted again as additional full-model parameters.  The coefficient count for the baseline model is $3N+(N-1)+N(N-1)+N_R$: two diagonal viscosity families, heat conduction, $N-1$ independent $d_i^T$, the two pairwise interaction families, and one kinetic rate function per reaction. The table fixes $\gamma_a=1$ and uses the near-balanced reaction coefficients for comparison. These are counts of scalar constitutive functions, excluding the thermodynamic potential and partial-pressure data. If the reaction rank is smaller, the independent scalar count decreases. Retaining dependent reaction labels gives the separate formal count above. Without reactions,  the coefficient counts for the fully coupled models are $15$, $36$, and $66$, respectively.

To limit material-data requirements, the baseline model retains couplings required by the class-II limit, supported by an elementary-process or interaction argument, or identifiable from data. The formal completion above remains confined to the stated isotropic reciprocal co-factor representation.

\section{Detailed internal-state ionization derivation}
\label{sec:saha}

A state-resolved ionization network illustrates the thermochemical sources of \cref{sec:closed-model}. The stationary formulas describe local population kinetics in a homogeneous common-velocity reference state. Electromagnetic field evolution, its feedback on transport, and collective plasma-force closure are excluded. No self-consistent spatial plasma model is claimed. The net reaction
\begin{equation}
  A^z\rightleftharpoons A^{z+1}+\mathrm e^-
\label{eq:ionization-reaction}
\end{equation}
is an aggregate description. Electron- and heavy-particle-impact ionization, excitation, de-excitation, and their reverse processes act on internal states and exchange energy with different thermal groups. Each resolved electronic level is therefore a thermodynamic species satisfying the Class-III mass, momentum, and nonkinetic-energy balances, without an additional ``excitation temperature''.

Internal partition functions are derived rather than prescribed in aggregate potentials. Reservoir assignments follow from resolved processes. The construction also covers competing colliders for which no single stationary Saha formula exists.

\subsection{\texorpdfstring{{Internal-state species and the two-temperature reference manifold}}{Internal-state species and the two-temperature reference manifold}}
\label{subsec:internal-state-species}

For each charge state $z$, let
\begin{equation}
  \mathcal I_z=\{0,1,\ldots,L_z\}
\label{eq:resolved-level-set}
\end{equation}
be a finite set of resolved electronic levels. The species associated with level $\alpha\in\mathcal I_z$ is denoted by $A^z_\alpha$, with number density $n_{z\alpha}$, degeneracy $g_{z\alpha}$, molar mass $M_{z\alpha}$, and molar energy of the resolved internal state
\begin{equation}
  E_{z\alpha}^{\mathrm m,0}
  =E_{z0}^{\mathrm m,0}+\epsilon_{z\alpha}^{\mathrm m},
  \qquad
  \epsilon_{z0}^{\mathrm m}=0.
\label{eq:level-reference-energy}
\end{equation}
Excitation and ionization energies belong to the nonkinetic internal energies. We use fixed nonrelativistic model masses $M_{z\alpha}=M_z$, neglecting mass defects of order $I_z^{\mathrm m}/c^2$, and impose $M_z=M_{z+1}+M_{\mathrm e}$ exactly. Every excitation, ionization, and recombination process thus conserves model mass. The later neglect of electron mass in a heavy-species wavelength ratio is a prefactor approximation only, never a mass-balance approximation. The aggregate charge-state density is the diagnostic sum,
\begin{equation}
  n_z:=\sum_{\alpha\in\mathcal I_z}n_{z\alpha},
  \qquad
  \rho_z:=\sum_{\alpha\in\mathcal I_z}\rho_{z\alpha},
\label{eq:aggregate-charge-state-density}
\end{equation}
not an additional dynamical variable.

In the exact class-III model each $A^z_\alpha$ may have a distinct velocity and temperature. Let $\mathcal G_{\mathrm h}$ contain all resolved heavy internal-state species and heavy colliders, and let $\mathcal G_{\mathrm e}=\{\mathrm e^-\}$. For comparison with two-temperature ionization formulas, we use the grouped-temperature manifold
\begin{equation}
  T_i=T_{\mathrm h}\quad(i\in\mathcal G_{\mathrm h}),
  \qquad
  T_i=T_{\mathrm e}\quad(i\in\mathcal G_{\mathrm e}),
  \qquad T_{\mathrm e}\ \text{independent of }T_{\mathrm h}.
\label{eq:state-resolved-two-temperature-manifold}
\end{equation}

The level-population balances remain distinct. Below we evaluate the network at prescribed positive $T_{\mathrm h},T_{\mathrm e}$ in a homogeneous common-velocity reference state, $\bu_i=\boldsymbol0$; the full Class-III model retains individual velocities. Stationary population sources need not satisfy stationary partial-energy balances. An autonomous two-temperature reduction additionally requires group-energy projection and reduced closures. No invariance or dynamical convergence of this reduction is asserted here.

Exact temperature grouping requires constraint reactions or a justified singular relaxation limit: rapid finite intragroup relaxation generally gives only approximate grouping, and the grouped manifold need not be invariant under the unconstrained system. The electron remains separate. Radiation requires an explicit photon reservoir or radiation balance.

For dilute Maxwell--Boltzmann translational statistics, the state-species chemical potentials on \cref{eq:state-resolved-two-temperature-manifold} are
\begin{align}
  \mu_{z\alpha}^{\mathrm m}
  & =E_{z\alpha}^{\mathrm m,0}
  +RT_{\mathrm h}\log\!\Bigl(
  \frac{n_{z\alpha}\lambda_z(T_{\mathrm h})^3}{g_{z\alpha}}
  \Bigr),
\label{eq:state-resolved-heavy-chemical-potential}
\\
  \mu_{\mathrm e}^{\mathrm m}
  & =E_{\mathrm e}^{\mathrm m,0}
  +RT_{\mathrm e}\log\!\Bigl(
  \frac{n_{\mathrm e}\lambda_{\mathrm e}(T_{\mathrm e})^3}{g_{\mathrm e}}
  \Bigr),
  \qquad g_{\mathrm e}=2,
\label{eq:state-resolved-electron-chemical-potential}
\end{align}
where
\begin{equation}
  \lambda_i(T):=\frac{h}{\sqrt{2\pi m_i k_{\mathrm B}T}}
\label{eq:thermal-wavelength}
\end{equation}
is the thermal de Broglie wavelength, with particle mass $m_i=M_i/N_{\mathrm A}$ and constants $h$ (Planck), $k_{\mathrm B}$ (Boltzmann), and $N_{\mathrm A}$ (Avogadro). Henceforth choose $E_{\mathrm e}^{\mathrm m,0}=0$. Since $n_{z\alpha}$ are resolved state variables, \cref{eq:state-resolved-heavy-chemical-potential} contains no internal partition function. The Boltzmann partition function arises only upon imposing detailed-balanced bound--bound relations or a justified fast-relaxation limit to that manifold.
\begingroup

To realize Specification~(M2), take every resolved species, including colliders, as ideal and structureless apart from its discrete degeneracy. Before temperature grouping, set $n_i=N_{\mathrm A}\rho_i/M_i$ and
\begin{align}
 z_i:=\rho_i e_i
 &=\frac{n_i E_i^{\mathrm m,0}}{N_{\mathrm A}}
   +\frac32 n_i k_{\mathrm B}T_i,
 &p_i&=n_i k_{\mathrm B}T_i,
 \label{eq:state-ideal-caloric-realization}\\
 \widehat{\rho s}
 &=\sum_i n_i k_{\mathrm B}
 \Bigl[\frac52-\log\!\bigl(\frac{n_i\lambda_i(T_i)^3}{g_i}\bigr)\Bigr].
 \label{eq:state-ideal-entropy-realization}
\end{align}
Solve the first relation for $T_i$ on $n_i>0$, $a_i:=z_i-\rho_i E_i^{\mathrm m,0}/M_i>0$. Differentiation gives \cref{eq:entropy-conjugates} and the ideal chemical potentials; the pressures satisfy \cref{eq:weighted-euler}. Each entropy summand has Hessian in $(a_i,n_i)$
\begin{equation}
 k_{\mathrm B}\begin{pmatrix}
 -3n_i/(2a_i^2)&3/(2a_i)\\
 3/(2a_i)&-5/(2n_i)
 \end{pmatrix},
 \qquad \det=\frac{3k_{\mathrm B}^2}{2a_i^2}>0.
 \label{eq:state-ideal-entropy-hessian}
\end{equation}
The negative first diagonal entry and positive determinant imply negative definiteness, preserved by the invertible linear change from $(z_i,\rho_i)$ to $(a_i,n_i)$. Together with $\partial_{z_i}\widehat{\rho s}=1/T_i>0$, this verifies the entropy assumptions. Unresolved rotational or vibrational modes require consistent additional caloric and entropy terms.
\endgroup

\subsection{Closure for resolved internal-state processes}
\label{subsec:pathway-ionization}

Let $\ell$ denote an elementary bound--bound or ionization process among the internal-state species. Its stoichiometric coefficients are $\nu_i^{\ell}$, with
\begin{equation}
  \sum_iM_i\nu_i^{\ell}=0,
  \qquad
  \Acal_{\ell}:=\sum_iM_i\nu_i^{\ell}\frac{\mu_i}{T_i}.
\label{eq:state-event-affinity}
\end{equation}

Prescribe molar objective barycentric-frame energy increments with the resolved forward/reverse interpretation of \cref{subsec:general-event-channels},
\begin{equation}
  \boldsymbol h^{\ell}=(h_i^{\ell})_{i=1}^{N},
  \qquad
  \sum_i h_i^{\ell}=0.
\label{eq:state-event-energy-allocation}
\end{equation}
Internal-state energy differences in the chemical potentials specify stored energies; $\boldsymbol h^{\ell}$ assigns the process's threshold loss and gain to constituent objective-energy productions. Their distinction determines the reciprocal-temperature weighting of the process entropy change.

Let $W_{\ell}^f,W_{\ell}^b>0$ be one-way molar process rates and set
\begin{equation}
  \chi_{\ell}
  :=\frac1R\Bigl[
  \sum_i\frac{h_i^{\ell}}{T_i}-\Acal_{\ell}
  \Bigr],
  \qquad
  \log\frac{W_{\ell}^f}{W_{\ell}^b}=\chi_{\ell},
  \qquad
  W_{\ell}:=W_{\ell}^f-W_{\ell}^b.
\label{eq:state-event-local-detailed-balance}
\end{equation}
For a collection $\mathcal L$ of resolved processes, define
\begin{equation}
  r_i^{\mathrm{st}}
  =M_i\sum_{\ell\in\mathcal L}\nu_i^{\ell}W_{\ell},
  \qquad
  \varepsilon_i^{\mathrm{st}}
  =\sum_{\ell\in\mathcal L}h_i^{\ell}W_{\ell}.
\label{eq:state-event-productions}
\end{equation}

\begin{proposition}[Entropy-admissible resolved-process representation]
\label[proposition]{prop:state-event-admissibility}
The productions in \cref{eq:state-event-productions} obey the zero-sum mass and objective-energy constraints. These do not imply separate thermal-energy conservation, whose balance also contains mechanical exchange. Their thermochemical entropy production is
\begin{equation}
  \sum_i\frac1{T_i}\,\varepsilon_i^{\mathrm{st}}
  -\sum_{\ell\in\mathcal L}W_{\ell}\Acal_{\ell}
   =R\sum_{\ell\in\mathcal L}W_{\ell}\chi_{\ell}
\label{eq:state-event-dissipation}
   =R\sum_{\ell\in\mathcal L}
  (W_{\ell}^f-W_{\ell}^b)
  \log\frac{W_{\ell}^f}{W_{\ell}^b}
  \ge0.
\end{equation}
One positive one-way rate and the specified zero-sum energy increments therefore determine the reverse rate and an entropy-admissible source.
\end{proposition}

\begin{proof}
The constraints in \cref{eq:state-event-affinity,eq:state-event-energy-allocation} give $\sum_i r_i^{\mathrm{st}}=0$ and $\sum_i\varepsilon_i^{\mathrm{st}}=0$. Substituting \cref{eq:state-event-productions,eq:state-event-local-detailed-balance} gives \cref{eq:state-event-dissipation}, nonnegative process by process by monotonicity of the logarithm.
\end{proof}

\begin{corollary}[Embedding in the closed class-III model]
\label[corollary]{cor:pathway-closed-model}
Replacing the matching separated thermal/reaction block in \cref{sec:closed-model} by the sources of \cref{prop:state-event-admissibility} preserves \cref{thm:entropy-admissibility}, with \cref{eq:state-event-dissipation} replacing that block's entropy production. Appending it to already counted mechanisms is inadmissible. Independent thermal mechanisms require an explicit nonoverlapping source split. For resolved momentum inheritance, impose \cref{eq:resolved-event-total-admissibility} on the same process source, including its mechanism and balance-anchoring requirements, or use the nonoverlapping energy-source construction given there. Applying \cref{prop:chemical-joint-condition} to an energy pairing already used in \cref{eq:state-event-dissipation} is insufficient.
\end{corollary}

\begingroup
For the stationary level and Saha formulas, split the full objective-energy increment per mole of progress, $\boldsymbol h^{\ell}$ in \cref{eq:state-event-energy-allocation}, into threshold and residual parts, imposing
\begin{equation}
 \boldsymbol h_{\mathrm{full}}^\ell
 =\boldsymbol h_{\mathrm{th}}^\ell+\boldsymbol h_{\mathrm{res}}^\ell,
 \qquad \sum_i h_{\mathrm{res},i}^\ell=0,
 \qquad \sum_i\frac{h_{\mathrm{res},i}^\ell}{T_i}=0.
\label{eq:full-threshold-event-allocation}
\end{equation}
The last condition makes the residual entropy neutral. It is an additional constitutive restriction on the grouped-temperature manifold, not a consequence of conservation. With one electron constituent and $T_{\mathrm e}\ne T_{\mathrm h}$, it implies $h_{\mathrm{res},\mathrm e}^\ell=0$ and $\sum_{i\in\mathcal G_{\mathrm h}}h_{\mathrm{res},i}^\ell=0$: intragroup sensible-energy redistribution is allowed, but net transfer between the groups is excluded. This residual differs from the mass-production column $\boldsymbol g^\ell$ in \cref{eq:event-species-production}.

Sections~\ref{subsec:state-resolved-excitation}--\ref{subsec:state-resolved-ionization} display $\boldsymbol h_{\mathrm{th}}^\ell$. A dynamical model must supply the collision-kinematic residual and use the full vector in \cref{eq:state-event-productions}. The threshold vectors alone need not preserve the grouped-temperature manifold under net conversion. A nonzero pairing $Q_\ell:=\sum_i h_{\mathrm{res},i}^\ell/T_i$ multiplies each subsequent stationary level-ratio or level-ionization right-hand side by $\exp(Q_\ell/R)$, requiring corresponding modification of the threshold-only formulas and their aggregate reduction.
\endgroup

\subsection{Bound--bound transitions and derivation of internal partition functions}
\label{subsec:state-resolved-excitation}

Let $\alpha,\beta\in\mathcal I_z$ with $\epsilon_{z\beta}^{\mathrm m}>\epsilon_{z\alpha}^{\mathrm m}$ and define
\begin{equation}
  \Delta E_{z,\alpha\beta}^{\mathrm m}
  :=E_{z\beta}^{\mathrm m,0}-E_{z\alpha}^{\mathrm m,0}>0.
\label{eq:level-gap}
\end{equation}
Let $\boldsymbol\delta^{\,i}$ denote the canonical unit vector associated with constituent $i$ in the array of objective energy productions.

For electron-impact excitation,
\begin{equation}
  A^z_\alpha+\mathrm e^-
  \rightleftharpoons
  A^z_\beta+\mathrm e^-,
\label{eq:electron-impact-excitation}
\end{equation}
choose the threshold energy-source contribution
\begin{equation}
  \boldsymbol h_{\mathrm{th},z,\alpha\beta}^{\mathrm e}
  =\Delta E_{z,\alpha\beta}^{\mathrm m}
  \bigl(\boldsymbol\delta^{\,z\beta}
  -\boldsymbol\delta^{\,\mathrm e}\bigr).
\label{eq:electron-excitation-allocation}
\end{equation}
At detailed equilibrium of each resolved process, \cref{eq:state-event-local-detailed-balance} gives
\begin{equation}
  \frac{\mu_{z\beta}^{\mathrm m}-\mu_{z\alpha}^{\mathrm m}}{T_{\mathrm h}}
  =\Delta E_{z,\alpha\beta}^{\mathrm m}
  \Bigl(\frac1{T_{\mathrm h}}-\frac1{T_{\mathrm e}}\Bigr).
\label{eq:electron-excitation-affinity}
\end{equation}
Using \cref{eq:state-resolved-heavy-chemical-potential} yields
\begin{equation}
  \frac{n_{z\beta}}{n_{z\alpha}}
  =\frac{g_{z\beta}}{g_{z\alpha}}
  \exp\!\Bigl(-\frac{\Delta E_{z,\alpha\beta}^{\mathrm m}}{RT_{\mathrm e}}\Bigr).
\label{eq:electron-controlled-level-ratio}
\end{equation}
Although both heavy state species translate at $T_{\mathrm h}$, their population ratio is controlled by $T_{\mathrm e}$ because the electron constituent supplies the excitation energy.

For excitation by a heavy collider $B$,
\begin{equation}
  A^z_\alpha+B
  \rightleftharpoons
  A^z_\beta+B,
  \qquad
  \boldsymbol h_{\mathrm{th},z,\alpha\beta}^{B}
  =\Delta E_{z,\alpha\beta}^{\mathrm m}
  \bigl(\boldsymbol\delta^{\,z\beta}
  -\boldsymbol\delta^{\,B}\bigr),
\label{eq:heavy-excitation-event}
\end{equation}
with $T_B=T_{\mathrm h}$, the energy-source contribution has zero reciprocal-temperature pairing. Detailed equilibrium therefore gives
\begin{equation}
  \frac{n_{z\beta}}{n_{z\alpha}}
  =\frac{g_{z\beta}}{g_{z\alpha}}
  \exp\!\Bigl(-\frac{\Delta E_{z,\alpha\beta}^{\mathrm m}}{RT_{\mathrm h}}\Bigr).
\label{eq:heavy-controlled-level-ratio}
\end{equation}

\begin{proposition}[Reduction of a connected bound--bound network]
\label[proposition]{prop:state-level-partition}

\begingroup
Assume the ideal potentials above and the threshold energy-source contributions with residuals satisfying \cref{eq:full-threshold-event-allocation}. Let the resolved bound--bound graph in charge state $z$ be connected, with all active processes individually detailed-balanced by one collider bath at temperature $T_c$. Then
\endgroup

\begin{equation}
  n_{z\alpha}
  =n_z\,
  \frac{g_{z\alpha}
  \exp[-\epsilon_{z\alpha}^{\mathrm m}/(RT_c)]}{U_z(T_c)},
  \qquad
  U_z(T_c)
  :=\sum_{\gamma\in\mathcal I_z}
  g_{z\gamma}
  \exp\!\Bigl[-\frac{\epsilon_{z\gamma}^{\mathrm m}}{RT_c}\Bigr].
\label{eq:derived-internal-partition-function}
\end{equation}
For electron-controlled excitation $T_c=T_{\mathrm e}$; for heavy-controlled excitation $T_c=T_{\mathrm h}$.
\end{proposition}

\begin{proof}
{\Cref{eq:electron-controlled-level-ratio,eq:heavy-controlled-level-ratio} give} the ratio of every pair joined by an edge. Connectedness propagates the corresponding Boltzmann ratio through the graph. Normalization by \cref{eq:aggregate-charge-state-density} yields \cref{eq:derived-internal-partition-function}.
\end{proof}

The partition function thus reduces a detailed-balanced state network. Metastable groups not rapidly connected must remain separate dynamical species. Combining them would discard a slow variable.

\subsection{Level-resolved ionization and the two stationary Saha relations}
\label{subsec:state-resolved-ionization}

For $\alpha\in\mathcal I_z$ and $\beta\in\mathcal I_{z+1}$ define the level-to-level molar ionization threshold
\begin{equation}
  I_{z,\alpha\beta}^{\mathrm m}
  :=E_{z+1,\beta}^{\mathrm m,0}
  +E_{\mathrm e}^{\mathrm m,0}
  -E_{z\alpha}^{\mathrm m,0}
  =I_z^{\mathrm m}
  +\epsilon_{z+1,\beta}^{\mathrm m}
  -\epsilon_{z\alpha}^{\mathrm m},
\label{eq:level-ionization-threshold}
\end{equation}
where $I_z^{\mathrm m}$ is the ground-state molar ionization energy in the chosen reference convention.

Electron-impact ionization is
\begin{equation}
  A^z_\alpha+\mathrm e^-
  \rightleftharpoons
  A^{z+1}_\beta+2\mathrm e^-,
  \qquad
  \boldsymbol h_{\mathrm{th},z,\alpha\beta}^{\mathrm{ion,e}}
  =I_{z,\alpha\beta}^{\mathrm m}
  \bigl(\boldsymbol\delta^{\,(z+1,\beta)}
  -\boldsymbol\delta^{\,\mathrm e}\bigr).
\label{eq:electron-impact-ionization-event}
\end{equation}
The electron occurs on both sides of the reaction equation and has net stoichiometric coefficient one. Detailed equilibrium gives
\begin{equation}
  \frac{n_{z+1,\beta}n_{\mathrm e}}{n_{z\alpha}}
  =\frac{g_{z+1,\beta}g_{\mathrm e}}{g_{z\alpha}}
  \frac{\lambda_z(T_{\mathrm h})^3}
       {\lambda_{z+1}(T_{\mathrm h})^3
        \lambda_{\mathrm e}(T_{\mathrm e})^3}
  \exp\!\Bigl[-\frac{I_{z,\alpha\beta}^{\mathrm m}}{RT_{\mathrm e}}\Bigr].
\label{eq:state-electron-ionization-relation}
\end{equation}

For ionization by a heavy collider,
\begin{equation}
  A^z_\alpha+B
  \rightleftharpoons
  A^{z+1}_\beta+\mathrm e^-+B,
  \qquad
  \boldsymbol h_{\mathrm{th},z,\alpha\beta}^{\mathrm{ion},B}
  =I_{z,\alpha\beta}^{\mathrm m}
  \bigl(\boldsymbol\delta^{\,(z+1,\beta)}
  -\boldsymbol\delta^{\,B}\bigr),
\label{eq:heavy-impact-ionization-event}
\end{equation}
detailed equilibrium gives
\begin{equation}
  \frac{n_{z+1,\beta}n_{\mathrm e}}{n_{z\alpha}}
  =\frac{g_{z+1,\beta}g_{\mathrm e}}{g_{z\alpha}}
  \frac{\lambda_z(T_{\mathrm h})^3}
       {\lambda_{z+1}(T_{\mathrm h})^3
        \lambda_{\mathrm e}(T_{\mathrm e})^3}
  \exp\!\Bigl[-\frac{I_{z,\alpha\beta}^{\mathrm m}}{RT_{\mathrm h}}\Bigr].
\label{eq:state-heavy-ionization-relation}
\end{equation}

The electron translational phase-space factor remains evaluated at $T_{\mathrm e}$, while the threshold is weighted by the temperature of the collider bath that supplies it.

\begingroup
The heavy-impact threshold-only formula uses an idealized source closure. The threshold vector has zero electron component and, for $T_{\mathrm e}\ne T_{\mathrm h}$, the residual restriction in \cref{eq:full-threshold-event-allocation} makes the full electron-group objective-energy increment zero. Outgoing-electron kinetic energy, recoil, and other intergroup excess-energy transfers generally require an entropy-active residual or a separately resolved nonoverlapping mechanism. A molar electron-group transfer $q_{\mathrm e}$ with compensating heavy transfer multiplies the stationary right-hand side by $\exp[q_{\mathrm e}(1/T_{\mathrm e}-1/T_{\mathrm h})/R]$.
\endgroup

Specify one positive forward molar rate. In the following number-density examples, the displayed coefficients $k$ are the conventional particle-based collision coefficients divided by $N_{\mathrm A}$. This conversion is already included in $k$, so the displayed $W^{f,b}$ are molar progress rates:
\begin{align}
  W_{z\alpha\beta,\mathrm e}^f
  & =k_{z\alpha\beta,\mathrm e}^{\mathrm{ion}}(T_{\mathrm e})
  n_{z\alpha}n_{\mathrm e},
  &
  W_{z\alpha\beta,\mathrm e}^b
  & =k_{z\alpha\beta,\mathrm e}^{3\mathrm b}
  n_{z+1,\beta}n_{\mathrm e}^2,
\label{eq:state-electron-ionization-rates}
\\
  W_{z\alpha\beta,B}^f
  & =k_{z\alpha\beta,B}^{\mathrm{ion}}(T_B)
  n_{z\alpha}n_B,
  &
  W_{z\alpha\beta,B}^b
  & =k_{z\alpha\beta,B}^{\mathrm{rec}}
  n_{z+1,\beta}n_{\mathrm e}n_B.
\label{eq:state-heavy-ionization-rates}
\end{align}

These temperature-only forward coefficients illustrate common-velocity closures. Full Class-III coefficients may depend on all participating temperatures and objective relative velocities. Given the forward coefficient and constituent energy increments, \cref{eq:state-event-local-detailed-balance} fixes the reverse coefficient.

\begin{proposition}[Aggregate two-temperature Saha relation]
\label[proposition]{prop:state-resolved-saha}

\begingroup
Assume that the bound--bound networks of charge states $z$ and $z+1$ satisfy \cref{eq:derived-internal-partition-function} at one controlling temperature $T_c$. Suppose at least one ionization process connecting these charge states is individually detailed-balanced, uses the same threshold-supplying bath and the displayed threshold energy-source contribution, and has a residual satisfying \cref{eq:full-threshold-event-allocation}. Then
\endgroup

\begin{equation}
  \frac{n_{z+1}n_{\mathrm e}}{n_z}
  =\frac{g_{\mathrm e}}{\lambda_{\mathrm e}(T_{\mathrm e})^3}
  \frac{\lambda_z(T_{\mathrm h})^3}
       {\lambda_{z+1}(T_{\mathrm h})^3}
  \frac{U_{z+1}(T_c)}{U_z(T_c)}
  \exp\!\Bigl(-\frac{I_z^{\mathrm m}}{RT_c}\Bigr).
\label{eq:state-resolved-generalized-saha}
\end{equation}
Electron-controlled excitation and ionization give $T_c=T_{\mathrm e}$; heavy-controlled excitation and ionization give $T_c=T_{\mathrm h}$.
\end{proposition}

\begin{proof}
Insert \cref{eq:derived-internal-partition-function} into either \cref{eq:state-electron-ionization-relation} or \cref{eq:state-heavy-ionization-relation}. The level degeneracies and the factors containing $\epsilon_{z\alpha}^{\mathrm m}$ and $\epsilon_{z+1,\beta}^{\mathrm m}$ cancel because the same $T_c$ controls the level populations and the threshold. The remaining relation is independent of $\alpha$ and $\beta$ and gives \cref{eq:state-resolved-generalized-saha}.
\end{proof}

The electron- and heavy-controlled cases of \cref{prop:state-resolved-saha} permit a mechanism-resolved comparison with the multi-reservoir setting of Ridenti and Silva~\cite{RidentiSilva2026}. Here reservoir assignments arise from state-resolved networks and their specified energy-source contributions, rather than temperatures assigned to modes within one species. Identification with another two-temperature Saha convention requires matching reservoir constraints, energy zeros, and population definitions.

\begin{proposition}[Classical Saha--Eggert limit]
\label[proposition]{prop:saha-limit}
{Under the hypotheses of \cref{prop:state-resolved-saha}, impose the common-temperature specialization $T_{\mathrm e}=T_{\mathrm h}=T_c=T$ and let $\chi_z=I_z^{\mathrm m}/N_{\mathrm A}$. Then}
\begin{equation}
  \frac{n_{z+1}n_{\mathrm e}}{n_z}
  =\frac{U_{z+1}(T)g_{\mathrm e}}{U_z(T)}
  \frac{\lambda_z(T)^3}
       {\lambda_{z+1}(T)^3\lambda_{\mathrm e}(T)^3}
  \exp\!\Bigl(-\frac{\chi_z}{k_{\mathrm B}T}\Bigr).
\label{eq:exact-saha}
\end{equation}

\begingroup
With $g_{\mathrm e}=2$ and the approximation $\lambda_z(T)^3/\lambda_{z+1}(T)^3=(m_{z+1}/m_z)^{3/2}\simeq1$ in the prefactor only,
\begin{equation}
  \frac{n_{z+1}n_{\mathrm e}}{n_z}
  \simeq\frac{2}{\lambda_{\mathrm e}(T)^3}
  \frac{U_{z+1}(T)}{U_z(T)}
  \exp\!\Bigl(-\frac{\chi_z}{k_{\mathrm B}T}\Bigr).
\label{eq:saha-eggert}
\end{equation}
\endgroup

\end{proposition}

At common temperature, $T^{-1}\sum_i h_i^{\ell}=0$ for every zero-sum process increment, so the classical Saha--Eggert limit is collider independent. Unequal-temperature analogues depend on the reaction mechanism.

\subsection{Competing pathways, stationary populations, and dissipative cycles}
\label{subsec:state-resolved-cycles}

With simultaneous electron and heavy collisions, the state-resolved model remains defined but need not admit one internal partition function. For one bound--bound transition, simultaneous detailed equilibrium of \cref{eq:electron-impact-excitation,eq:heavy-excitation-event} requires
\begin{equation}
  \Delta E_{z,\alpha\beta}^{\mathrm m}
  \Bigl(\frac1{T_{\mathrm h}}-\frac1{T_{\mathrm e}}\Bigr)=0.
\label{eq:excitation-pathway-compatibility}
\end{equation}
For a nondegenerate transition and $T_{\mathrm e}\ne T_{\mathrm h}$, the two processes cannot both be individually detailed-balanced. This is the state-resolved form of the general compatibility condition
\begin{equation}
  \sum_i\frac{h_i^{\ell}-h_i^m}{T_i}=0
\label{eq:pathway-compatibility}
\end{equation}
for two resolved processes with the same net stoichiometry.

A stationary level population is nevertheless possible. Let $W_{\alpha\beta}^{\mathrm e}$ and $W_{\alpha\beta}^{B}$ denote the net electron- and heavy-impact excitation rates. If
\begin{equation}
  W_{\alpha\beta}^{\mathrm e}
  =-W_{\alpha\beta}^{B}=:J\ne0,
\label{eq:state-cycle-zero-population-source}
\end{equation}
then the population source of both internal-state species vanishes. The threshold contribution to the objective-energy source is
\begin{equation}
  \ensuremath{\boldsymbol\varepsilon_{\mathrm{th}}^{\mathrm{cyc}}}
  =J\bigl(
  \boldsymbol h_{\mathrm{th},z,\alpha\beta}^{\mathrm e}
  -\boldsymbol h_{\mathrm{th},z,\alpha\beta}^{B}
  \bigr)
  =\Delta E_{z,\alpha\beta}^{\mathrm m}J
  \bigl(\boldsymbol\delta^{\,B}
  -\boldsymbol\delta^{\,\mathrm e}\bigr).
\label{eq:state-excitation-cycle-energy}
\end{equation}
The contributions to creation of the excited state cancel, leaving an energy transfer between thermal groups. The full cycle source is $\boldsymbol\varepsilon^{\mathrm{cyc}}=\boldsymbol\varepsilon_{\mathrm{th}}^{\mathrm{cyc}}+J(\boldsymbol h_{\mathrm{res}}^{\mathrm e}-\boldsymbol h_{\mathrm{res}}^{B})$. By \cref{eq:full-threshold-event-allocation}, the residual difference contributes no entropy production and, at unequal bath temperatures, no groupwise energy transfer. The cycle entropy production is

\begin{equation}
  \zeta_{\mathrm{cyc}}
  =\Delta E_{z,\alpha\beta}^{\mathrm m}J
  \Bigl(\frac1{T_B}-\frac1{T_{\mathrm e}}\Bigr)>0
\label{eq:state-excitation-cycle-dissipation}
\end{equation}
for the rate direction selected by the local-detailed-balance relations. {An analogous ionization--recombination cycle can maintain stationary $n_z$ and $n_{z+1}$ while transferring} the level-dependent threshold $I_{z,\alpha\beta}^{\mathrm m}$ between the two baths.

\begingroup
Thus zero population sources can coexist with dissipative cycles. In the homogeneous common-velocity state, the excitation cycle supplies $-\Delta E_{z,\alpha\beta}^{\mathrm m}J$ to electrons and the opposite amount to the heavy group. Fixed temperatures require compensating reservoir or transport fluxes; isolated group energies evolve. Full nonequilibrium stationarity also requires stationary energy and momentum balances. Neither an aggregate charge-state equation nor a fitted excitation temperature determines the pathway fluxes.
\endgroup

\paragraph{An explicit stationary two-bath calculation.}
For a two-level system with fixed collider densities, set $c_\alpha=n_{z\alpha}/N_{\mathrm A}$ and $c_\beta=n_{z\beta}/N_{\mathrm A}$, and let $c_*=c_\alpha+c_\beta>0$. Choose constant positive pseudo-first-order coefficients at prescribed bath temperatures,
\[
 W^c=k_c^+c_\alpha-k_c^-c_\beta,
 \qquad
 \frac{k_c^+}{k_c^-}=K_c
 :=\frac{g_{z\beta}}{g_{z\alpha}}
       \exp\!\left(-\frac{\Delta E^{\mathrm m}}{RT_c}\right),
 \qquad c\in\{\mathrm e,B\},
\]
where $T_B=T_{\mathrm h}$ and $\Delta E^{\mathrm m}=\Delta E_{z,\alpha\beta}^{\mathrm m}>0$. Each $k_c^\pm$ has units $\mathrm{s}^{-1}$ and includes the prescribed collider density. These one-way rates obey the threshold-only local-detailed-balance relations. The population equation is
\[
 \dot c_\beta=(k_{\mathrm e}^++k_B^+)c_*
              -\Lambda c_\beta,
 \qquad
 \Lambda=k_{\mathrm e}^++k_{\mathrm e}^-+k_B^++k_B^->0.
\]
It has the unique interior stationary population and cycle current
\begin{align}
 \frac{c_\beta^{\mathrm{st}}}{c_*}
   &=\frac{k_{\mathrm e}^++k_B^+}{\Lambda},
 &
 J&=\frac{c_*(k_{\mathrm e}^+k_B^--k_{\mathrm e}^-k_B^+)}{\Lambda}
   =\frac{c_*k_{\mathrm e}^-k_B^-(K_{\mathrm e}-K_B)}{\Lambda}.
 \label{eq:explicit-two-bath-stationary-cycle}
\end{align}
Moreover, $c_\beta(t)-c_\beta^{\mathrm{st}}=
(c_\beta(0)-c_\beta^{\mathrm{st}})e^{-\Lambda t}$. Since $K_c$ increases with $T_c$, $J$ has the sign of $T_{\mathrm e}-T_{\mathrm h}$. Substitution into \cref{eq:state-excitation-cycle-dissipation} proves strictly positive production whenever the bath temperatures differ, despite vanishing population sources. This relaxation statement concerns only the prescribed-bath two-level equation, not the full Class-III PDE.

\paragraph{Calculation and reproducibility.}
Table~\ref{tab:two-bath-cycle} is obtained by direct evaluation of \cref{eq:explicit-two-bath-stationary-cycle,eq:state-excitation-cycle-dissipation} with the parameters stated in Section~\ref{sec:saha-overview}. ChatGPT (OpenAI; 28 October 2026) assisted with the two-level calculation and auxiliary consistency checks.

\subsection{Kinetic input, truncation, and relation to collisional--radiative models}
\label{subsec:ionization-kinetic-input}

The balances constrain source increments through conservation, and the entropy inequality constrains their entropy pairing. Once the increments and the local-detailed-balance law \cref{eq:state-event-local-detailed-balance} are specified, one one-way rate fixes the other. The finite level set, forward state-to-state coefficients, residual energy increments, and momentum inheritance require collision theory, state-resolved kinetics, experiment, or collisional--radiative calculations. Three-temperature models retaining metastable and excited populations, multiple ionization mechanisms, reverse recombination, and energy exchange provide such parameterization data~\cite{AikenBoydAdamovich2025}. Annaloro et al.~\cite{AnnaloroEtAl2017} found that generalized two-temperature mass-action relations are nonunique and examined the effect of kinetic pathways on the calculated compositions.

A finite level set or justified occupation-probability regularization is necessary. Otherwise an atomic partition sum may diverge. Slowly connected metastable groups should remain resolved. Strongly non-Maxwellian electrons require distribution-dependent kinetic data~\cite{ShenEtAl2025}. Changing a forward coefficient alone does not extend the Maxwell--Boltzmann derivation. The distribution or reservoir state, reverse rates, and entropy accounting must also be specified consistently, and the displayed Saha formulas are not asserted for that extension. Radiation requires its own balance or a prescribed photon reservoir. Within these restrictions, the construction gives an entropy-admissible thermochemical model and reduction to stationary Saha relations.

\section{\texorpdfstring{{Mass--energy Liu consistency calculation for the selected Class-III state space}}{Mass--energy Liu consistency calculation for the selected Class-III state space}}
\label{app:liu-consistency}

This a posteriori check uses the classical multiplier method~\cite{Liu1972} with the variables, thermodynamic state, balances, and mechanism selection prescribed in \cref{sec:balances,sec:entropy-principle}. Set
\begin{equation}
  z_i:=\rho_i e_i,
  \qquad
  \eta:=\rho s=\widehat{\rho s}(z_1,\ldots,z_N,\rho_1,\ldots,\rho_N),
\label{eq:liu-selected-state}
\end{equation}
and use the barycentric derivative $\mathrm D/\mathrm Dt$ of \cref{eq:barycentric-material-derivative}. The calculation checks that the Liu augmentation reproduces the conjugates and residual inequality of \cref{sec:entropy-exploitation}. Admissible entropy flux, physical mechanism representation in Principle~(IV), and balance anchoring remain prior requirements.

\subsection{Mass--energy Liu equations}

The partial mass and objective nonkinetic-energy balances, equivalently expressed by \cref{eq:mass-derivative-elimination,eq:energy-derivative-elimination}, are
\begin{align}
  \mathcal C_i^{\rho}
  :={}&\frac{\mathrm D\rho_i}{\mathrm Dt}+\rho_i\Div\bv+\Div(\rho_i\bu_i)-r_i=0,
\label{eq:liu-mass-constraint}
\\
  \mathcal C_i^{e}
  :={}&\frac{\mathrm Dz_i}{\mathrm Dt}+z_i\Div\bv+\Div(z_i\bu_i+\bq_i)
  -\bS_i:\grad\bv_i-\varepsilon_i+\bu_i\cdot\bmom_i
  -\half r_i\bu_i^2=0.
\label{eq:liu-energy-constraint}
\end{align}

\begingroup
The second relation follows from the mass--momentum--total-energy system by \cref{eq:partial-internal-energy-objective}. Equivalent smooth-process constraint sets retain the momentum balances. Mass and internal energy alone do not suffice. In primitive variables $(\rho_i,z_i,\bv_i)$, entropy is independent of $\bv_i$, and the constitutive class has no dependence on velocity storage rates. Their coefficients in the full augmentation therefore set the momentum multipliers to zero. The mass--energy augmentation is the remaining reduced calculation, not an independent balance.
\endgroup
  This consistency check uses primitive variables and a particular balance representation. Its vanishing momentum multipliers therefore do not imply vanishing momentum components of the conservative main-field block in~\cite{BotheRuggeriPartII}. It neither establishes spatial flux-potential compatibility nor determines the complete main field, which also includes multipliers for zero-storage constitutive equations.

{Introduce Liu multipliers $a_i$ and $b_i$ for these balance equations and consider}
\begin{equation}
  \mathcal I_{\mathrm L}
  :=\frac{\mathrm D\eta}{\mathrm Dt}+\eta\Div\bv+\Div\bPhi
  -\sum_i a_i\mathcal C_i^{\rho}
  -\sum_i b_i\mathcal C_i^{e}\ge0.
\label{eq:liu-augmented-inequality}
\end{equation}

\begingroup
Vary storage rates independently before imposing balances, holding state and spatial derivatives fixed; this does not assert independence along solutions. For \cref{eq:liu-selected-state}, canceling the coefficients of $\mathrm D\rho_i/\mathrm Dt$ and $\mathrm Dz_i/\mathrm Dt$ gives
\endgroup

\begin{equation}
  a_i=\frac{\partial\eta}{\partial\rho_i}
      =-\frac{\mu_i}{T_i},
  \qquad
  b_i=\frac{\partial\eta}{\partial z_i}
      =\frac1{T_i}.
\label{eq:liu-basic-multipliers}
\end{equation}
These are the conjugates of \cref{eq:entropy-conjugates}. Substituting \cref{eq:liu-basic-multipliers} into \cref{eq:liu-augmented-inequality} cancels storage rates and leaves
\begin{align}
\mathcal I_{\mathrm L}={}&\Div\bPhi
+\Bigl(\eta-\sum_i\frac{z_i}{T_i}+\sum_i\frac{\rho_i\mu_i}{T_i}\Bigr)\Div\bv
+\sum_i\frac1{T_i}\bS_i:\grad\bv_i
\nonumber\\
&-\sum_i\frac1{T_i}\Div(z_i\bu_i+\bq_i)
+\sum_i\frac{\mu_i}{T_i}\Div(\rho_i\bu_i)
\nonumber\\
&+\sum_i\Bigl[
\frac1{T_i}\,\varepsilon_i-\frac1{T_i}\bu_i\cdot\bmom_i
+\frac{r_i}{2T_i}\bu_i^2-\frac{\mu_i}{T_i}r_i
\Bigr].
\label{eq:liu-residual-before-products}
\end{align}
This is \cref{eq:entropy-before-product-rules}. The same two product rules give \cref{eq:entropy-first-representation} identically. The residual entropy inequality is thus recovered under the prior state assumption \cref{eq:liu-selected-state}.

\section*{Declaration of competing interests}
The authors declare no known competing interests.

\section*{Data availability}
No empirical data were used.

\section*{Declaration of generative AI and AI-assisted technologies in manuscript preparation}
During preparation of this manuscript, the authors used ChatGPT (OpenAI) to assist with editing, literature searches, and mathematical consistency checks, including the illustrative two-level calculation. The authors take full responsibility for the content of the manuscript.

\end{document}